\documentstyle[epsfig,12pt,twoside]{article}

\begin{document}

\setcounter{page}{1}

\begin{center}
{\Large {\bf
Microscopic Studies on Two-Phonon
Giant Resonances
}}\\[1cm]

{\Large
  C.A. Bertulani$^{a}$ and V.Yu. Ponomarev$^{b}$} \\
  \bigskip
$^a$Instituto de F\'\i sica,
Universidade Federal do Rio de Janeiro 
21945-970 Rio de Janeiro, RJ, Brazil, 
(E-mail: bertu@if.ufrj.br)\\
$^b$Bogoliubov Lab. of Theor.
Phys., 
Joint Inst. for Nucl. Res., 141980, Dubna, Russia,\\
(E-mail: vlad@thsun1.jinr.ru) \\
  \bigskip
  \bigskip
\end{center}

\centerline{\bf ABSTRACT}
\begin{quotation}
\vspace{-0.10in}
{\small
A new class of giant resonances in nuclei, namely double giant
resonances, is discussed. They are giant resonances built on top of
other giant resonances.
Investigation on their properties, together with similar studies
on low-lying two-phonon states, should give an answer on how far
the harmonic picture of boson-type excitations holds in the finite fermion
systems like atomic nuclei.

The main attention in this review is paid to double
giant dipole resonances (DGDR) which are observed in relativistic heavy
ion collisions with very large cross sections. A great
experimental and theoretical effort is underway to
understand the reaction mechanism which leads to the excitation of these
states in nuclei, as well as the better microscopic understanding of their
properties.
The Coulomb mechanism of the excitation of single and double giant
resonances in heavy ion collision at different projectile energies is
discussed in details. A contribution of the nuclear excitation to
the total cross section of the reaction is also considered.
The Coulomb excitation of double resonances is described within both,
the second-order perturbation theory approach and in coupled-channels
calculation.
The properties of single and double resonances are considered within
the phenomenologic harmonic vibrator model and microscopic
quasiparticle-RPA approach. For the last we use the Quasiparticle-Phonon
Model (QPM) the basic ideas and formalism of which are presented.
The QPM predictions of the DGDR properties (energy centroids, widths,
strength distributions, anharmonicities and excitation cross sections)
are compared to predictions of harmonic vibrator model, results of other
microscopic calculations and experimental data available. }
\end{quotation}

\noindent
{\it Pacs:} 24.30.Cz, 25.70.De, 21.60.-n  

\tableofcontents

\newpage

\vspace*{-5mm}

\section{\bf Introduction}

The phenomenon of a Giant Resonance (GR) in a nucleus is now known for more
than 60 years. The first article on this subject was published
in 1937 by Bothe and Gentner
\cite{BG37} who observed an unexpectedly large absorption of 17.6 MeV
photons (from the $^7$Li$(p,\ \gamma )$ reaction) in some targets.
They noticed that the cross section for $^{63}$Cu was surprisingly high
and they suggested that this might be due to a resonance phenomenon. These
observations were later confirmed by Baldwin and Klaiber (1947) with photons
from a betatron. In 1948 Goldhaber and Teller \cite{GT48} interpreted these
resonances (named by isovector giant dipole resonances) with a
hydrodynamical model in which rigid proton and neutron fluids vibrate
against each other, the restoring force resulting from the surface energy.
Steinwendel and Jensen \cite{SJ50} later developed the model, considering
compressible neutron and proton fluids vibrating in opposite phase in a
common fixed sphere, the restoring force resulting from the volume symmetry
energy. The standard microscopic basis for the description of giant
resonances is the Random Phase Approximation (RPA) in which giant resonances
appear as coherent superpositions of one-particle one-hole ($1p1h$)
excitations in closed shell nuclei or two quasi-particle excitations in open
shell nuclei (for a review of these techniques, see, e.g., Ref.
\cite{SW91}).

The isoscalar quadrupole resonances were discovered in inelastic electron
scattering by Pitthan and Walcher (1971) and in proton scattering by Lewis
and Bertrand \cite{Ber76}. Giant monopole resonances were found later
and their
properties are closely related to the compression modulus of nuclear matter.
Following these, other resonances of higher multipolarities and giant
magnetic resonances were investigated. Typical probes for giant resonance
studies are (a) $\gamma $'s and electrons for the excitation of GDR
(isovector giant dipole resonance), (b) $\alpha $-particles and electrons
for the excitation of isoscalar GMR (giant monopole resonance) and GQR
(giant quadrupole resonance), and (c) ($p,n$), or ($^3$He, t), for
Gamow-Teller resonances, respectively.

Relativistic Coulomb Excitation (RCE) is a well established tool
to unravel interesting aspects of nuclear structure \cite{BB88}.
Examples are the studies of multiphonon resonances
in the SIS accelerator at the GSI facility, in Darmstadt,
Germany \cite{Em94,CF95,Aum98}. Important properties of
nuclei far from stability \cite{Ha93,BCH93} have also been studied
with this method.

Inelastic scattering studies with heavy ion beams have opened new
possibilities in the field (for a review the experimental developments, see
Ref. \cite{Em94,Aum98}). A striking feature was observed when either the beam
energy was increased, or heavier projectiles were used, or both
\cite{Bar88}. This is displayed in Fig. \ref{a-f1}, where the
excitation of the GDR in
$^{208}$Pb was observed in the inelastic scattering of $^{17}$O at
22A MeV and 84A MeV, respectively, and $^{36}$Ar at 95A MeV
\cite{Bee90,Bee93}. What one clearly sees is that the ``bump''
corresponding to the GDR at 13.5 MeV is appreciably enhanced. This feature
is solely due to one agent: the electromagnetic interaction between the
nuclei. This interaction is more effective at higher energies, and for
increasing charge of the projectile.

Baur and Bertulani showed in Ref.~\cite{BB86}  that the excitation
probabilities of the
GDR in heavy ion collisions approach unity at grazing impact parameters. They
also showed that, if double GDR resonance (i.e. a GDR excited on a GDR
state) exists then the cross sections for their excitation in heavy ion
collisions at relativistic energies are of order of a few hundred of 
milibarns.
These calculations motivated experimentalists at the GSI \cite{Em94,Aum98} and
elsewhere \cite{Bee90,Bee93} to look for the signatures of the DGDR in the
laboratory. This has by now become a very active field in nuclear physics
with a great theoretical and experimental interest \cite{Em94,CF95,Aum98}.

In the first part of this review, we study the reactions mechanism
in RCE of giant resonances in several collisions between
heavy ions. In section 2 we start with the description of the
semiclassical theory for relativistic Coulomb excitation, then we consider
the effects of recoil and later we describe the fully quantum mechanical
approach. The role of nuclear excitation in relativistic heavy ion
collisions is also discussed here. We demonstrate that the experimental
data on the excitation and decay of single giant resonances are well
described by these formalisms.
In section 3 the process of the excitation of multi-phonon resonances
in relativistic heavy ion collisions is considered within the
second-order perturbation theory and in coupled-channels calculations.
Giant resonances are treated in this section within the phenomenologic
harmonic vibrator model. Some general arguments for the width of
multi-phonon resonances are discussed here as well as an influence of
giant resonances width on the total cross section of their excitation.
A good part of this report (sections 4 and 5) is dedicated to a
review of the microscopic properties on the giant resonances in the
Quasiparticle-Phonon Model. In section 4 we present the main ideas and
formalism of this model. The particle-hole modes of nuclear excitation
are projected into the space of quasi-bosons, phonons, and matrix
elements of interaction between one- and multi-phonon configurations are
calculated on
microscopic footing within this approach. In section 5 we use this model
as a basis for a detailed investigation of the interplay between
excitation mechanisms and the nuclear structure in the excitation of the
DGDR. Different aspects related to the physical properties of the DGDR
in heavy nuclei (energy centroids, widths, strength distributions,
anharmonicities and excitation cross sections) as predicted by
microscopic studies are discussed in this section and compared to
experimental data.

\section{\bf Heavy ion excitation of giant resonances}

\subsection{Coulomb excitation at relativistic energies}

In relativistic heavy ion collisions, the wavelength associated to the
projectile-target separation is much smaller than the
characteristic lengths of system. It is, therefore, a reasonable
approximation to treat ${\bf r}$ as a classical variable ${\bf
r}(t)$, given at each instant by the trajectory followed by the
relative motion. At high energies, is also a good approximation
to replace this trajectory by a straight line.  The intrinsic
dynamics can then be handled as a quantum mechanics problem with
a time dependent Hamiltonian. This treatment is discussed in full
details by Alder and Winther in Refs.~\cite{AW66,AW75,WA79}. We will
describe
next the formalism developed by Canto {\it et al.} \cite{BCH96}, which
explicitly gives the time-dependence of the multipole fields, useful for
a coupled-channels calculation.

The intrinsic state $\vert \psi(t)>$ satisfies the Schr\"odinger equation
\begin{equation}
\left[ h\ +\ V({\bf r}{\scriptstyle (t)}) \right]\ \vert \psi(t)> =
i \hbar {\partial \vert \psi(t)>
\over \partial t}\; . \label{eqS}
\end{equation}
Above, $h$ is the intrinsic Hamiltonian and $V$ is the
channel-coupling interaction.

Expanding the wave function in the set $\{ \vert m>;\ m=0,N\}$ of eigenstates
of $h$, where $N$ is the number of excited states included in the
Coupled-Channels problem, we obtain
\begin{equation}
\vert \psi (t)> = \sum_{m=0}^N\ a_m(t)\ \vert m>\ \exp \Big(-i
E_m t /\hbar\Big) \; , \label{expan}
\end{equation}
where $E_m$ is the energy of the state $\vert m>$. Taking scalar product
with each of the states $<n \vert $, we get the set of coupled equations
\begin{equation}
i \hbar\ {\dot a}_n(t) = \sum_{m=0}^N\ < n\vert V\vert m >
\;
e^{i(E_n-E_m) t /\hbar}\; a_m(t)
\qquad \qquad n =0\;\; {\rm to}\;\;  N \,.\label{AW}
\end{equation}
It should be remarked that the amplitudes depend also
on the impact parameter $b$ specifying the classical trajectory followed
by the system. For the sake of keeping the notation simple, we do not
indicate this dependence explicitly. We write, therefore, $a_n(t)$ instead
of $a_n(b,t)$. Since the interaction $V$ vanishes as $t\rightarrow \pm
\infty$, the amplitudes have as initial condition $a_n(t\rightarrow -\infty)
= \delta(n,0)$ and they tend to constant values as
$t\rightarrow \infty$. Therefore, the excitation probability of an intrinsic
state $\vert n> $ in a collision with impact parameter $b$ is given as
\begin{equation}
P_n(b) = \vert a_n(\infty)\vert ^2\; .\label{Pn}
\end{equation}
The total cross section for excitation of the state $\vert n>$ can
be approximated by the classical expression
\begin{equation}
\sigma_n=2\pi\ \int\ P_n(b)\ b db \;. \label{sigman}
\end{equation}

Since we are interested in the excitation of
specific nuclear states, with good angular momentum and parity quantum
numbers, it is appropriate to develop the time-dependent coupling interaction
$V(t)$ into multipoles. In Ref.~\cite{WA79}, a multipole expansion of the
electromagnetic excitation amplitudes in relativistic heavy ion collisions
was carried out. This work used first order perturbation theory and the
semiclassical approximation. The time-dependence of the multipole
interactions was not explicitly given. This was accomplished in
Ref. \cite{BCH96}, which we describe next.

We consider a nucleus 2 which is at rest and a relativistic
nucleus 1 which moves along the $z$-axis and is excited from the
initial state $|I_iM_i>$  to the state $|I_fM_f>$ by the electromagnetic
field of nucleus 1.  The nuclear states are specified by the
spin quantum numbers $I_i$, $I_f$ and by the corresponding
magnetic quantum numbers $M_i$, $M_f$, respectively. We
assume that the relativistic nucleus 1 moves along a
straight-line trajectory with impact parameter $b$, which is
therefore also the distance of the closest approach between the
center of mass of the two nuclei at the time $t=0$. We shall
consider the situation where $b$ is larger than the sum of the
two nuclear radii, such that the charge  distributions of the
two nuclei do not strongly overlap at any time.  The
electromagnetic field of the nucleus 1 in the reference frame of
nucleus 2 is given by the usual Lorentz transformation
\cite{Ja75} of the scalar potential $\phi({\bf r})= Z_1e/|{\bf
r}|$, i.e.,
\begin{eqnarray}
\phi({\bf r}', t) &=& \gamma \ \phi\Big[{\bf b'-b}, \ \gamma (z'-vt)\Big],
\nonumber \\
{\bf A}({\bf r}', t) &=&
{{\bf v}\over c}\ \gamma \ \phi\Big[{\bf b'-b}, \ \gamma (z'-vt)\Big].
\label{Aphi}
\end{eqnarray}
Here ${\bf b}$ (impact parameter) and ${\bf b}'$ are the components
of the radius-vectors ${\bf r}$  and
${\bf r}'$ transverse to ${\bf v}$.

The time-dependent matrix element for electromagnetic excitation
is of the form
\begin{equation}
V_{fi}(t)=<I_fM_f| \Big[ \rho({\bf r}')-{{\bf v}\over c^2} \cdot {\bf J}
({\bf r}') \Big] \ \phi({\bf r}', \ t) |I_iM_i> \ ,\label{Vfi}
\end{equation}
where $\rho$ (${\bf J}$) is the nuclear transition density (current).

A Taylor-series expansion of the Li\'enard-Wiechert
potential around ${\bf r}'=0$ yields
\begin{equation}
\phi({\bf r}', \ t)=\gamma \phi[{\bf r}(t)]+\gamma \nabla \phi[{\bf
r}(t)]\cdot {\bf r}'
+ \cdots\label{phiexp}
\end{equation}
where  ${\bf r}=({\bf b}, \ \gamma v t)$, and the following
simplifying notation is used:
\begin{eqnarray}
\nabla \phi[{\bf r}]& \equiv& \nabla' \phi({\bf r}', \ t)
\bigg|_{{\bf r}'=0}
\nonumber\\
&=& - \nabla_{\bf b} \phi({\bf r}) - {\partial \over \partial (vt)}
\phi({\bf r}) \ \hat{\bf z} =
- \nabla_{\bf b} \phi({\bf r}) - {{\bf v} \over  c^2} \
{\partial \over \partial t} \phi({\bf r})\ .\label{nphi}
\end{eqnarray}
Thus,
\begin{equation}
V_{fi}(t)=<I_fM_f| \Big[ \rho({\bf r}')-{{\bf v}\over c^2} \cdot {\bf J}
({\bf r}') \Big] \
\Big[ \gamma \phi({\bf r})+\gamma {\bf r}'\cdot \nabla \phi({\bf r})
\Big]|I_iM_i> \ .\label{Vfi1}
\end{equation}
Using the continuity equation
\begin{equation}
\nabla \cdot {\bf J}=-i \ \omega \ \rho\ ,\label{ce}
\end{equation}
where $\omega=(E_f-E_i)/\hbar$, and integrating by parts,
\begin{equation}
V_{fi}(t)=<I_fM_f| \Big[ {\bf J}({\bf r}) \cdot \Big[
{\nabla' \over i\omega} -{{\bf v} \over c^2}\Big]
\
\Big[ \gamma \phi({\bf r}+\gamma {\bf r}'\cdot \nabla \phi({\bf r})
\Big]|I_iM_i> \ .\label{Vfi2}
\end{equation}

In spherical coordinates
\begin{equation}
{\bf r}' \cdot \nabla \phi={\sqrt{4\pi}\over 3} \
\sum_{\mu=-1}^1 \alpha_\mu \ r'\ Y_{1\mu}^*
\ ,\label{rgrad}
\end{equation}
where
\begin{equation}
\alpha_\mu=\hat{\bf e}_\mu \cdot \nabla \phi\label{alpha},
\end{equation}
and $\hat{\bf e}_\mu$ are the spherical unit vectors
$$
\hat{\bf e}_\pm=
\mp {1 \over \sqrt{2}} \Big( \hat{\bf e}_X\pm\hat{\bf e}_Y\Big),
\ \ \ \ \ \hat{\bf e}_0=\hat{\bf e}_Z \ .
$$
We will use the relations
\begin{equation}
{{\bf r}\over c^2}={v\over c^2} \hat{\bf e}_0 =
{v\over c^2} \ \sqrt{4\pi \over 3} \
\nabla (rY_{10}^*) \label{rc2}
\end{equation}
and
\begin{equation}
\nabla \times {\bf L} (r^k Y_{lm})=i(k+1) \nabla
(r^kY_{lm})\label{nL}
\end{equation}
where ${\bf L} = -i \nabla \times {\bf r}$.

Then, one can write
\begin{eqnarray}
{\bf J}\cdot \Big({\nabla \over i\omega} -{{\bf v}\over c^2}\Big)
\ \Big[ \gamma \phi +\gamma {\bf r}' \cdot \nabla \phi\Big]&=&
- \gamma {\bf J} \cdot \bigg[
{{\bf v}\over c^2} (\nabla \phi \cdot {\bf r}')
\ -\ \sqrt{4\pi \over 3} \nonumber\\
&\times& \Big\{
\sum_{\mu=-1}^1 {\alpha_\mu\over i\omega} \
\nabla' (r'Y_{1\mu})-{v\over c^2} \
\phi \ \nabla' (r'Y_{10}^*)\Big\}\
\bigg]
\ .\label{J}
\end{eqnarray}
The last term in the above equation can be rewritten as
\begin{eqnarray}
\Big({\bf J}\cdot {{\bf v} \over c^2}
\Big) \ \Big({\bf r}'\cdot \nabla \phi\Big)
&=& {v\over 2c^2} \ {\bf J} \cdot
\Big[ \hat{\bf e}_0 ({\bf r}'\cdot \nabla \phi)+({\bf r}'\cdot \hat{\bf
e}_0)\nabla \phi \Big]
\nonumber \\
&+& {v\over 2c^2} \ {\bf J} .
\Big[ \hat{\bf e}_0 ({\bf r}'\cdot \nabla \phi)-({\bf r}'\cdot \hat{\bf
e}_0)\nabla \phi
\Big]
\ .\label{Jvc}
\end{eqnarray}
The first term in this equation is symmetric under parity inversion,
and contributes to the electric quadrupole ($E2$) excitation amplitudes,
since
\begin{equation}
 {v\over 2c^2} \ {\bf J} \cdot
\Big[ \hat{\bf e}_0 ({\bf r}'\cdot \nabla \phi)+({\bf r}'\cdot \hat{\bf
e}_0)\nabla \phi
\Big]=
 {v\over 2c^2} \ {\bf J} \cdot \nabla' \Big[ z' ({\bf r}' \cdot \nabla \phi)
 \Big]
 \ .\label{v2c2}
 \end{equation}
 The second term in Eq.~(\ref{Jvc}) is antisymmetric in $\bf J$ and $\bf r'$,
 and leads to magnetic dipole ($M1$) excitations. Indeed, using
 Eqs.~(\ref{rgrad}--\ref{nL}), one finds
 \begin{equation}
 {v\over 2c^2} \ {\bf J} \cdot
\Big[ \hat{\bf e}_0 ({\bf r}'\cdot \nabla \phi)-({\bf r}'\cdot \hat{\bf
e}_0)\nabla \phi
\Big]=
 {v\over 2c^2} \ {\bf J} \cdot  \Bigg[
\sqrt{4\pi \over 3} \sum_{\mu=-1}^1 \alpha_\mu (-1)^\mu \
{\bf L} \Big({\bf r} Y_{1,-\mu}\Big)
 \Bigg]
 \ .\label{v2c3}
 \end{equation}
Thus, only the first two terms on the right-hand-side
of Eq.~(\ref{J}) contribute to the
electric dipole ($E1$) excitations. Inserting them into
Eq.~(\ref{Vfi2}), we get 
\begin{equation}
V^{(E1)}_{fi}(t)=\gamma \ \sqrt{4\pi\over 3} \
\sum_{\mu=-1}^1
(-1)^\mu \ \beta_\mu \
<I_fM_f| {\cal M} (E1, -\mu) |I_iM_i>
\ ,\label{Vfi3}
\end{equation}
where
\begin{equation}
{\cal M} (E1, -\mu) = {i\over \omega} \ \int d^3r {\bf J}({\bf r})
. \nabla \Big(rY_{1\mu}\Big)= \int d^3 r \ \rho({\bf r}) \ r \ Y_{1\mu}
({\bf r})
\ ,\label{ME1}
\end{equation}
and
\begin{eqnarray}
\beta_{\pm}&=&-\alpha_\mu = - \big( \nabla \phi
\cdot \hat{\bf e}_\mu \big) = \hat{\bf e}_\mu \cdot {\partial \phi \over
\partial {\bf b}} \nonumber \\
\beta_0 &=& -\alpha_0 - i {\omega v \over c^2} \phi
\ .\label{beta}
\end{eqnarray}
The derivatives of the potential $\phi$ are explicitly
given by
\begin{eqnarray}
{\partial \phi \over \partial {\bf b}_x} &\equiv&
\nabla_{{\bf b}_x} \phi \Big|_{{\bf r}'=0}= - \hat{\bf x}  \
b_x {Z_1e \over [b^2+\gamma^2v^2t^2]^{3/2}}\nonumber \\
\nabla_z \phi\Big|_{{\bf r}'=0}&=& - \hat{\bf z} \ \gamma^2 v t \
{Z_1e\over [b^2+\gamma^2v^2t^2]^{3/2}}
\ .\label{deriv}
\end{eqnarray}

Using the results above, we get for the electric dipole
potential
\begin{eqnarray}
V^{(E1)}_{fi}(t)&=& \sqrt{2\pi\over 3}\ \gamma \
\Bigg\{ {\cal E}_1 (\tau) \ \Big[ {\cal M}_{fi} (E1, -1) -
{\cal M}_{fi} (E1, 1)\Big]
\nonumber \\
&+&\sqrt{2} \ \gamma \ \ \bigg[ \tau{\cal E}_1(\tau)
-
i{\omega v \over \gamma c^2}
\ {\cal E}_2(\tau)
\bigg] \ {\cal M}_{fi} (E1, 0) \Bigg\}
\ ,\label{Vfi4}
\end{eqnarray}
where $\tau=\gamma v/b$, and
\begin{equation}
{\cal E}_1(\tau)={ Z_1 e  \over
b^2 \ [1+\tau^2]^{3/2}} \
\ \ \ \
{\rm and} \  \ \
{\cal E}_2(\tau)={ Z_1 e  \over b \ [1+\tau^2]^{1/2}} \label{E1}
\end{equation}
are the transverse and longitudinal electric fields
generated the relativistic nucleus with charge $Z_1e$, respectively.
 From the definition
\begin{equation}
{\cal M}_{fi} (M1, \mu)= - {i\over 2c}\ \int d^3r \
{\bf J}
({\bf r}) . {\bf L} \Big( r Y_{1\mu}\Big)
\ ,\label{Mfi}
\end{equation}
and Eq.~(\ref{v2c2}),  we find
\begin{equation}
V^{(M1)}_{fi}(t)= i \ \sqrt{2\pi\over 3}\ {v\over c} \gamma \
{\cal E}_1 (\tau) \ \Big[ {\cal M}_{fi} (M1, 1)
+
{\cal M}_{fi} (M1, -1)\Big]
\ .\label{Vfi5}
\end{equation}
The current ${\bf J}$ in Eq. (\ref{Mfi}) is made up of the usual convective
part and a magnetization part, proportional to the intrinsic (Dirac and
anomalous) magnetic moment of the nucleons.

To obtain the electric quadrupole ($E2$) potential we use the
third term in the Taylor expansion of Eq.~(\ref{phiexp}). Using the
continuity equation, a part of this term will contribute to $E3$
and $M2$ excitations, which we neglect. We then find that
\begin{eqnarray}
V^{(E2)}_{fi}(\tau)&=& -\sqrt{\pi\over 30}\ \gamma \
\bigg\{ 3 \ {\cal E}_3 (\tau) \
\Big[ {\cal M}_{fi} (E2, 2) +
{\cal M}_{fi} (E2, - 2)\Big] \nonumber \\
&+&\gamma \Big[ 6\ \tau  {\cal E}_3(\tau) - i \
{\omega v\over \gamma c^2}
\ {\cal E}_1(\tau) \Big]
\ \Big[ {\cal M}_{fi} (E2, -1) +
{\cal M}_{fi} (E2, 1)\Big]\nonumber \\
&+&
\sqrt{6} \ \gamma^2 \ \Big[ \Big( 2 \tau^2-1\Big)
\ {\cal E}_3(\tau) - i \ {\omega v \over \gamma c^2} \
\tau \  {\cal E}_1(\tau)\Big]
{\cal M}_{fi} (E2, 0)
\bigg\}
\ ,\label{Vfi6}
\end{eqnarray}
where
${\cal E}_3(\tau)$ is
the quadrupole electric field of nucleus 1, given by
\begin{equation}
{\cal E}_3(\tau)={ Z_1 e
\over b^3\ [1+\tau^2]^{5/2}}
\ .\label{E3}
\end{equation}

The fields ${\cal E}_i(\tau)$ peak around $\tau=0$, and
decrease fastly within an interval
$\Delta \tau \simeq 1$.
This corresponds to a collisional
time $\Delta t \simeq b / \gamma v$.
This means that, numerically one needs to integrate
the Coupled-Channels equations (Eq.~(\ref{AW})) only in a
time interval within a range $n\times \Delta \tau$ around
$\tau=0$, with
$n$ equal to a small integer number.

\subsubsection{First-order perturbation theory}

In most cases, the first-order perturbation theory is a good
approximation to calculate the amplitudes for relativistic
Coulomb excitation. It amounts to using $a_k=\delta_{k0}$ on the right
hand side of
Eq.~(\ref{AW}). The time integrals can be evaluated analytically
for the $V_{Ei}(t)$ perturbations, given by Eqs. (\ref{Vfi4}),
(\ref{Vfi5}), and (\ref{Vfi6}).  The result is
\begin{eqnarray}
a^{(E1)}_{1st}&=& - i \sqrt{8\pi\over 3}\ {Z_1 e \over
\hbar v b} \ \xi \ \
\Bigg\{ K_1(\xi) \ \Big[ {\cal M}_{fi} (E1, -1) -
{\cal M}_{fi} (E1, 1)\Big]
\nonumber \\
&+& i {\sqrt{2} \over \gamma} \ K_0(\xi) \
{\cal M}_{fi} (E1, 0) \Bigg\}
\ ,\label{a1st}
\end{eqnarray}
where $K_1$ ($K_2$) is the modified Bessel function of first (second) degree,
and $\xi=\omega b/\gamma v$. For the $E2$ and $M1$ multipolarities, we
obtain respectively,
\begin{eqnarray}
a^{(E2)}_{1st}&=& 2 i \ \sqrt{\pi\over 30}\ {Z_1 e \over
\gamma \hbar vb^2} \ \xi^2
\bigg\{ K_2(\xi)\
\Big[ {\cal M}_{fi} (E2, 2) +
{\cal M}_{fi} (E2, - 2)\Big] \nonumber \\
&+&i \gamma \ \Big(2-{v^2\over c^2}\Big) \ K_1(\xi)
\ \Big[ {\cal M}_{fi} (E2, -1) +
{\cal M}_{fi} (E2, 1)\Big]\nonumber \\
&-&
\sqrt{6} \
K_0(\xi) \
{\cal M}_{fi} (E2, 0)
\bigg\}
\ ,\label{aE2}
\end{eqnarray}
and
\begin{equation}
a^{(M1)}_{1st}= \sqrt{8\pi\over 3}\ {Z_1e\over \hbar c b} \ \xi
\ K_1(\xi) \ \Big[ {\cal M}_{fi} (M1, 1)
-
{\cal M}_{fi} (M1, -1)\Big]
\ .\label{aM1}
\end{equation}

These expressions are the same as those obtained from the
formulae deduced in Ref. \cite{WA79}. We note that the multipole
decomposition developed by those authors is accomplished by a
different approach, i.e., using recurrence relations for the
Gegenbauer polynomials, after the integral on time is performed.
Therefore, the above results present a good check for the
time-dependence of the multipole fields deduced here.

The formulas above have been derived under the assumption of the
long-wavelength
approximation. When this approximation is not valid the matrix elements
given by Eqs. (\ref{ME1},\ref{Mfi}) are to be replaced by the non-approximated
matrix-elements for electromagnetic excitations \cite{AW66}, i.e.,
\begin{equation}
{\cal M} (E\lambda, \mu)= {(2\lambda+1)!! \over
\kappa^{\lambda+1}c(\lambda+1)}\int {\bf J}({\bf r})
\cdot \nabla \times {\bf L}\Big[j_\lambda (\kappa r)
Y_{\lambda \mu} (\hat{\bf r}) \Big]d^3r
\ ,\label{ME1lm}
\end{equation}
\begin{equation}
{\cal M} (M\lambda, \mu)= -i{(2\lambda+1)!! \over
\kappa^\lambda c(\lambda+1)}\int {\bf J}({\bf r})
\cdot {\bf L}\Big[j_\lambda (\kappa r)
Y_{\lambda \mu} (\hat{\bf r}) \Big]d^3r
\ ,\label{MM1lm}
\end{equation}
for electric and magnetic excitations ($\kappa=\omega/c$), respectively.
However, the other factors do not change (see, e.g., \cite{WA79}).

\subsubsection{Excitation probabilities and virtual photon numbers}

The square modulus of Eqs. (\ref{a1st},\ref{aE2},\ref{aM1})
gives the probability of exciting the target
nucleus from the
initial state $\mid I_i M_i>$ to the final state
$\mid I_f M_f >$ in a collision
with impact parameter  $b$. If the orientation of the initial
state is not specified, the probability for exciting the nuclear
state of energy $E_f$ and spin $I_f$ is
\begin{equation}
P_{i \rightarrow f} =  {1 \over 2I_i +1} \; \sum_{M_i , M_f }
\mid a_{fi} \mid^2
\, .\label {(4.1)}
\end{equation}
Integration of (\ref{(4.1)}) over all energy transfers $\varepsilon = \hbar
\omega$,
and summation over all possible final states of the projectile
nucleus (making use of  the Wigner-Eckart theorem and the orthogonality
of the properties of the
Clebsch-Gordan coefficients) leads to the Coulomb excitation probability in
a collision with impact parameter $b$:
\begin{equation}
P_C  = \sum_f \int P_{i
\rightarrow f} (b) \;
\rho_f (\varepsilon ) \; d\epsilon \label {(4.4)}
\end{equation}
where $\rho_f (\varepsilon )$ is the density of final states of the target
with energy $E_f = E_i + \varepsilon$.

Inserting (\ref{a1st},\ref{aE2},\ref{aM1}) into (\ref{(4.4)}) one finds
\begin{equation}
P_C (b,\varepsilon )= \sum_{\pi \lambda} \;
P_{\pi \lambda }  (b,\varepsilon ) = \sum_{\pi \lambda}
\int {d\varepsilon  \over \varepsilon } \; n_{\pi \lambda}
(b,\varepsilon ) \; \sigma_\gamma^{\pi \lambda}(\varepsilon )
\label {(4.5a)}
\end{equation}
where
\begin{equation}
\sigma_\gamma^{\pi \lambda} (\varepsilon )
= { (2 \pi )^3 (\lambda +1) \over \lambda \left\lbrack (2 \lambda +1 )!!
\right\rbrack^2 } \; \sum_f \rho_f (\varepsilon ) \, \kappa^{2 \lambda -1 }
\; B(\pi \lambda, I_i \rightarrow I_f )
\label {(4.5b)}
\end{equation}
are the photonuclear absorption cross sections for a given multipolarity
$\pi \lambda$. The total photonuclear cross section is a sum of all
these multipolarities,
\begin{equation}
\sigma_\gamma = \sum_{\pi \lambda} \sigma_\gamma^{\pi \lambda}
(\varepsilon ) \, .\label {(4.5c)}
\end{equation}
The functions $n_{\pi \lambda} (\varepsilon)$ are called the
{\it virtual photon numbers}, and are given by

\begin{equation}
n_{E1}(b,\varepsilon ) =
{Z_1^2 \alpha \over \pi^2} \, {\xi^2 \over b^2} \, ({c \over v})^2
\, \left\{K_1^2 + {1\over \gamma^2 } \, K_0^2 \right\}
\label {(A.3a)}
\end{equation}
\begin{equation}
n_{E2}(b,\varepsilon )  =
{Z_1^2 \alpha \over \pi^2 b^2} \, ({c \over v})^4
\, \left\{ {4 \over \gamma^2} \, \left\lbrack
K_1^2 + \xi K_0 K_1 + \xi^2 K_0^2 \right\rbrack
+ \xi^2 (2-v^2/c^2 )^2 K_1^2 \right\} \label {(A.3b)}
\end{equation}
and
\begin{equation}
n_{M1}(b,\varepsilon ) =
{Z_1^2 \alpha \over \pi^2} \, {\xi^2 \over b^2} \,
K_1^2
\label {(A.3c)}
\end{equation}
where all $K_\mu$'s are functions of
$\xi (b) ={\omega b \over \gamma v}$.

Since all nuclear excitation dynamics is contained in the photoabsorption
cross section, the virtual photon numbers (\ref{(A.3a)},
\ref{(A.3b)},
\ref{(A.3c)}) do not
depend on the nuclear structure. They are kinematical factors, depending
on the orbital motion. They may be interpreted as the number of equivalent
(virtual) photons that hit the target per unit
area.
These expressions show that Coulomb excitation probabilities are
exactly directly proportional to the photonuclear cross sections,
although the exchanged photons are off-shell. This arises from the
condition that the reaction is peripheral and the nuclear charge
distributions of each nuclei do not overlap during the collision.
This result can be proved from first principles, and has been
shown in some textbooks (see, eg. \cite{EG87}).

The usefulness of Coulomb excitation, even in first order processes,
is displayed in Eq. (\ref{(4.5a)}). The field
of a real photon contains all multipolarities with the same weight and the
photonuclear cross section (\ref{(4.5c)}) is a mixture of the contributions
from all multipolarities, although only a few contribute in most processes.
In the case of Coulomb excitation the total cross section is weighted by
kinematical factors which are different for each projectile or bombarding
energy.
This allows one to disentangle the multipolarities when several ones are
involved in the excitation process, except for the very high bombarding
energies $\gamma \gg 1$ for which all virtual photon numbers can be
shown to be the same \cite{BB85}.

\subsubsection{Cross sections and total virtual photon numbers}

The cross section is obtained by the impact parameter integral of the
excitation probabilities. Eq. ({\ref{(4.5a)}) shows that we
only need to integrate the
number of virtual photons over impact parameter. One has to introduce a
minimum impact parameter $b_0$ in the integration. Impact parameters
smaller than $b_0$ are dominated by nuclear fragmentation processes.
One finds
\begin{equation}d \sigma_C  = \sum_{\pi \lambda} \;
\sigma_{\pi \lambda }  = \sum_{\pi \lambda}
\int {d\varepsilon  \over \varepsilon } \; N_{\pi \lambda}
(\varepsilon ) \; \sigma_\gamma^{\pi \lambda}(\varepsilon )
\label{(4.5e)}
\, ,\end{equation}
where the {\it total virtual photon numbers}
$N_{\pi \lambda}(\varepsilon )=2\pi\int db \ b\ n(b,\varepsilon)$
are given analytically by
\begin{equation}
N_{E1}(\varepsilon ) =
{2Z_1^2 \alpha \over \pi} \, ({c \over v})^2
\, \left\lbrack
\xi K_0K_1 - {v^2\xi^2\over 2c^2 } \, (K_1^2-K_0^2) \right\rbrack
\, ,
\label {(A.6a)}
\end{equation}
\begin{eqnarray}
N_{E2}(\varepsilon )  &=&
{2Z_1^2 \alpha \over \pi} \, ({c \over v})^4
\, \left\lbrack 2(1-{v^2 \over c^2})K_1^2
+ \xi (1-{v^2 \over c^2})^2 K_0 K_1 + {\xi^2v^4\over 2c^4}
(K_1^2-K_0^2) \right. \nonumber\\
&+&  \left. \xi^2 (2-v^2/c^2 )^2 K_1^2 \right\rbrack
\label {(A.6b)}
\end{eqnarray}
and
\begin{equation}
N_{M1}(\varepsilon ) =
{2Z_1^2 \alpha \over \pi} \, {\xi^2 \over b^2}
\, \left\lbrack
\xi K_0K_1 - {\xi^2\over 2 } \, (K_1^2-K_0^2) \right\rbrack
\label {(A.6c)}
\end{equation}
where all $K_\mu$'s are now functions of
$\xi (b) ={\omega b_0 \over \gamma v}$.

\subsection{Coulomb excitation at intermediate energies}
\subsubsection{Classical trajectory: recoil and retardation corrections}

The semiclassical theory of Coulomb excitation in low energy collisions
accounts for the Rutherford bending of the trajectory, but relativistic
retardation effects are neglected \cite{AW75}. On the other hand,
in the theory of
relativistic Coulomb excitation \cite{WA79} recoil effects on the trajectory
are
neglected (one assumes straight-line motion) but retardation is handled
correctly.  In fact, the onset of retardation brings new important
effects such as the steady increase of the excitation cross sections with
bombarding energy.  In a heavy ion collision around
100A MeV the Lorentz factor $\gamma$ is about 1.1.  Since this
factor enters the excitation cross sections in many ways, like in
the adiabaticity parameter
\begin{equation}
\xi(R) = {\omega_{fi}R \over \gamma v}
 \label{3.3}
\ ,
\end{equation}
one expects that some sizable
modifications in the theory of relativistic Coulomb excitation should
occur \cite{AB89}.  Recoil corrections are not negligible either, and
the
relativistic calculations based on the straight-line parametrization should
not be completely appropriate to describe the excitation probabilities
and cross sections. The Coulomb recoil in a single collision is
of the order of
\begin{equation}
a_o = {Z_1Z_2e^2 \over m_ov^2} \ , \label{3.5}
\end{equation}
which is {\it half-distance of closest approach} in a
head-on collision, with m$_o$ equal to the reduced mass of the colliding
nuclei.  Although this recoil is small for intermediate energy collisions,
the excitation probabilities are quite sensitive to it.  This is
important for example in the excitation of giant resonances because the
adiabaticity parameter is of the order of one (see, Eq. (\ref{3.3})).
When $\xi(b)
\ll 1$, the excitation probabilities depends on $b$
approximately like $1/b^2$,
while when $\xi(b)$ becomes greater than one they decrease approximately
as $e^{-2\pi\xi(b)}/b^2$.
Therefore, when $\xi \simeq 1$
a slight change of $b$ may vary appreciably the excitation
probabilities.

In the semiclassical theory of Coulomb excitation the nuclei are assumed
to follow classical trajectories and the excitation probabilities are
calculated in time-dependent perturbation theory. At low energies one
assumes Rutherford trajectories for the relative motion
while at relativistic energies one assumes
straight-line motion. In intermediate energy
collisions, where one wants to account for recoil and retardation
simultaneously, one should solve the general classical problem of the
motion of two relativistic charged particles. But, even if radiation is
neglected, this problem can only be solved if one particle has infinite
mass \cite{LL79}. This approximation should be sufficient if we take, e.g.,
the
collision $^{16}$O + $^{208}$Pb as our system. An improved solution
may be obtained by use of the reduced mass, as we show next, in a
formalism developed by Aleixo and Bertulani \cite{AB89}.

In the classical one-body problem, one starts with the relativistic Lagrangian
\begin{equation}
{\cal L} = -m_o c^2 \left\{ 1 - {1 \over c^2 } \,
({\dot r}^2 + r^2 {\dot \phi}^2 ) \right\}^{1/2}
- {Z_1 Z_2 e^2 \over r} \label{3.6} \ ,
\end{equation}
where ${\dot r}$ and ${\dot \phi}$ are the radial and the angular
velocity of the particle, respectively (see Fig. \ref{a-f2}).
Using the Euler-Lagrange equations one finds
three kinds of solutions, depending on the sign of the charges
and the angular momentum in the collision. In the case of our
interest, the appropriate solution relating the collisional angle
$\phi$ and the distance r between the nuclei is \cite{LL79}
\begin{equation}
{1 \over r} = A \left\lbrack  \epsilon \; \cos ( W\phi)
- 1 \right\rbrack
\label{3.7}
\end{equation}
where
\begin{equation}
W = \left\lbrack 1 - ({Z_1Z_2e^2 \over c{L}_0})^2 \right\rbrack ^{1/2}\, ,
\label{3.8a}
\end{equation}
\begin{equation}
A= {Z_1Z_2e^2E \over c^2{L}^2_0 W^2}\, , \label{3.8b}
\end{equation}
\begin{equation}
\epsilon = {c{L}_0 \over Z_1Z_2e^2E}\, \left\lbrack E^2-m^2_o c^4 +
({m_o cZ_1Z_2e^2 \over {L}_0})^2 \right\rbrack ^{1/2} \, .\label{3.8c}
\end{equation}
$E$ is the total bombarding energy in MeV,
$m_o$ is the mass of the particle and ${L}_0$
its angular momentum. In terms of the Lorentz factor $\gamma$
and of the impact parameter $b$, $E=\gamma m_oc^2$ and ${L}_0
=\gamma m_o v b$.
The above solution is valid if ${L}_0 > Z_1Z_2e^2/c$.
In heavy ion collisions at intermediate energies
one has $L_0 \gg Z_1Z_2e^2/c$ for impact parameters
that do not lead to strong interactions.
It is
also easy to show that, from the magnitudes of the
parameters involved in heavy ion collisions at intermediate energies,
the trajectory (\ref{3.7}) can be very well described by approximating
\begin{equation}
W = 1 \, , \qquad A = {a_o \over \gamma b^2} \, , \qquad \epsilon =
\sqrt {{b^2 \gamma^2 \over a^2_0} + 1 }\ ,  \label{3.9}
\end{equation}
where $a_o$ is half the distance of closest approach
in a head on collision (if
the nuclei were pointlike
and if non-relativistic kinematics were used),
and $\epsilon$ is the eccentricity parameter.
In the approximation (\ref{3.9}) $\epsilon$ is related to the deflection
angle $\vartheta$ by $\epsilon = (a_o/\gamma ) \cot \vartheta$.

The time dependence for a particle moving along the trajectory (\ref{3.7})
may be directly obtained by solving the equation of angular momentum
conservation.
Introducing the paramet\-ri\-zation
\begin{equation}
r(\chi )  = {a_0 \over \gamma} \,
\left\lbrack \epsilon \, \cosh \chi + 1 \right\rbrack \label{3.10}
\end{equation}
we find
\begin{equation}
t = {a_0 \over \gamma v} \, \left\lbrack \chi + \epsilon \, \sinh
\chi \right\rbrack \, .\label{3.11}
\end{equation}
Using the scattering plane perpendicular to the Z-axis, one finds that the
corresponding components of ${\bf r}$ may be written as
\begin{equation} x = a \, \left\lbrack \cosh \chi  + \epsilon \right\rbrack
\label{3.12b}
\ ,
\end{equation}
\begin{equation} y = a\, \sqrt{\epsilon ^2 -1}\, \sinh \chi \label{3.12a}
\ ,\end{equation}
\begin{equation} z = 0  \label{3.12c} \ ,
\end{equation}
where $a=a_0 / \gamma$.
This parametrization is of
the same form as commonly used in the non-relativistic case \cite{AW75},
except that $a_o$ substituted by $a_o / \gamma \equiv a$.

In the limit of  straight-line motion $\epsilon \simeq b/a \gg 1$, and the
equations above
reduce to the simple parametrization
\begin{equation}
y=vt \ , \ \ \ \ \ \ \ x=b \ , \ \ \ {\rm and} \ \ \ z=0 \ .
\end{equation}

As we quoted before, the classical solution for the relative motion of
two relativistic charges interacting electromagnetically can only be
solved analytically
if one of the particles has infinite mass.  Non-relativistically
the two-body problem is solvable by introduction of center of mass and
relative
motion coordinates. Then, the result is equivalent to that of a particle
with reduced mass
$ m_o = m_Pm_T / (m_P + m_T) $
under the action of the same potential.
The particle with reduced mass m$_o$  is lighter than those with mass
m$_P$ and m$_T$, and this accounts for the
simultaneous recoil of them.  An exact
relativistic solution should reproduce this behavior as the relative
motion energy is lowered. We shall use the reduced mass
definition of m$_o$ as usual in the parametrization of the
classical trajectory of Coulomb excitation in intermediate energy
collisions, as outlined above.  In a $^{16}$O + $^{208}$Pb collision
this is not a too serious approximation. For heavier systems like
U+U it would be the simplest way to overcome this difficulty. But, as
energy increases, this approximation is again unimportant since the
trajectories will be straight-lines parametrized by an impact parameter
b.  A more exact result can be obtained 
numerically using the Darwin Lagrangian
to determine the classical trajectory in collisions
at intermediate energies \cite{AAB89}.
But, the parametrization of the classical trajectory as given
by Eqs. (\ref{3.12b},\ref{3.12a},\ref{3.12c}) with a reduced mass particle,
besides reproducing both the
non-relativistic and the relativistic energies,  gives a reasonable
solution to the kind of collisions we want to study.

\subsubsection{Excitation amplitudes}

Including retardation, the amplitude for Coulomb excitation of a target from
the initial state
$\mid i >$ to the final state $\mid f>$ is given in first order
time-dependent perturbation theory by
\begin{equation} a_{fi} = {1 \over i \hbar} \int \Bigl\{ \rho _{fi} ({\bf r})
\, \phi (\omega , \, {\bf r}) +
{1 \over c} \; {\bf J}_{fi} ({\bf r}) \cdot {\bf A} (\omega ,\,
{\bf r}) \Bigr\} \, d^3r
\label{3.13} \end{equation}
where $\rho_{fi} \; ({\bf J}_{fi})$ is the nuclear transition
density (current) and
\begin{equation}
\phi (\omega, \, {\bf r}) = Z_1 e \int ^\infty_{-\infty}
e^{i\omega t}\,
{e^{i\kappa \mid{\bf r-
r'}(t)\mid} \over \mid {\bf r -  r'} (t) \mid} \, dt \label{3.14a}
\end{equation}
\begin{equation}
{\bf A} (\omega , \, {\bf r}) = {Z_1e \over c} \int ^\infty _{-\infty}
{\bf v'} (t) \, e^{iwt}\, { e^{i\kappa \mid {\bf r -  r'} (t) \mid} \over
\mid {\bf r  -  r'} (t) \mid} \, dt \label{3.14b}
\end{equation}
are the retarded potentials generated by a projectile with charge Z$_2$
following a Coulomb trajectory,
and $\kappa = \omega /c$. When the magnitude of the
amplitudes (\ref{3.13}) is small compared to unity, the
use of first order perturbation theory is
justified.

We now use the expansion
\begin{equation}
{ e^{i\kappa \mid {\bf r -  r'}  \mid} \over
\mid {\bf r  -  r'}  \mid}
=4\pi \ i\kappa \sum_{\lambda\mu} j_\lambda(\kappa r_<)Y^*_{\lambda\mu}(\hat
{\bf r}_<)
h_\lambda(\kappa r_>)
\, Y_{\lambda\mu}(\hat{\bf r}_>) \ ,
\label{3.15b}
\end{equation}
where $j_\lambda$ ($h_\lambda$) denotes the spherical Bessel (Hankel)
functions
(of first kind),
${\bf r}_>$ (${\bf r}_<$) refers to whichever of ${\bf r}$ and ${\bf r}'$ has
the larger
(smaller) magnitude. Assuming that the projectile does not
penetrate the target, we use ${\bf r}_>$ (${\bf r}_<$) for the projectile
(target)
coordinates. At collision energies above the Coulomb barrier this assumption
only
applies for impact parameters larger than a certain minimum, below which the
nuclei
penetrate each other.

Using the continuity equation (\ref{ce}) for the nuclear transition current
(we changed the notation: $\rho\equiv \rho_{fi}$,
${\bf J}\equiv {\bf j}_{fi}$),
we can show that the expansion (\ref{3.15b}) can be expressed in terms of
spherical tensors
(see, e.g., Ref. \cite{EG87}, Vol. II)
and Eq. (\ref{3.13}) becomes
\begin{eqnarray}
a_{fi} = {Z_1e \over i \hbar} \sum_{\lambda \mu} \, {4\pi
\over
2\lambda + 1} \, (-1)^\mu \, \biggl\{
S(E\lambda, \, \mu) {\bf \cal M}_{fi}(E\lambda, \, -\mu)
+ S(M\lambda, \, \mu) {\bf \cal M}_{fi}
(M\lambda, \, -\mu) \biggr\} \nonumber \\
\label{3.17}
\end{eqnarray}
where ${\bf \cal M} (\pi \lambda, \mu)$ are the matrix elements for
electromagnetic transitions, as defined in (\ref{ME1lm},\ref{MM1lm}).

The orbital integrals $S(\pi \lambda, \mu$) are given by
\begin{eqnarray}
S(E\lambda, \mu) & = &- {i\kappa^{\lambda+1}
\over \lambda(2 \lambda-1) !! }
\, \int ^\infty_{-\infty} {\partial \over \partial r'} \left\{ r'
(t)h_\lambda \left\lbrack \kappa r' (t) \right\rbrack \right\}
\, Y_{\lambda \mu}
\left\lbrack \theta ' (t), \phi ' (t) \right\rbrack \, e^{i\omega t} \, dt
\nonumber \\
& - &{\kappa^{\lambda + 2} \over c \lambda (2 \lambda -1) !! } \,
\int ^\infty _{-\infty} {\bf v'} (t) \cdot {\bf r} ' (t) \, h_\lambda
[\kappa r'(t)]\, Y_{\lambda \mu} [ \theta '(t), \phi ' (t)
]\, e^{i\omega t}\, dt \nonumber \\ \label{3.19a}
\end{eqnarray}
and
\begin{equation}
S(M\lambda, \mu) =- {i \over \gamma m_oc} \,
{\kappa^{\lambda +1} \over
\lambda (2 \lambda -1) !! } \; {\bf L}_o \cdot \int ^\infty _{-\infty}
\nabla ' \left\{ h_\lambda [ \kappa r'(t) ]
\,Y_{\lambda \mu} [\theta '(t), \phi ' (t) ] \right\}
\, e^{i\omega t} \, dt \label{3.19b}
\end{equation}
where ${\bf L}_o$ is the angular momentum of relative motion, which
is constant:
\begin{equation} L_o = \gamma a m_o v \cot {\vartheta \over 2} \label{3.20}
\end{equation}
with $\vartheta$ equal to the (center-of-mass)
scattering angle.

In non-relativistic collisions
\begin{equation} \kappa r' = {\omega r' \over c} =
{v \over c} {\omega r'\over v} < {v \over c}
\ll 1 \label{3.20b} \end{equation}
because when the relative distance $r'$ obeys the relations $\omega
r'/ v
\geq 1$ the interaction becomes adiabatic. Then one uses the
limiting form of $h_\lambda$ for small values of its argument \cite{AS64}
to show that
\begin{equation} S^{NR}(E \lambda,\mu) \simeq \int ^\infty _{-\infty}
{r'}^{-\lambda-1}
(t) \, Y_{\lambda \mu} \left\{ \theta ' (t), \phi' (t) \right\} \,
e^{i\omega t} \, dt
\label{3.21a} \end{equation}
and
\begin{equation}
S^{NR}(M\lambda, \mu) \simeq - {1 \over \lambda m_oc}
\, {\bf L}_o \cdot \int_{-\infty}^\infty \nabla ' \left\{ {r'}^{-\lambda-1}
(t)\,
Y_{\lambda \mu} \left\lbrack \theta ' (t), \phi ' (t) \right\rbrack
\right\}\,
e^{i\omega t} \, dt
\label{3.21b}
\end{equation}
which are the usual orbital integrals in the non-relativistic Coulomb
excitation theory with hyperbolic trajectories (see
Eqs. (II.A.43) of Ref. \cite{AW75}).

In the intermediate energy case the relation (\ref{3.20}) is partially relaxed
(of course, for relativistic energies, v $\sim$ c, it is not valid)
and one has to keep the more complex forms (\ref{3.19a},\ref{3.19b}) for the
orbital
integrals.

Using the $Z$-axis perpendicular to the
trajectory plane, the recursion relations
for the spherical Hankel functions
and the
identity
\begin{equation}{\bf v} \cdot {\bf r}
= {d\chi \over dt} \otimes {d {\bf r} \over d\chi} \cdot {\bf r}
=a\epsilon v \, \sinh \chi \, , \label{3.22}\end{equation}
we can rewrite the orbital
integrals, in terms of the parametrization (\ref{3.12a},\ref{3.21b}), as
\begin{eqnarray}
S(E \lambda, \mu)
&=& - {i\kappa^\lambda \eta \over c \lambda (2 \lambda -1)!! }
\;  {\cal C}_{\lambda \mu} \int ^\infty _{-\infty} d \chi \;
e^{i \eta
(\epsilon \sinh \chi + \chi)}\nonumber \\
&\times& \, {( \epsilon + \cosh   \chi
+ i \sqrt {\epsilon^2 - 1}\, \sinh   \chi)^\mu \over
(\epsilon \cosh   \chi + 1)^{ \mu-1}}\nonumber \\
&\times& \left\lbrack (\lambda + 1)\, h_ \lambda - {v \eta \over c}
\, (\epsilon \cosh   \chi+1)\,
h_{\lambda+1} + i ({v \over c})^2 \, \eta \epsilon \, \sinh
\chi \cdot h_ \lambda \right\rbrack\nonumber \\
\label {3.23}
\end{eqnarray}
where
\begin{equation}
{\cal C}_{\lambda \mu} =
\left\{ \begin{array}{ll}
\sqrt{2\lambda+1 \over 4\pi }
\, {\sqrt {(\lambda - \mu)! (\lambda + \mu)!} \over (\lambda - \mu) !!
(\lambda
+ \mu)!! } \;
(-1)^{(\lambda + \mu)/2}, &
\mbox{for  $ \lambda + \mu $ = even\ ,}\\
0, & \mbox{ for   $\lambda +\mu $ = odd\ ,}
\end{array}
\right.
\end{equation}
and
\begin{equation}
\eta = {\omega a \over v} = {\omega a_o \over \gamma v} \, ,
\label{3.25}
\end{equation}
and with all $h_ \lambda$'s as functions of
$ (v/c) \eta \, (\epsilon \cosh
\chi +1)$.

For convenience, we define
\begin{equation}
I(E\lambda, \mu) = {va^ \lambda \over {\cal C}_{\lambda \mu}}
\; S(E_\lambda, \mu) \label{3.26}
\end{equation}
and we translate the path of integration by an amount $i \pi / 2$
to avoid strong oscillations of the integral.  We find,
\begin{eqnarray}
I(E \lambda, \mu)
&=& -i ({v \eta \over c}) ^{\lambda+1} \, {1 \over \lambda (2\lambda-1)!!}
\, e^{- \pi \eta / 2  }\,
\int^\infty _ {-\infty} d\chi \; e^{-\eta \epsilon \cosh
\chi}\, e^{i \eta \chi} \nonumber \\
& \times& {(\epsilon + i \sinh
\chi - \sqrt {\epsilon^2 -1} \cosh   \chi)^ \mu \over
(i \epsilon \sinh   \chi + 1) ^{ \mu -1}}\nonumber \\
& \times& \left\lbrack ( \lambda + 1) \, h_ \lambda -
z \, h_{\lambda + 1}
-({v \over c})^2 \epsilon \, \eta \,  \cosh
\chi \cdot h_ \lambda \right\rbrack
\label{3.27}
\end{eqnarray}
where all $h_\lambda$'s are now functions of
\begin{equation}
z = {v \over c} \, \eta \, (i\epsilon \sinh
\chi +1) \, .\label{3.28}
\end{equation}

In the case of magnetic excitations, one may explore the fact that ${\bf L}_0$
is perpendicular to the scattering plane to show that
\begin{eqnarray}
{1 \over m_o} {\bf L}_0 \cdot \nabla \left\{ h_ \lambda
(\kappa r) \; Y_{\lambda \mu} ({\pi \over 2}, \phi) \right\}
&=& \gamma \, {av \over r} \cot {\vartheta \over 2} \, \sqrt {2 \lambda + 1
\over
2 \lambda +3}
\, \sqrt {(\lambda + 1)^2 - \mu ^2} \; {\cal C}_{\lambda + 1, \mu}\nonumber \\
&\times& e^{i \mu \phi}\;
h_\lambda (\kappa r) \, .
\label{3.29}\end{eqnarray}
The magnetic orbital integrals become
\begin{eqnarray}
S(M \lambda, \mu) &=&-ia {v \over c}\, {\kappa^
{ \lambda + 1} \over
\lambda (2 \lambda - 1)!!} \, \sqrt {2 \lambda + 1 \over 2 \lambda + 3}
\, \sqrt {(\lambda + 1)^2 - \mu ^2 } \nonumber \\
& \times &{\cal C}_{\lambda+1, \mu} \; \cot {\vartheta \over 2}\;
\int^ \infty _{-\infty} h_ \lambda \left\lbrack \kappa r'(t) \right\rbrack
\, {1\over
r'(t)} \, e^{i\mu \phi ' (t)} \, e^{i\omega t} \, dt \, . \label{3.30}
\end{eqnarray}
Defining,
\begin{equation} I(M \lambda, \mu) = - { {\lambda ca^ \lambda S(M \lambda, \mu
)
} \over
{{\cal C}_{\lambda + 1, \mu }  \, \cot \vartheta / 2}}
\, \left\{ [(2 \lambda + 1)/ (2 \lambda + 3)]\,
[(\lambda + 1)^2 - \mu^2] \right\} ^{-1/2}
\label{3.31}\end{equation}
we obtain, using the parametrization
(\ref{3.12b},\ref{3.12a},\ref{3.12c}), and
translating the integral path by $i \pi / 2$,
\begin{eqnarray}
I(M\lambda , \mu ) &=&
{i(v \eta /c)^{\lambda +1} \over (2\lambda
-1)!!} \, e^{-\pi \eta /2} \,
\, \int^\infty _{-\infty} d \chi \; h_\lambda (z) \;
e^{- \eta \epsilon \, \cosh   \chi }\nonumber \\
& \times&  e^{i \eta \chi}\,
{ (\epsilon + i \sinh   \chi - \sqrt { \epsilon ^2 -1} \, \cosh
\chi)^\mu
\over (i \epsilon \sinh   \chi + 1)^\mu } \, .
\label{3.19}
\end{eqnarray}

Generally, the most important magnetic excitation has $M1$
multipolarity.
The orbital integrals (\ref{3.27},\ref{3.19}) can only be solved numerically.

\subsubsection{Cross sections and equivalent photon numbers}

In the high-energy limit the classical trajectory reduces to a straight-line.
One can show that using the approximation $\epsilon=b/a \gg1$ the orbital
integrals (\ref{3.27}) and (\ref{3.19}) can be expressed in terms of  simple
analytical
functions. However it is instructive and useful to deduce the excitation
amplitudes from the first principles again.

The square modulus of Eq. (\ref{3.17}) gives the probability of exciting the
target
nucleus from the
initial state $\mid I_i M_i>$ to the final state
$\mid I_f M_f >$ in a collision
with c.m. scattering angle $\vartheta$. If the orientation of the initial
state is not specified, the cross section for exciting the nuclear
state of spin $I_f$ is
\begin{equation}d\sigma_{i \rightarrow f} = {a^2 \epsilon^4 \over
4} \; {1 \over 2I_i +1} \; \sum_{M_i , M_f } \mid a_{fi} \mid^2 \, d\Omega
\, ,\label{4.1}\end{equation}
where $a^2 \epsilon^4 d\Omega /4$ is the elastic (Rutherford) cross section.
Using the Wigner-Eckart theorem and the orthogonality properties of the
Clebsch-Gordan coefficients, one can show that
\begin{equation}{d \sigma_{i \rightarrow f} \over d\Omega} =
{4 \pi^2 Z_1^2e^2 \over \hbar^2} \; a^2 \epsilon^4 \; \sum_{\lambda \mu}
{B(\pi \lambda, I_i \rightarrow I_f )
\over (2\lambda + 1)^3}  \; \mid S(\pi \lambda, \mu) \mid^2\, ,
\label{4.2}\end{equation}
where $\pi =E $ or $M$ stands for the electric or magnetic multipolarity, and
the reduced transition probability is given by
\begin{eqnarray}
B(\pi\lambda; I_i \longrightarrow I_f )
&=& {1\over 2I_i+1}\
\sum_{M_i M_f} \Big| <I_iM_i| {\cal M} (\pi\lambda, \ \mu)
|I_fM_f>\Big|^2 \nonumber \\
&=&
{1\over 2I_i+1}\ \Big| <I_i|| {\cal M} (\pi\lambda)
||I_f>\Big|^2 \ .\label{Bl}
\end{eqnarray}

Integration of (\ref{4.2}) over all energy transfers $\varepsilon = \hbar
\omega$,
and summation over all possible final states of the projectile
nucleus leads to
\begin{equation}{d \sigma_C \over d\Omega} = \sum_f \int {d \sigma_{i
\rightarrow f} \over d\Omega} \;
\rho_f (\varepsilon ) \; d\epsilon \, ,\label{4.4}\end{equation}
where $\rho_f (\varepsilon )$ is the density of final states of the target
with energy $E_f = E_i + \varepsilon$.
Inserting (\ref{4.2}) into (\ref{4.4}) one finds
\begin{equation}{d \sigma_C \over d\Omega} = \sum_{\pi \lambda} \;
{d \sigma_{\pi \lambda } \over d\Omega} = \sum_{\pi \lambda}
\int {d\varepsilon  \over \varepsilon } \; {d n_{\pi \lambda} \over
d\Omega}(\varepsilon ) \; \sigma_\gamma^{\pi \lambda}(\varepsilon )\, ,
\label{4.5a}\end{equation}
where $\sigma_\gamma^{\pi \lambda}$
are the photonuclear absorption cross sections for a given multipolarity
$\pi \lambda$.
The {\it virtual photon numbers}, $n_{\pi \lambda} (\varepsilon)$, are given
by
\begin{equation}{d n_{\pi \lambda} \over
d\Omega} = {Z_1^2\alpha \over 2 \pi}\; {\lambda \left\lbrack
(2\lambda +1 )!! \right\rbrack^2 \over
(\lambda+1)(2\lambda+1)^3} \;
{c^2a^2\epsilon^4 \over \kappa^{2(\lambda-1)}}
\sum_\mu \mid S(\pi \lambda , \mu) \mid^2 \, .\label{4.6}\end{equation}

In terms of the orbital integrals $I(E\lambda, \mu)$, given by (\ref{3.28}),
and using the Eq. (\ref{4.6}), we find for the electric multipolarities
\begin{eqnarray}
{d n_{E \lambda} \over
d\Omega} &= & {Z_1^2\alpha \over 8 \pi^2}\; ({c \over v})^{2\lambda}\;
 {\lambda \left\lbrack
(2\lambda +1 )!! \right\rbrack^2 \over
(\lambda+1)(2\lambda+1)^2} \;
\epsilon^4 \, \eta^{-2\lambda+2}\nonumber \\
& \times& \sum_{\mu \atop {\lambda + \mu =  even}}
{(\lambda - \mu)! (\lambda +\mu)! \over
\left\lbrack (\lambda - \mu)!! (\lambda +\mu)!! \right\rbrack^2} \;
\mid I(E \lambda , \mu) \mid^2 \, .\label{4.7}
\end{eqnarray}
In the case of magnetic excitations we find
\begin{eqnarray}
{d n_{M \lambda} \over
d\Omega} &= &{Z_1^2\alpha \over 8 \pi^2}\;  ({c \over v})^{2(\lambda-1)}\;
{\left\lbrack
(2\lambda +1 )!! \right\rbrack^2 \over
\lambda (\lambda+1)(2\lambda+1)^2} \;
\eta^{-2\lambda+2}\; \epsilon^4 \;( \epsilon^2 - 1)\nonumber \\
& \times & \sum_{\mu \atop {\lambda + \mu = odd}}
{\left\lbrack (\lambda+1)^2 -\mu^2 \right\rbrack \,
(\lambda+1 - \mu)! (\lambda+1 +\mu)! \over
\left\lbrack (\lambda +1 - \mu)!! (\lambda +1 +\mu)!! \right\rbrack^2} \;
\mid I(M \lambda , \mu) \mid^2 \, .\nonumber \\
\label{4.8}
\end{eqnarray}
Only for the $E1$ multipolarity the integrals can be performed
analytically and we get closed expression
\begin{equation}
{d n_{E1} \over d \Omega}={Z_1^2 \alpha \over 4 \pi^2}
 \; \bigl({c\over v} \bigr)^2
 \; {\epsilon^4 \; \zeta^2 }\;
\; e^{-\pi \zeta } \; \biggl\{ {1 \over \gamma^2} \;
{\epsilon^2 -1 \over \epsilon^2} \;
\bigl[ K_{i\zeta} ({\epsilon\zeta}) \bigr]^2
+ \bigl[ K'_{i\zeta} ({\epsilon\zeta})\bigr]^2 \biggr\} \, ,
\label{nE1ann}
\end{equation}
where $\epsilon=1/\sin (\theta/2)$, $\alpha=1/137$,
$\zeta=\omega a_0/\gamma v$, $a_0=Z_1Z_2 e^2/2 E_{Lab}$, $K_{i\zeta}$
is the modified Bessel function with imaginary index, $K'_{i\zeta}$ is
the derivative with respect to its argument.

Since the impact parameter is related to the scattering angle
by $b= a \cot \vartheta / 2$, we can also write
\begin{equation}
n_{\pi \lambda}(\varepsilon , b)\equiv {d n_{\pi \lambda} \over
2 \pi b db} = {4 \over a^2 \epsilon^4 }\;
{d n_{\pi \lambda} \over
d\Omega}
\label{4.9}\end{equation}
which are interpreted as the number of equivalent photons of energy
$\varepsilon=\hbar\omega$,
incident on the target per unit area, in a collision
with impact parameter $b$, in analogy with the results obtained in section
2.1.2.

Again we stress the usefulness of the concept of virtual photon numbers,
especially in
relativistic collisions. In these collisions the momentum
and the energy transfer
due to the Coulomb interaction are related by $\Delta p = \Delta E/v \simeq
\Delta E/ c$. This means that the virtual photons are almost real. One
usually explores this fact to extract information about real photon processes
from the reactions induced by relativistic charges, and vice-versa.
This is the basis of the Weizs\"acker-Williams method, commonly used
to calculate cross sections for Coulomb excitation, particle production,
Bremsstrahlung, etc (see, e.g., Ref. \cite{BB88}).
In the case of Coulomb excitation, even at low energies, although
the equivalent photon numbers should not be interpreted as (almost) real
ones, the cross sections can still be written as a product of them and the
cross sections induced by real photons, as we have shown above.

\subsection{Comparison of Coulomb excitation of GR's at low energies and at
relativistic energies}

Inserting the non-relativistic orbital integrals
into Eq. (\ref{4.6}), we get the following relation for the
non-relativistic equivalent photon numbers ($NR$)
\begin{equation}{dn_{\pi\lambda}^{(NR)} \over d\Omega} =
Z_1^2 \alpha \, {\lambda [(2 \lambda +1)!!]^2 \over (2\pi)^3
(\lambda +1)} \, \zeta^{-2\lambda+2} \, ({c\over v})^{2(\lambda+\delta )}
\, {df_{\pi\lambda} \over d\Omega}(\vartheta , \zeta ) \, ,
\label {A.1}
\end{equation}
where $\delta = 0$ for electric, and $\delta = -1$ for magnetic
multipolarities, and $\zeta = \omega a_o / v$.
The non-relativistic
Coulomb excitation functions $f_{\pi \lambda } (\vartheta , \zeta )$
are very well known and, e.g., are tabulated in Ref. \cite{AW75},
or maybe calculated numerically.

Using the Eqs. (\ref{4.7}), (\ref{4.8}) and (\ref{nE1ann}), we make an
analysis
of Coulomb excitation extending from low to high energy collisions.
As an example, we study the excitations induced by $^{208}$Pb in
$^{16}$O + $^{208}$Pb collisions. Since the expression
(\ref{4.6}) is quite general, valid for all energies, under
the assumption  that the nuclei do not overlap,
the equivalent photon numbers contain all information
about the differences among the low and the high energy scattering.
In Figs. \ref{a-f3}, \ref{a-f4} and \ref{a-f5} we
show $dn_{\pi \lambda, \varepsilon}$, for the $E1$ (Fig. \ref{a-f3}),
$E2$ (Fig. \ref{a-f4}), and $M1$ (Fig. \ref{a-f5}) multipolarities, and
for the collision
$^{16}$O + $^{208}$Pb with an impact parameter $b=15 \;$fm. They are
the equivalent photon numbers with frequency $\omega = 10 \;$MeV$/\hbar$
incident on $^{208}$Pb.
The dotted lines are obtained by using the non-relativistic
Eq. (\ref{A.1}), while the dashed lines correspond
to the relativistic expressions (\ref{(A.3a)},\ref{(A.3b)},\ref{(A.3c)}).
One observes that the relativistic expressions
overestimate the equivalent photon numbers
at low energies, while the non-relativistic
expressions underestimate them at high energies.
The most correct values are  given by the solid lines, calculated according to
Eqs. (\ref{4.7}) and (\ref{4.8}). They reproduce the low and the high energy
limits, giving an improved interpolation between these limits at
intermediate energies. These discrepancies are more apparent  for the $E1$
and the $E2$ multipolarities. In the energy interval around 100A MeV
neither the low energy theory nor the high energy one can reproduce well the
correct values. This energy interval is indeed very sensitive to the
effects of retardation and of Coulomb recoil.

At these bombarding energies, the differences between the
magnitude of the
non-relativis\-tic and the correct
relativistic  virtual photon numbers
are kept at a constant value, of about 20\%, for excitation energies
$\varepsilon=\hbar \omega < 10$ MeV. However, they increase sharply
when one reaches the excitation energy of
$\varepsilon=\hbar \omega > 10 \; $MeV.
The reason is that, for such excitation energies,
the adiabaticity factor  becomes greater than
unity ($\xi >1$). This means that excitation energies of order
of $10 \; $MeV (like in the case of giant resonance excitation)
are in the transition region from a constant behavior of the
equivalent photon numbers to that of an exponential ($\sim e^{-\pi \xi}$)
decay. This is more transparent in Fig. \ref{a-f6} where we plot the
equivalent
photon numbers for $E_{lab} = $100A MeV, $b=15 \; $fm, and
as a function of $\hbar \omega$. One also
observes from  this figure that the $E2$ multipolarity component
of the electromagnetic field dominates at low frequencies. Nonetheless,
over the range of $\hbar \omega$ up to some tens of MeV, the $E2$
matrix elements
of excitation are much smaller than the $E1$ elements for most nuclei, and the
$E2$ effects become unimportant. However, such effects are relevant
for the excitation of the isoscalar $E2$ giant resonance (GQR$_{is}$)
which have large matrix elements.

As an application of the semiclassical approach to
Coulomb excitation in intermediate energy collisions, we study
the excitation of giant isovector dipole resonances ($E1$) and of giant
isoscalar quadrupole resonances ($E2$) in $^{208}$Pb by means of
the Coulomb interaction with
a $^{16}$O projectile. At 100A MeV the maximum
scattering angle which still leads to a pure Coulomb scattering
(assuming a sharp cut-off at an impact parameter $b=R_P +R_T$)
is $3.9^\circ$. The cross sections are calculated by assuming a Lorentzian
shape for the photonuclear cross sections:
\begin{equation}\sigma^{\pi\lambda}_\gamma =\sigma_{m} \;
{\epsilon^2 \Gamma^2 \over (\epsilon^2 - E_{m}^2 )^2
+ \epsilon^2 \Gamma^2} \label{5.3}\end{equation}
with $\sigma_{m}$ chosen to reproduce the Thomas-Reiche-Kuhn
sum rule for $E1$ excitations,
\begin{equation}
\int \sigma_\gamma^{E1} (\epsilon ) \, d\epsilon
\simeq 60 {NZ \over A} \; \mbox{MeV} \, \mbox{mb}
\label{5.4a}
\end{equation}
and the energy-weighted sum rule for the
quadrupole mode,
\begin{equation}
\int \sigma_\gamma^{E2} (\epsilon ) \, {d\epsilon
\over \epsilon^2}
\simeq 0.22 \, ZA^{2/3} \; \mu \mbox{b} / \mbox{MeV} \, .
\label{5.4b}
\end{equation}
The resonance energies are approximately given by $E_{\mbox{\tiny GDR}}
\simeq 77 \cdot A^{-1/3} \; $MeV and
$E_{\mbox{\tiny GQR}} \simeq 63 \cdot A^{-1/3} \;$MeV. We use the
widths $\Gamma_{\mbox{\tiny GDR}} = 4 \; $MeV and $\Gamma_{\mbox{\tiny
GQR}} = 2.2 \;$MeV for $^{208}$Pb.

We will discuss the differential cross sections as a function of the
scattering angle later, when we introduce the effects of
strong absorption.  To obtain the total cross sections,
one has to integrate the equivalent photon numbers
in (\ref{4.7}) and (\ref{4.8}) from $0^\circ$ to a maximum scattering
angle $\theta_{max}$, where the nuclear absorption
sets in, or equivalently, one can integrate over the
impact parameter, from a minimum value $b_{mim}$ up to infinity.
In Fig. \ref{a-f7} is shown
the total cross section for the excitation of giant dipole and
of giant quadrupole resonances in $^{208}$Pb in a collision
with $^{16}$O as a function of the laboratory energy per nucleon.
The same average behavior of the photonuclear cross sections, as
assumed in Eqs. (\ref{5.3}) and (\ref{5.4a}), is used.

Only for the $E1$ multipolarity the angular integration can be performed
analytically. One obtains\footnote{We observe that the original
formula for the dipole case appearing in \cite{BB88} has a misprinted
sign in one of its terms.}
\begin{eqnarray}
{}{}
N_{E1} &=&
{2\over \pi}
 \ Z_1^2\alpha \;e^{-\pi \zeta }\;(c/v)^2\;\left\{ -\xi
K_{i\zeta }K_{i\zeta }^{\prime }- {1 \over 2}
(c/v)^2\xi ^2\right. \nonumber \\
&&\times \left. \left[
(\zeta/\xi)^2K_{i\zeta}^2+K_{i\zeta}^{\prime 2}
-K_{i\zeta }^2-\frac i{\epsilon _0}
\left( K_{i\zeta }\left(
{{\partial K_\mu ^{\prime }}\over
{\partial \mu }}
\right) _{\mu =i\zeta }-
K_{i\zeta }^{\prime }
\left( {{\partial K_\mu }\over
{\partial \mu }}
\right) _{\mu =i\zeta }\right)
\right. \right\} \ , \label{Ne1an}
\end{eqnarray}
where
\begin{equation}
\epsilon_0 =
\left\{ \begin{array}{ll}
1,     & \mbox{for    $ 2a>b_{mim} $ ,}\\
R/a-1, & \mbox{for    $ 2a<b_{mim} $ ,}
\end{array}
\right.
\end{equation}
and $\xi=\epsilon_0 \zeta=\omega b_{mim}/\gamma v$.

It is easy to see that this equation reduces to Eq. (\ref{(A.6a)})
in the relativistic limit, when $\zeta \longrightarrow 0$,
$\epsilon_0 \longrightarrow \infty$.

The cross sections increase very rapidly
to large values, which are already attained at intermediate energies.
A salient feature is that the cross section for the excitation of
giant quadrupole
modes is very large at low and intermediate energies, decreasing in
importance (about $10$\% of the $E1$ cross section) as the energy increases
above 1A GeV. This occurs because the equivalent photon
number for the $E2$ multipolarity is much larger than that for the $E1$
multipolarity at low collision energies. That is, $n_{E2} \gg n_{E1}$,
for $v\ll c$.
This has a simple explanation.
Pictorially, as seen from an observator at rest,
when a charged particle moves at low energies
the lines of force of its corresponding electric field are
isotropic, diverging from its center in all directions. This means that
the field carries a large amount of tidal ($E2$) components.
On the other hand, when the particle
moves very fast its lines of force appear contracted in the
direction perpendicular to its motion due to Lorentz contraction.
For the observator this field looks like a pulse of plane waves of light.
But plane waves contain all multipolarities with the same weight, and
the equivalent photon numbers become all approximately equal, i.e.,
$n_{E1} \simeq n_{E2} \simeq n_{M1}$, and increase logarithmically
with the energy for $\gamma \gg 1$. The difference in the cross sections
when $\gamma \gg 1$ are then approximately equal to the difference
in the relative strength of the two giant resonances $\sigma_\gamma^{E2}
/\sigma^{E1}_\gamma < 0.1$. The excitation of giant magnetic monopole
resonances is of less importance, since for low energies $n_{M1} \ll
n_{E1}$ ($n_{M1} \simeq (v/c)^2 n_{E1}$), whereas for high energies, where
$n_{M1} \simeq n_{E1}$, it will be also much smaller than the excitation
of electric dipole resonances since their relative strength
$\sigma^{M1}_\gamma/\sigma^{E1}_\gamma$ is much smaller than unity.

At very large energies the cross sections for the Coulomb excitation
of giant resonances overcome the nuclear geometrical cross sections.
Since these resonances decay mostly through particle emission or fission,
this indicates that Coulomb excitation of giant resonances is
a very important process to be considered in relativistic heavy
ion collisions
and fragmentation processes,
especially in heavy ion colliders.
At intermediate energies the cross sections are also
large and this offers good possibilities to establish
and study the properties of giant resonances.

\subsection{Quantum description of Coulomb excitation at high energies}

Inelastic scattering of heavy ions at intermediate energy collisions is an
important tool to investigate the structure of stable and unstable nuclei.
Laboratories like GANIL/France, GSI/Germany, RIKEN/Japan, and MSU/USA,
frequently use this technique.
The angular distribution of the
inelastically scattered fragments are
particularly useful to identify unambiguously
the multipolarity of the interaction, and consequently the spin and parities
of the excited states. In previous sections we have shown that
recoil and retardation effects, are important at this energy regime.
However, as shown by Bertulani and Nathan \cite{BN93},
in order to describe correctly the angular distribution, absorption
and diffraction effects have to be included properly. Next
we show how quantum mechanical effects show up in the differential
cross sections.

\subsubsection{Inelastic amplitudes and virtual photon numbers}
Defining {\bf r} as the separation between the centers
of mass of the two nuclei and {\bf r$^{\prime}$} to be the intrinsic
coordinate of the target nucleus
to first-order the inelastic scattering amplitude is
given by
\begin{equation}
f(\theta)= {i k \over 2\pi \hbar v} \ \int d^3r \ d^3r'
<\Phi_{\bf k'}^{(-)} ({\bf r}) \ \phi_f({\bf r}')
\ \vert \ V_{int} ({\bf r}, \ {\bf r}') \ \vert \
\Phi_{\bf k}^{(+)} ({\bf r}) \ \phi_i({\bf r}') >
\; , \label{fth}
\end{equation}
where $\Phi_{\bf k'}^{(-)}({\bf r})$ and $\Phi_{\bf k}^{(+)}({\bf r})$
are the incoming and outgoing distorted waves, respectively,
for the scattering of the center of mass of the nuclei,
and $\phi({\bf r}')$ is the intrinsic nuclear wavefunction
of the target nucleus.

At intermediate energies, $\Delta E/E_{lab}\ll 1$, and forward
angles, $\theta \ll 1$,
we can use eikonal wavefunctions for the distorted waves; i.e.,
\begin{equation}
\Phi_{\bf k'}^{(-)*}({\bf r}) \ \Phi_{\bf k}^{(+)}({\bf r})
=  \exp \Biggl\{ - i{\bf q.r}+ i\chi(b)\Biggr\}
\; , \label{peik}
\end{equation}
where
\begin{equation}
\chi(b)={i \over \hbar v} \int_{-\infty}^\infty
U_N^{opt} (z', \ b) \ dz' +i \psi_C(b)
 \label{peikb}
\end{equation}
is the eikonal-phase, ${\bf q}={\bf k}'-{\bf k}$,
$U_N^{opt}$ is the nuclear optical potential, and $\psi_C(b)$ is the
Coulomb eikonal phase.
 We have defined the impact parameter {\bf b} by
${\bf b}=\vert {\bf r}\times \hat{\bf z} \vert$.

For light nuclei, one can assume Gaussian nuclear densities,
and the Coulomb phase is given by
\begin{equation}
\psi_C(b)=2 \ {Z_1Z_2e^2 \over \hbar v}\ \biggl\{ ln (kb) + {1\over 2}
E_1 \bigl( {b^2 \over R_G^2}\bigr) \biggr\}\; ,
\label{phc}
\end{equation}
with $R_G^{(i)}$ equal to the size
parameter of each Gaussian matter density,
$R_G^2=[R_G^{(1)}]^2+[R_G^{(2)}]^2$,
and
\begin{equation}
E_1(x)=\int_x^\infty {e^{-t} \over t} \ dt
\; .
\label{E1x}
\end{equation}
The first term in Eq. (\ref{phc}) is the contribution to the Coulomb phase
of a point-like charge distribution.  It reproduces the elastic Coulomb
amplitude when introduced into the eikonal expression for the elastic
scattering
amplitude. The
second term in Eq. (\ref{phc}) is a correction due to the extended
Gaussian charge distribution. It eliminates
the divergence of the Coulomb phase at $b=0$, so that
\begin{equation}
\psi_C(0)=2 {Z_1 Z_2 e^2 \over \hbar v} \, \Bigl[ \ln (kR_G)-C\Bigr]
\end{equation}
where $C=0.577$ is the Euler constant.

For heavy nuclei a "black-sphere" absorption model is more appropriate.
Assuming an
absorption radius $R_0$,
the Coulomb phase is given by
\begin{eqnarray}
\chi_C(b)&=&2 {Z_aZ_A e^2 \over \hbar v} \ \biggl\{
\Theta(b-R_0) \ ln (kb)
+\Theta(R_0-b)\ \Bigl[ ln(kR_0)\nonumber \\
&+&ln\Bigl(1+(1-b^2/R_0^2)^{1/2}
\Bigr)-(1-b^2/R_0^2)^{1/2}-{1\over 3} \ (1-b^2/R_0^2)^{3/2}
\Bigr]\biggr\} \; .
\nonumber \\
\end{eqnarray}

Again, the first term inside the parentheses is  the Coulomb
eikonal phase for pointlike charge
distributions. The second term accounts for the finite
extension of the charge distributions.

For high energy collisions, the optical potential $U(r)$
can be constructed by using the t-$\rho\rho$ approximation
\cite{HRB91}. One gets
\begin{equation}
U(r)=-{\hbar v \over 2} \ \sigma_{NN} \ (\alpha_{NN}+i)
\ \int \rho_1({\bf r}')\ \rho_2({\bf r-r'})\ d^3 r'
\ ,\label{Ur}
\end{equation}
where $\sigma_{NN}$ is the nucleon-nucleon cross section, and
$\alpha_{NN}$ is the real-to-imaginary ratio of the forward
($\theta=0^\circ$) nucleon-nucleon scattering amplitude.
A set of the experimental values of these quantities,
useful for our purposes, is given in Table \ref{a-t1}.

\begin{table}[tb]
\caption[ ]
{Parameters \protect\cite{Ra79} for the nucleon-nucleon amplitude,
$f_{NN}(\theta=0^\circ)=
(k_{NN}/4\pi)$ $\sigma_{NN} \ (i+\alpha_{NN})$.
\label{a-t1}}
\begin{center}
\begin{tabular}{ l l l l l l l r } \hline
$E$ [A MeV]&
$\sigma_{NN}$ [fm$^2$]&
$\alpha_{NN}$   \\ \hline
85&6.1&1 \\
94&5.5&1.07 \\
120&4.5&0.7 \\
200&3.2&0.6 \\
342.5&2.84&0.26 \\
425&3.2&0.36 \\
550&3.62&0.04 \\
650&4.0&-0.095 \\
800&4.26&-0.075 \\
1000&4.32&-0.275 \\
2200&4.33&-0.33 \\ \hline
\end{tabular}
\end{center}
\end{table}

In Eq. (\ref{fth})  the interaction potential,
assumed to be purely Coulomb, is given  by
\begin{equation}
V_{int}({\bf r}, \ {\bf r}')= {v^\mu \over c^2} \ j_\mu ({\bf r}')
\ {e^{i\kappa \vert {\bf r} - {\bf r}' \vert} \over
\vert {\bf r} - {\bf r}' \vert} \; ,
\label{Vint}
\end{equation}
where $v^\mu=(c, \ {\bf v})$, with ${\bf v}$ equal to the projectile
velocity, $\kappa=\omega/c$, and $j_\mu ({\bf r}')$ is the charge
four-current for the intrinsic excitation of nucleus 1 by an energy
of $\hbar \omega$.
Inserting (\ref{peikb}) and (\ref{Vint}) in (\ref{fth})
and following the same steps as in Ref. \cite{BB88},
one finds
\begin{eqnarray}
f({\theta})&=&i {Z_1 e k \over \gamma \hbar v}
\sum_{\pi \lambda m} i^m \biggl( {\omega \over c} \biggr)^\lambda
\sqrt{2\lambda+1} \ e^{-im\phi} \nonumber \\
&\times&\Omega_m(q)  \ G_{\pi \lambda m} \biggl( {c \over v} \biggr)
<I_f\ M_f \ \vert\ {\cal M} (\pi \lambda, \ -m ) \ \vert
I_i\ M_i >
\label{fth2}
\end{eqnarray}
where $\pi \lambda m$ denotes the multipolarity, $G_{\pi \lambda m}$ are the
Winther-Alder relativistic functions \cite{WA79},
and $<I_f\ M_f  \vert {\cal M} (\pi \lambda,
-m )  \vert
I_i\ M_i >$ is the matrix element for the electromagnetic transition
of multipolarity $\pi \lambda m$ from
$\vert  I_i\ M_i >$ to $\vert  I_f\ M_f >$, with
$E_f-E_i=\hbar \omega$.
The function $\Omega_m(q)$ is given by
\begin{equation}
\Omega_m(q)=\int_0^\infty db \ b \ J_m(qb) \ K_m \biggl( {\omega b
\over \gamma v } \biggr) \ \exp\bigl\{i\chi(b)\bigr\} \; ,
\label{Omeg}
\end{equation}
where $q=2k \sin (\theta/2)$ is the momentum transfer, $\theta$ and
$\phi$ are the polar and azimuthal scattering angles, respectively.

For the $E1$, $E2$ and $M1$ multipolarity, the functions $G_{\pi\lambda
m}(c/v)$ are given by \cite{WA79}
\begin{eqnarray}
G_{E11}(x)&=&-G_{E1-1}(x)=x\sqrt{8\pi}/3; \ \ \
G_{E10}(x)=-i4\sqrt{\pi(x^2-1)}/3;
\nonumber \\
G_{M11}(x)&=&G_{M1-1}(x)=-i\sqrt{8\pi}/3; \ \ \
G_{M10}(x)=-i4\sqrt{\pi(x^2-1)}/3;
\nonumber \\
G_{E22}(x)&=&G_{E2-2}(x)=-2x\sqrt{\pi(x^2-1)/6};
\nonumber \\
G_{E21}(x)&=&-G_{E2-1}(x)=-i2\sqrt{\pi/8}(2x^2-1)/5;
\ \ \ G_{E20}(x)=2\sqrt{\pi(x^2-1)}/5 \ .
\label{Glm}
\end{eqnarray}

Using the Wigner-Eckart theorem, one can calculate the
inelastic differential cross section from (\ref{fth2}), using techniques
similar to those discussed in previous sections.
One obtains
\begin{equation}
{d^2\sigma_C \over d \Omega\, dE_\gamma} \ \bigl( E_\gamma \bigl)
={1 \over E_\gamma} \ \sum_{\pi \lambda}
{d n_{\pi \lambda} \over d\Omega} \ \sigma_\gamma^{\pi \lambda}
\ \bigl( E_\gamma \bigl)
\label{dsig}
\end{equation}
where $\sigma_\gamma^{\pi \lambda}
\bigl( E_\gamma \bigl)$ is the photonuclear cross section for the
absorption of a real photon with energy $E_\gamma$ by nucleus 2, and
$d n_{\pi \lambda} / d\Omega$ is the virtual photon
number, which is given by \cite{BN93}
\begin{equation}
{d n_{\pi \lambda} \over d\Omega} = Z_1^2 \alpha \ \biggl( {\omega k \over
\gamma v }\biggr)^2 {\lambda \bigl[ (2\lambda+1)!! \bigr]^2 \over (2\pi)^3
\ (\lambda+1)} \ \sum_m \ \vert G_{\pi \lambda m} \vert^2 \ \vert \Omega_m (q)
\vert^2 \; ,
\label{dn}
\end{equation}
where $\alpha=e^2/\hbar c$.

The total cross section for Coulomb excitation can be obtained from Eqs.
(\ref{dsig}) and (\ref{dn}), using the approximation $d\Omega \simeq
2 \pi q dq  /k^2$, valid for small scattering angles and small energy
losses. Using the closure relation for the Bessel functions, we obtain
\begin{equation}
{d\sigma_C\over dE_\gamma} \ \bigl( E_\gamma \bigl)
={1 \over E_\gamma} \ \sum_{\pi \lambda}
n_{\pi \lambda}  \bigl( E_\gamma \bigl)\ \sigma_\gamma^{\pi \lambda}
\ \bigl( E_\gamma \bigl)
\; ,
\label{sig}
\end{equation}
where
\begin{equation}
n_{\pi \lambda} (\omega ) = Z_1^2 \alpha \
{\lambda \bigl[ (2\lambda+1)!! \bigr]^2 \over (2\pi)^3
\ (\lambda+1)} \ \sum_m \ \vert G_{\pi \lambda m} \vert^2 \ g_m (\omega)
\; ,
\label{n}
\end{equation}
and
\begin{equation}
g_m(\omega)= 2\pi \ \biggl( {\omega \over \gamma v} \biggr)^2
\ \int db \ b  \ K_m^2\biggl( {\omega b \over \gamma v}  \biggr) \ \exp
\bigl\{ -2\ \chi_I (b) \bigr\}
\; ,
\label{g}
\end{equation}
where $\chi_I(b)$ is the imaginary part of $\chi(b)$, which is obtained from
Eq.~(\ref{peikb}) and the imaginary part of the optical potential.

Before proceeding further,
it is worthwhile to mention that the present calculations differ from
those of previous sections  by the proper inclusion
of absorption. To reproduce the angular distributions of
the  cross sections, it is essential to include the nuclear transparency.
In the limit of a black-disk approximation,
the above formulas  reproduce the results presented in Ref. \cite{BB88}.
One also observes that the Coulomb phase in the distorted waves, which is
necessary for the quantitative reproduction of the experimental
angular distributions, is
not important for the total cross section in high energy collisions.
This fact explains why
semiclassical and quantum methods give the same result for the
total cross section for Coulomb excitation at relativistic energies
\cite{BB88}. At
intermediate energies, however, it is just this important phase which
reproduces
the semiclassical limit for the scattering of large-$Z$ ions, as we shall see
next. Using the semiclassical terminology, for $E_{lab}
\approx 100$A MeV or less,
the recoil in the Coulomb trajectory is relevant.
At the distance of closest approach, when the Coulomb field is most effective
at inducing  the excitation, the ions are displaced farther from each other
due to the Coulomb recoil. As we discussed before,
this effect can be accounted for approximately by using the effective
impact parameter $b_{eff}=b+\pi Z_1Z_2e^2/4E_{lab}$ in the semiclassical
calculations. This recoil approximation can also be used in Eq. (\ref{g}),
replacing $b$ by $b_{eff}$ in the Bessel function and the nuclear phase,
in order to obtain the total cross section. Since the
modified Bessel function is a rapidly decreasing function of its argument,
this modification leads to sizable modifications of the total cross section
at intermediate energy collisions.

Finally, we point out that for very light heavy ion partners,
the distortion of the scattering
wavefunctions
caused by the nuclear field
is not important. This distortion is manifested in the diffraction
peaks of the angular distributions,
characteristic of strong absorption processes. If $Z_1Z_2 \alpha \gg 1$,
one can neglect the diffraction peaks in the inelastic scattering cross
sections and a purely Coulomb excitation process emerges. One can gain insight
into the excitation mechanism by looking at how the semiclassical limit of the
excitation amplitudes emerges
from the general result (\ref{dn}). We do this next.

\subsubsection{Semiclassical limit of the excitation amplitudes}
If we assume that Coulomb scattering is dominant and neglect the
nuclear phase in Eq.~(\ref{peikb}), we get
\begin{equation}
\Omega_m(q)\simeq \int_0^\infty db \ b \ J_m(qb) \ K_m \biggl( {\omega b
\over \gamma v } \biggr) \ \exp\bigl\{ i\psi_C(b) \bigr\} \; .
\label{Omegc}
\end{equation}

This integral can be done analytically by rewriting it as
\begin{equation}
\Omega_m(q)=\int_0^\infty db \ b^{1+i2\eta}
\ J_m(qb) \ K_m \biggl( {\omega b
\over \gamma v } \biggr) \; ,
\label{Oma1}
\end{equation}
where we used the simple form $\psi_C(b)=2\eta \ ln(kb)$,
with $\eta=Z_1Z_2e^2/\hbar v$.
Using standard techniques found in
Ref. \cite{GR80}, we find
\begin{eqnarray}
\Omega_m(q)&=&2^{2i\eta}\ {1\over m!} \ \Gamma (1+m+i\eta ) \Gamma(1+i\eta)
\nonumber \\
&\times& \ \Lambda^m \ \biggl( {\gamma v \over \omega} \biggr)^{2+2i\eta}
\ F\biggl( 1+m+i\eta;\ 1+i\eta; \ 1+m; -\Lambda^2 \biggr)
\; ,
\label{Oma2}
\end{eqnarray}
where
\begin{equation}
\Lambda={q\gamma v \over \omega }\; ,
\label{Lamb}
\end{equation}
and
$F$ is the hypergeometric function \cite{GR80}.

The connection with the semiclassical results may be obtained by using
the low momentum transfer limit
\begin{eqnarray}
J_m(qb)&\simeq&
\sqrt{ 2\over \pi q b} \ \cos \biggl( qb - {\pi m \over 2} - {\pi
\over 4} \biggr) \nonumber \\
&=& { 1 \over \sqrt{2 \pi q b}} \
\biggl\{ e^{iqb}\ e^{-i\pi (m+1/2)/2} +
e^{-iqb}\ e^{i\pi (m+1/2)/2} \biggr\}
\; ,
\label{Bes}
\end{eqnarray}
and using the stationary phase method, i.e.,
\begin{equation}
\int G(x) \ e^{i\phi(x)} \ dx \simeq \Biggl(
{2 \pi i \over \phi'' (x_0)} \Biggr)^{1/2} \ G(x_0) \ e^{i\phi(x_0)}
\; ,
\label{stat}
\end{equation}
where
\begin{equation}
{d \phi \over dx} (x_0) =0 \qquad {\rm and} \qquad
\phi''(x_0)={d^2 \phi \over dx^2} (x_0)
\; .
\label{stat2}
\end{equation}
This result is valid for a slowly varying function $G(x)$.

Only the second term in brackets of Eq.~(\ref{Bes}) will have a positive
($b=b_0>0$) stationary point, and
\begin{equation}
\Omega_m(q) \simeq
{1 \over \sqrt{2 \pi q}} \ \biggl( {2 \pi i \over \phi''(b_0)}
\biggr)^{1/2} \ \sqrt{b_0} \ K_m \biggl( {\omega b_0 \over
\gamma v} \biggr) \ \exp\biggl\{ i\phi (b_0)+i {\pi (m+1/2)\over 2} \biggr\}
\; ,
\label{Oma7}
\end{equation}
where
\begin{equation}
\phi(b)=-qb+2\eta \ ln (kb)
\; .
\label{phi}
\end{equation}
The condition $\phi'(b_0)=0$ implies
\begin{equation}
b_0={2 \eta \over q}={a_0 \over \sin (\theta/2)}
\; ,
\label{b0}
\end{equation}
where $a_0=Z_1Z_2e^2 /\mu v^2$ is half the distance of closest approach
in a classical head-on collision.

We observe that the relation (\ref{b0}) is the same [with $\cot
(\theta/2) \sim
\sin^{-1} (\theta/2)$] as that between impact parameter and deflection
angle of a particle following a classical Rutherford trajectory.
Also,
\begin{equation}
\phi''(b_0)=-{2 \eta \over b_0^2}=- {q^2 \over 2\eta}
\; ,
\label{phi2}
\end{equation}
which implies that in the semiclassical limit
\begin{eqnarray}
\vert \Omega_m(q) \vert^2_{s.c.} &=& {4 \eta^2 \over q^4}
\ K_m^2 \biggl( {2 \omega \eta \over \gamma v q} \biggr) \nonumber \\
&=&{1 \over k^2} \ \biggl( {d\sigma \over d \Omega} \biggr)_{Ruth}
\ K_m^2 \biggl( {\omega a_0 \over \gamma v \sin (\theta/2)} \biggr)
\; .
\label{phi3}
\end{eqnarray}
Using the above results, Eq.~(\ref{dn}) becomes
\begin{equation}
{d n_{\pi \lambda} \over d\Omega} = \biggl( {d \sigma \over d \Omega}
\biggr)_{Ruth}
Z_1^2 \alpha \ \biggl( {\omega  \over \gamma
v }\biggr)^2 {\lambda \bigl[ (2\lambda+1)!! \bigr]^2 \over (2\pi)^3
\ (\lambda+1)} \ \sum_m \ \vert G_{\pi \lambda m} \vert^2 \
K_m^2 \biggl( {\omega a_0 \over \gamma v \sin (\theta/2)} \biggr)
\; .
\label{dn2}
\end{equation}

If strong absorption is not relevant, the above formula can be used to
calculate the equivalent photon numbers. The stationary value given by
Eq. (\ref{b0}) means that the important values of $b$ which contribute
to $\Omega_m(q)$ are those close to the classical impact parameter.
Dropping the index $0$ from Eq. (\ref{b0}), we can also rewrite (\ref{dn2}) as
\begin{equation}
{d n_{\pi \lambda} \over 2 \pi b \ db} =
Z_1^2 \alpha \ \biggl( {\omega \over \gamma
v }\biggr)^2 {\lambda \bigl[ (2\lambda+1)!! \bigr]^2 \over (2\pi)^3
\ (\lambda+1)} \ \sum_m \ \vert G_{\pi \lambda m} \vert^2 \
K_m^2 \biggl( {\omega b \over \gamma v} \biggr)
\; ,
\label{dn3}
\end{equation}
which is equal to the semi-classical
expression given in Ref. \cite{AB89}, Eq. (A.2).

For very forward scattering angles, such that
$\Lambda <<1$,
a further approximation can be made by setting the hypergeometric function
in Eq. (\ref{Oma2}) equal to unity \cite{GR80}, and we obtain
\begin{equation}
\Omega_m(q)=2^{2i\eta}\ {1\over m!} \ \Gamma (1+m+i\eta ) \ \Gamma(1+i\eta)
\ \Lambda^m \ \biggl( {\gamma v \over \omega }\biggr)^{2+2i\eta}
\; .
\label{Oma3}
\end{equation}

The main value of $m$ in this case will be $m=0$, for which one gets
\begin{eqnarray}
\Omega_0(q)&\simeq&2^{2i\eta} \ \Gamma (1+i\eta ) \ \Gamma(1+i\eta) \
\biggl( {\gamma v \over \omega} \biggr)^{2+2i\eta} \nonumber \\
&=& - \eta^2 \ 2^{2i\eta} \ \Gamma (i\eta ) \ \Gamma(i\eta) \
\biggl( {\gamma v \over \omega }\biggr)^{2+2i\eta}
\; ,
\label{Oma4}
\end{eqnarray}
and
\begin{equation}
\vert \Omega_0(q)\vert^2=
\eta^4 \ \biggl( {\gamma v \over \omega }\biggr)^4 \ {\pi^2 \over
\eta^2 \ sinh^2 (\pi \eta)}
\; ,
\label{Oma5}
\end{equation}
which, for $\eta \gg 1$, results in
\begin{equation}
\vert \Omega_0(q)\vert^2=
4 \pi^2 \ \eta^2 \ \biggl( {\gamma v \over \omega } \biggr)^4 \
e^{-2 \pi \eta}
\; .
\label{Oma6}
\end{equation}

This result shows that in the absence of strong absorption and
for $\eta \gg 1$, Coulomb excitation is strongly suppressed at
$\theta=0$. This also follows from semiclassical arguments, since
$\theta \rightarrow 0$ means large impact parameters, $b\gg 1$, for which
the action of the Coulomb field is weak.

\subsection{Singles spectra in Coulomb excitation of GDR}

In this section, we apply the formalism developed in previous sections
in the analysis of the data
of Ref. \cite{BB90}, in which a projectile of $^{17}$O with an energy
of $E_{lab}=84$A MeV excites the target nucleus $^{208}$Pb to the
GDR.  We first seek parameters
of the optical potential which fits the elastic scattering data.
We use the eikonal approximation for the elastic amplitude in the form
given by
\begin{equation}
f_{el}(\theta )=i k \int J_0(qb) \ \biggl\{ 1 - \exp \bigl[ i \chi(b)
\bigr] \biggr\}\; b \, db
\; ,
\label{fel}
\end{equation}
where $J_0$ is the Bessel function of zeroth-order and
the phase $\chi(b)$ is given by Eq. (\ref{peikb}).
In Fig. \ref{a-f8} we compare the calculated
elastic scattering angular distribution to the data
from Ref. \cite{Bar88}.  The calculation utilized
Eq. (\ref{fel}), with $\chi(b)$ obtained from an
optical potential of a standard Woods-Saxon form with parameters
\begin{eqnarray}
V_0&=&50 \; \mbox{MeV} \, , \;\;\; W_0=58 \; \mbox{MeV}\, , \;\;\;
R_V=R_W=8.5 \; \mbox{fm} \;\;\; {\rm and} \nonumber \\
 a_V&=&a_W=0.85 \; \mbox{fm}
\; .
\label{par}
\end{eqnarray}

In order to calculate the inelastic cross section for the excitation of
the GDR, we use a Lorentzian parameterization for the photoabsorption
cross section of $^{208}$Pb \cite{Ve70}, assumed to be all $E1$, with
$E_{\mbox{\tiny GDR}} \, = \, 13.5 \, $MeV and
$\Gamma \, = \, 4.0 \, $MeV. Inserting this form into Eq. (\ref{sig})
and doing the calculations implicit in Eq. (\ref{dn}) for $dn_{E1}/d\Omega$,
we calculate the angular distribution and compare it with the data in
Fig. \ref{a-f9}.
The agreement with the data is excellent, provided we adjust the overall
normalization to a value corresponding to
$93\,\%$ of the energy weighted sum rule (EWSR) in the energy interval
$7-18.9\; $MeV.  Taking into account the $\pm 10\%$ uncertainty in the
absolute
cross sections quoted in Ref.~\cite{Bar88}, this is consistent with
photoabsorption cross section in that energy range, for which approximately
$110\,\%$ of the EWSR is exhausted.

To unravel the effects of relativistic corrections, we repeat the previous
calculations unplugging the factor $\gamma=(1-v^2/c^2)^{-1}$ which appears
in the expressions (\ref{n}) and (\ref{g})
and using the non-relativistic limit of the
functions $G_{E1m}$, as given in Eq. (\ref{Glm}).
These modifications eliminate
the relativistic corrections on the interaction potential. The result of this
calculation is shown in Fig. \ref{a-f10} (dotted curve). For
comparison, we also show the
result of a full calculation, keeping the relativistic corrections
(dashed curve).
We observe that the two results have approximately the same pattern, except
that the non-relativistic result is slightly smaller than the relativistic
one.
This fact may explain the discrepancy between the fit of Ref.
\cite{Bar88}
and ours as due to relativistic corrections not properly accounted for
in the ECIS code \cite{Ra81}. In fact, if we repeat
the calculation for the excitation of GDR$_{iv}$ using the
non-relativistic limit of Eqs. (\ref{n}) and (\ref{g}), we find
that the best fit to the data is obtained by exhausting $113 \, \%$ of the
EWSR. This value is very close to the $110\,\%$ obtained
by Barrette {\it et al} \cite{Bar88}.

In Fig. \ref{a-f10} we also show the result of a semiclassical
calculation (solid curve) for
the GDR$_{iv}$ excitation in lead, using Eq.~(\ref{dn2}) for the virtual
photon numbers. One observes that the semiclassical curve is not
able to fit the experimental data. This is mainly because diffraction
effects and strong absorption are not included. But the semiclassical
calculation displays the region of relevance for Coulomb excitation.
At small angles
the scattering is dominated by large impact parameters,
for which the Coulomb field is weak.  Therefore the Coulomb excitation
is small and the semiclassical approximation fails.
It also fails
in describing the large angle data (dark-side of the rainbow angle),
since absorption is not treated properly.  One sees that there is a ``window"
in the inelastic scattering data near $\theta=2-3^\circ$ in which
the semiclassical and full calculations give approximately the same
cross section.

In Fig. \ref{a-f11} we perform the same calculation, but for the
excitation of the GDR, the isoscalar giant quadrupole resonance
(GQR$_{is}$),
and the isovector quadrupole resonance (GQR$_{iv}$), in Pb for
the collision $^{208}$Pb + $^{208}$Pb at 640A MeV.
The solid (dotted) [dashed-dotted] line is the differential
cross section for
the excitation of the GDR (GQR$_{is}$) [GQR$_{iv}$]. The dashed line
is the result of a semiclassical calculation.
Here we see that a
purely semiclassical calculation, using Eq. (\ref{nE1ann}) is able to
reproduce the quantum results up to a maximum scattering angle $\theta_m$,
at which strong absorption sets in. This justifies the use of semiclassical
calculations for heavy systems, even to calculate angular distributions.
The cross sections increase rapidly with increasing scattering angle,
up to an approximately constant value as the maximum Coulomb scattering
angle is neared. This is explained as follows.
Very forward angles correspond to large impact
parameter collisions in which case
$\omega b/\gamma v > 1$ and the excitation of
giant resonances in the nuclei
is not achieved. As the impact parameter decreases, increasing the
scattering angle, this adiabaticity condition is fulfilled
and excitation occurs.

As discussed above, the semiclassical result works for large $Z$ nuclei
and for relativistic energies where the approximation of
Eq. (\ref{Omegc}) is justified. However, angular distributions are not
useful at relativistic energies since the scattering is concentrated at
extremely forward angles. The quantity of interest in this case is the
total inelastic cross  section. If we use a sharp-cutoff model for the
strong absorption, so that $\chi_I(b)=\infty$ for $b<b_{min}$ and 0 otherwise,
then Eqs. (\ref{n}) and (\ref{g}) yield the same result as an integration
of the semiclassical expression, Eq. (\ref{dn3}), from $b_{min}$ to $\infty$.
In fact, this result has been obtained earlier in Ref. \cite{BB88}.

\subsection{Excitation and photon decay of the GDR}

We now consider the excitation of the target nucleus
to the giant dipole resonance and the subsequent photon decay of that excited
nucleus, leaving the target in the ground state.  Experimentally, one detects
the inelastically scattered projectile in coincidence with the decay photon
and
demands that the energy lost by the projectile is equal to the energy of
the detected photon.
To the extent that the excitation mechanism is dominated by Coulomb
excitation,
with the exchange of a single virtual photon,
this reaction is very similar to the photon scattering reaction, except that
in the present case the incident photon is virtual rather than real.  In this
section, we investigate whether the connection between these two reactions
can be formalized.

We first review the excitation mechanism.
The physical situation is that of a heavy ion of energy $E$ incident on a
target.
The projectile loses an energy $\Delta E$ while scattering through an angle
$\theta$.  We have shown that, under the conditions $\Delta E/E \ll 1$,
the cross
section for excitation of the target nucleus partitions into the
following expression (we assume that the contribution of the
$E1$-multipolarity is dominant):
\begin{equation}
{d^{2}\sigma_C \over d \Omega d E_\gamma} \ \bigl( E_\gamma \bigl)
={1 \over E_\gamma} \
{d n_\gamma \over d\Omega}\ \bigl( E_\gamma \bigl) \ \sigma_\gamma
\ \bigl( E_\gamma \bigl),
\label{dsig1}
\end{equation}
where $\sigma_\gamma
\bigl( E_\gamma \bigl)$ is the photonuclear cross section for the
absorption of a real photon with energy $E_\gamma=\Delta E$ by
the target nucleus, and
the remaining terms on the right-hand-side
are collectively the number of virtual photons per
unit energy with energy
$E_\gamma$.  This latter quantity depends on the kinematics of the scattered
heavy ion and on the optical potential but is otherwise independent of the
target degrees of freedom.  This partitioning allows one to relate the
excitation cross section to the photoabsorption cross section.

Now, the usual way to write the cross section
$d^{2}\sigma_{C\gamma}/d\Omega dE_\gamma$
for the
excitation of the target followed by photon decay to the ground state
is simply to multiply
the above expression
by a branching ratio $R_\gamma$, which represents the probability
that the nucleus excited to an energy $E_\gamma$
will emit a photon leaving it in the
ground state \cite{Bee90}:
\begin{equation}
{d^{2}\sigma_{C\gamma} \over d \Omega d E_\gamma} \ \bigl( E_\gamma \bigl)
={1 \over E_\gamma} \
{d n_\gamma \over d\Omega} \ \bigl( E_\gamma \bigl)\ \sigma_\gamma \
\bigl( E_\gamma \bigl)
\ R_\gamma
\ \bigl( E_\gamma \bigl).
\label{dsig2}
\end{equation}
Instead, we propose the following expression, in complete analogy with
Eq.~(\ref{dsig1}):
\begin{equation}
{d^{2}\sigma_{C\gamma} \over d \Omega d E_\gamma} \ \bigl( E_\gamma \bigl)
={1 \over E_\gamma} \
{d n_\gamma \over d\Omega}\ \bigl( E_\gamma \bigl) \ \sigma_{\gamma\gamma}
\ \bigl( E_\gamma \bigl),
\label{dsig3}
\end{equation}
where $\sigma_{\gamma\gamma}
\bigl( E_\gamma \bigl)$ is the cross section
for the elastic scattering  of photons with
energy $E_\gamma$.  Formally, these expressions are equivalent in that
they simply define the quantity $R_\gamma$.  However, if one treats
$R_\gamma$ literally as a branching ratio, then
these expressions are
equivalent only if it were true that the photon scattering cross
section is just product of the
the photoabsorption cross section and the branching ratio.
In fact, it is well-known from the theory of photon scattering that the
relationship between the photoabsorption cross section and the photon
scattering cross section is more complicated \cite{Na91}.
In particular, it is not correct to think of
photon scattering as a two-step process consisting of
absorption, in which the target nucleus is excited to an intermediate state
of energy
$E_\gamma$, followed by emission, in which the emitted
photon has the same energy $E_\gamma$.
Since the intermediate state is not observable, one must sum over all
possible intermediate states and not just those allowed by conservation of
energy.  Now, if the energy $E_\gamma$ happens to coincide with a narrow
level, then that level will completely dominate in the sum over intermediate
states.  In that case, it is proper to regard the scattering as a two-step
process in the manner described above, and the two
expressions for the cross
section will be equal.  However, for $E_\gamma$ in the nuclear continuum
region (e.g., in the region of the GDR), this will not be the case, as
demonstrated in the following discussion.

We again consider the inelastic scattering of
$^{17}$O projectiles of energy $E_{lab}=84$  MeV/nu\-cleon from a
$^{208}$Pb nucleus at an angle of 2.5$^{\circ}$.  We use
Eq.~(\ref{dn}) to calculate the $E1$ virtual photon number and we use
a Lorentzian
parameterization of the GDR of $^{208}$Pb.  We calculate $R_\gamma$ and
$\sigma_{\gamma\gamma}$ according to the prescriptions of
Ref.~\cite{Bee90}
and Ref.~\cite{Na91}, respectively; in both cases we neglect the
statistical contribution to the photon decay.  The results are compared in
Fig. \ref{a-f12},
where is is very evident that they make very different predictions for
the cross section, especially in the wings of the GDR.

We next use our expression to compare directly with the data of Ref.
\cite{Bee90}.
For this purpose, we again calculate $\sigma_{\gamma\gamma}$ using the
formalism
of Ref.~\cite{Na91}, which relates $\sigma_{\gamma\gamma}$ to the total
photoabsorption.  For the latter, we use the numerically-defined data set of
Ref.~\cite{Ve70} rather than a Lorentzian parameterization.  The effect of
the underlying compound nuclear levels (i.e., the statistical contribution
to the photon scattering) is also included.  The calculation is compared to
the
data in Figs. \ref{a-f13} and
\ref{a-f14}.  Fig. \ref{a-f13} shows the cross section for the excitation of
the GDR without the detection of the decay photon.  The agreement with the
data
is excellent, giving us confidence that our calculation of the virtual photon
number as a function of $E_\gamma$ is correct.  Fig. \ref{a-f14} shows the
cross section for the excitation-decay process as a function of $E_\gamma$.
Although the qualitative trend of the data are well described,
the calculation systematically overpredicts
the cross section on the high-energy side of the GDR (solid curve).  If the
Thompson amplitude is not included in $\sigma_{\gamma\gamma}$, the calculation
is in significantly better agreement with the data (dashed curve).

\subsection{Nuclear excitation and strong absorption}
\medskip
Up to this point we have only considered the Coulomb excitation of the
nuclei, without accounting for nuclear excitation. But,
in peripheral collisions, the nuclear interaction between the ions can
also induce excitations. This can be easily calculated in a vibrational
model.
The amplitude for the
excitation of a vibrational mode
by the nuclear interaction in
relativistic heavy ion collisions
can be obtained assuming that
a residual interaction U between the projectile and the target exists,
and that it is weak. According to the Bohr-Mottelson
particle-vibrator
coupling model, the matrix element for the transition $i\longrightarrow
f$ is given by
\begin{equation}
V^{N(\lambda\mu)}_{fi} ({\bf r})
\equiv <I_fM_f| U |I_iM_i>=
{\delta_\lambda \over \sqrt{2\lambda +1}}
\ <I_fM_f| Y_{\lambda \mu}|I_iM_i>
\ Y_{\lambda \mu} (\hat {\bf r})
\ U_{\lambda} (r) \label{VfiN}
\end{equation}
where $\delta_\lambda=\beta_\lambda R$
is the vibrational amplitude, or {\it
deformation length},
$R$ is the nuclear radius, and $U_{\lambda}
(r)$ is the transition potential.

The deformation length $\delta_\lambda$ can be directly related to the
reduced matrix elements for electromagnetic transitions. Using well-known
sum-rules for these matrix elements one finds a relation between the
deformation length
and the nuclear masses and sizes.
For isoscalar excitations one gets \cite{Sa87}
\begin{equation}
\delta_0^2= 2 \pi \ {\hbar^2 \over m_N\ <r^2>} \
{1 \over A E_x} \ , \ \ \ \
\delta_{\lambda \geq 2}^2 = {2 \pi \over 3} \ {\hbar^2 \over m_N} \
\lambda \ (2\lambda +1) \ {1\over A E_x}\label{d0}
\end{equation}
where $A$ is the atomic
number, $<r^2>$ is the r.m.s. radius of the nucleus,
and $E_x$ is the excitation energy.

The transition potentials for nuclear excitations can be related to the
optical potential in the elastic channel.
The basic idea is that the interaction between the
projectile and the target induces surface vibrations in the target. Only
the contact region between the nuclei in grazing collisions is
of relevance. One thus expects that the interaction potential
is proportional to the derivatives of the optical potential in the
elastic channel, which peak at the surface.
This is discussed in details in Ref. \cite{Sa87}.
The transition potentials for isoscalar excitations are
\begin{equation}
U_0 (r) = 3 U_{opt} (r) + r {d U_{opt} (r) \over dr}
\ ,\label{U0}
\end{equation}
for monopole, and
\begin{equation}
U_2 (r)= {dU_{opt} (r) \over dr} \ , \label{U2}
\end{equation}
for quadrupole modes.

For dipole isovector excitations
\begin{equation}
\delta_1= {\pi \over 2} \ {\hbar^2 \over m_N}
\ {A \over NZ} \ {1\over E_x}\ , \label{d1}
\end{equation}
where $Z$ ($N$) the charge (neutron)  number. The transition potential
in this case is \cite{Sa87}
\begin{equation}
U_1(r)=\chi \ \Big( {N-Z \over A} \Big) \
\Big( {dU_{opt} \over dr} + {1\over 3} \ R_0 \ {d^2 U_{opt}
\over dr^2} \Big)
\ ,\label{U1}
\end{equation}
where the factor
$\chi$ depends on the difference between the proton and the neutron
matter radii  as
\begin{equation}
\chi {2(N-Z)\over 3A} = {R_n-R_p \over {1\over 2} \ (R_n+R_p)}
= {\Delta R_{np} \over R_0}
\ .\label{chi}
\end{equation}
Thus, the strength of isovector excitations increases with
the difference
between the neutron and the proton matter radii.
This difference is accentuated for neutron-rich
nuclei and should be a good test for the quantity $\Delta R_{np}$
which enters the above equations.

The time dependence of the matrix elements above can be obtained
by making a Lorentz boost. Since the potentials
$U_\lambda\Big[r(t)\Big]$ peak strongly
at $t=0$, we can safely approximate $\theta (t)
\simeq \theta(t=0) =\pi/2$ in the
spherical harmonic of Eq.~(\ref{VfiN}). One gets
\begin{eqnarray}
V^{N(\lambda\mu)}_{fi} ({\bf r})
&\equiv& <I_fM_f| U |I_iM_i>
\nonumber \\
&=&\gamma \
{\delta_\lambda \over \sqrt{2\lambda +1}}
\ <I_fM_f| Y_{\lambda \mu}|I_iM_i>
 Y_{\lambda \mu} \Big(\theta={\pi\over 2}\Big)
\ U_{\lambda} [r(t)]  \ ,\label{VfiN2}
\end{eqnarray}
where
$
r(t)=\sqrt{b^2+\gamma^2 v^2 t^2}
$.

Using the Wigner-Eckart theorem, the matrix element of the
spherical harmonics becomes
\begin{eqnarray}
<I_fM_f| Y_{\lambda \mu}|I_iM_i> =
(-1)^{I_f-M_f} \
\biggl[ {(2I_i+1)(2\lambda+1) \over 4 \pi (2I_f+1)}
\biggr]^{1/2} \
\biggl( {I_f \atop -M_f}{\lambda \atop \mu}
{I_i \atop M_i} \biggr)
\biggl( {I_f \atop 0}{\lambda \atop 0}
{I_i \atop 0} \biggr)
\ .\nonumber \\
\label{WE}
\end{eqnarray}

For high energy collisions, the optical potential $U(r)$
can be constructed by using the t-$\rho\rho$ approximation
\cite{HRB91}, as given by Eq. (\ref{Ur}).

We are not interested here in diffraction and
refraction effects in the scattering, but on the excitation
probabilities for a given impact parameter.
The strong absorption occurring in collisions with small
impact parameters can be included. This can be done
by using the eikonal approximation and the
optical potential, given by Eq.~(\ref{Ur}).
The practical result is that the excitation probabilities
for a given impact parameter $b$, including the sum of the
nuclear and the Coulomb contributions to the excitation,
are given by
\begin{equation}
P_{fi}(b)= \Big| a_{fi}^C(b)+
a_{fi}^N(b) \Big|^2 \ \exp\Big\{
- \sigma_{NN}
\ \int dz \int d^3r \rho_1({\bf r}')\
\rho_2({\bf r-r'})
\Big\}
\ ,\label{abs}
\end{equation}
where
$r=\sqrt{b^2+z^2}$. The corresponding excitation
cross sections are obtained by an integration of the
above equation over impact parameters.

\subsection{Nucleon removal in peripheral relativistic heavy ion collisions}

In Fig. \ref{a-f15} we plot the GDR excitation probability in Pb as a
function
of the impact parameter, for the system $^{208}$Pb + $^{208}$Pb at 640A
MeV.
We use 100\% of the sum rule to calculate the $B(E1)$-value for the
electromagnetic
excitation of an isolated GDR state at 13.5 MeV. In the solid line, we
consider absorption
according to Eq.~(\ref{abs}). In the construction of the optical
potential we used the g.s. densities calculated from the droplet model of
Myers and Swiatecki~\cite{MS}
in accordance
with Shen et al. \cite{shen}.
We will call it by soft-spheres model.

As shown in Ref.~\cite{abs}, this
parametrization yields the best agreement between experiment and theory.
The dashed line does not include absorption. To simulate strong absorption
at low impact parameters, we use $b_{min} = 15.1$~fm as a lower limit
in the impact parameter integration of Eq.~(\ref{sigman}). This value was
chosen such as to lead to the same cross section as that obtained from the
solid line. However, a more detailed comparison of the soft-sphere model for
strong
absorption and a simple semiclassical calculation, based on a single parameter
$b_{min}$, is described next \cite{abs}.

In Fig. \ref{a-f16}, we plot the nuclear contributions to the
excitation
probability as a function of the impact parameter. We study
the excitation of the isoscalar giant monopole resonance
(GMR$_{is}$), the GDR$_{iv}$, and the GQR$_{is}$,
in lead for the collision $^{208}$Pb + $^{208}$Pb at 640A MeV.
The GMR$_{is}$ in $^{208}$Pb is
located at 13.8 MeV.  As discussed previously, isovector
excitations are suppressed in nuclear excitation processes, due
to the approximate charge independence of the nuclear
interaction.  We use the formalism of this section, with the
deformation parameters such that  100\% of the sum rule is
exhausted. This corresponds to the monopole amplitude
$\alpha_0=0.054$. The GDR$_{iv}$ and GQR$_{is}$ deformation parameters
are $\delta_1=0.31$ fm and $\delta_2=0.625$ fm, respectively.  The
GQR$_{iv}$ excitation probability is much smaller than the other
excitation probabilities and is, therefore, not shown. The
nuclear excitation is peaked at the grazing impact parameter and
is only relevant within an impact parameter range of $\sim 2$
fm. Comparing to Fig. \ref{a-f15}, we see that these excitation
probabilities are orders of magnitude smaller than those for
Coulomb excitation. Consequently, the corresponding cross
sections are much smaller. We get 14.8 mb for the isoscalar GDR,
2.3 mb for the GQR$_{is}$, and 2.3 mb for the GDR$_{iv}$. The
interference
between the nuclear and the Coulomb excitation is also small and
can be neglected.

Since they are high lying states above the continuum, giant resonances mostly
decay
by particle emission (mainly neutron emission in heavy nuclei).
Therefore data on neutron removal in
relativistic heavy ion collisions is an appropriate comparison
between theory and experiment. As we have seen, above,  nuclear  excitation
of GR's
contribute very little to the cross section, as compared to Coulomb
excitation.
However, strong interactions at peripheral collisions also contribute to
"direct"
knockout (or stripping) of neutrons, and also should be considered. It has
been observed \cite{BB88}, however, that neutron removal cross sections
induced
by strong interactions scale with $A_1^{1/3}+A_2^{1/3}$, while the Coulomb
excitation
cross sections scale with the projectile's charge as $Z_2^2$, approximately.
One can thus
separate the nuclear contribution for the nucleon
removal of a target (or projectile) by measuring the cross sections
for different projectiles (or targets).

In the semiclassical approach the total cross section for
relativistic Coulomb excitation is obtained by integrating the
excitation probabilities over impact parameter, starting from a
minimum value $b_{min}$. It is assumed that below this minimum value
the interaction is exclusively due to the strong interaction
("sharp-cutoff" approximation).
It has been found that with this approximation
the Coulomb cross sections are very sensitive to
the parameterization of the minimum impact parameter
\cite{Ri93,Sc93,Au93,BZ93}.

One commonly used parameterization at relativistic energies
is  that of Benesh et al.\cite{BCV89},
fitted to Glauber-type calculations and reading
\begin{equation}
b_{min}^{BCV}=1.35\cdot( A_p^{1/3}+A_t^{1/3}-0.75
\cdot(A_p^{1/3}+A_t^{1/3}) \;\;\mbox{fm}
\label{b_bcv}
\end{equation}
which we refer to hereafter as "BCV".
In Ref. \cite{BCV89} a detailed study has been made
concerning the parameterization
procedure of the minimum impact parameter. It was also found that
the nuclear contribution to the neutron removal channels in
peripheral collisions has a negligible interference with the
Coulomb excitation mechanism. This is a very useful result since the
Coulomb and nuclear part of the cross sections may be treated
separately.

The other parametrization
is that of
Kox et al. \cite{Ko87} which reproduced well measured total reaction
cross sections of light and medium-mass systems:
\begin{equation}
b_{min}^{Kox}=1.1 \ \bigg( A_p^{1/3}+A_t^{1/3}+1.85
{A_p^{1/3}\ A_t^{1/3} \over {A_p^{1/3}+A_t^{1/3}}}
-1.9 \bigg) \;\; \mbox{fm}  \ .
\label{b_kox}
\end{equation}

This parametrization has been used previously \cite{Au93} and
a reasonable
agreement with the measured data for 1n cross sections was
found. It should be noted,
however, that the Kox parametrization of total interaction cross sections
has been derived mainly from experiments with light projectiles and that its
application to heavy systems involves an extrapolation into a region where
no data points are available.

To achieve a good a comparison with experimental data
on neutron-removal cross sections we will use
the experimental photo-neutron emission cross sections
from Refs. \cite{Ve70,saclay}.
A Lorentzian  fit to the $(\gamma,n)$-data is used to
parameterize the GDR in gold. The parameters are 13.72 MeV excitation
energy, a width of 4.61 MeV, and a strength of 128\% of the TRK sum rule.
The Lorentz
parameters for the isoscalar (isovector) GQR are taken as
10.8 (23.0) MeV for the excitation energy, 2.9 (7.0) MeV for the width,
and we assume 95\% exhaustion of the respective sum rules \cite{Ber76}.
With these parameters we calculate the excitation cross sections
$d\sigma(E)/dE$ for dipole- and quadrupole-excitations.
The respective neutron emission cross sections are given by
\begin{equation}
\sigma_{n} = \int {d\sigma(E)\over dE} f_{n}(E) dE,
\label{sigrem}
\end{equation}
where $f_{n}(E)$ is the probability to evaporate one neutron at
excitation energy $E$.
$f_{n}(E)$ is taken from the experimental
$(\gamma,n)$-data at low $E$ and from a statistical decay
calculation with the code HIVAP \cite{hivap}
for excitation energies above 20 MeV.
Since the three-neutron emission threshold in gold is above the energy
of the GDR state, this channel
is fed mainly by the two-phonon excitation mechanism,
while the 1n cross section is dominated by the excitation of the GDR.

We expect that the BCV
parametrization of $b_{min}$ should yield
similar results as the soft-spheres calculation since it was derived in
fitting the complementary process, the nuclear interaction, calculated
also with Glauber theory. Fig. \ref{a-f17} shows that this expectation
could be  verified: the soft-spheres calculation for 1n-removal from
$^{197}$Au by ED processes (upper full curve) is almost
indistinguishable from a
sharp-cutoff calculation using $b_{min}^{BCV}$ (upper dotted curve).
This remarkable agreement tells us that for practical purposes we can
avoid the extra numerical complication connected with the use of
a soft spheres model and corroborates the use of $b_{min}^{BCV}$
in sharp-cutoff calculations in earlier works \cite{LB92,hill}.
We also think that the soft-spheres calculation (and
the sharp-cutoff calculation using
$b_{min}^{BCV}$)
is physically better justified than the Kox parametrization
\cite{Ko87} since the former is derived from
realistic nuclear density distributions, whereas the latter is an
extrapolation of measured total reaction cross sections into a region
where no data points are available.

We will return later to discuss the other data points of
Fig. \ref{a-f17} when we treat the problem of the excitation of
multiphonon states. However, it is worthwhile noticing that a point
in the above curve, for uranium targets, is not well reproduced by
the theory. In fact, this has been observed in other experiments \cite{Ju93},
and deserves a special treatment.

\subsection{Excitation by a deformed nucleus}

Either by using the soft-sphere model, or by means of a semiclassical
calculation,
Coulomb excitation by a relativistic projectile, or target,
is well described
theoretically if the charge distribution of the projectile is
spherically symmetric \cite{norbury}.

However, there was found a discrepancy between theory and experiment with data
with deformed projectiles, as measured by Justice {\it et al}. \cite{Ju93} for
uranium projectiles. This problem was studied theoretically by Bertulani in
\cite{Bert93},
and we will briefly discuss it here.
To obtain a qualitative  insight of the effects we shall consider a prolate
deformed projectile with a variable deformation.

In the frame of reference of the projectile the Coulomb field
at a position ${\bf r}$ with respect to the center-of-charge of the
distribution is
given by
\begin{equation}
\phi({\bf r})=4 \pi \ \sum_{\lambda,\mu}
{1\over 2\lambda+1} \ {1\over r^{\lambda+1}}\  Y^*_{\lambda\mu}
(\theta,\ \phi) \ {\cal M} (E\lambda\mu ) \ ,
\label{phi_n}
\end{equation}
where
\begin{equation}
{\cal M} (E\lambda\mu )=\int \rho({\bf r'}) \ r'^l \ Y_{\lambda\mu}
(\hat {\bf r}')\ d^3 r'
\ , \label{Mlm}
\end{equation}
with $\rho({\bf r})$  equal to the ground state charge distribution of the
projectile. For simplicity we will consider a uniform spheroidal
charge distribution with the z-axis along the symmetry axis. The
charge distribution drops to zero for distances to the center greater
than the angle-dependent radius
\begin{equation}
R(\theta)=R_0\Bigl(1+\beta Y_{20}(\theta)\Bigr) \ .
\label{R}
\end{equation}

In lowest order in the multipole expansion, Eq. (\ref{phi_n}) becomes
\begin{equation}
\phi({\bf r})={Z_1e\over r}+\sqrt{\pi \over 5} \ {1\over r^3} \ Y_{20}
(\theta) \ Q^{(c)}_0
\label{phia}
\end{equation}
where $Q_0^{(c)}$ is the quadrupole moment of the charge distribution,
\begin{equation}
Q^{(c)}_0={3\over \sqrt{5 \pi}} \ Z_1 e \ R_0^2 \ \beta \ (1+0.16 \beta) +
{\cal O}(\beta^2) \ .
\label{Q}
\end{equation}

To obtain the (time-dependent) field in the frame of reference of the target
we perform a Lorentz transformation of Eq. (\ref{phia}). For a
straight-line trajectory one finds
\begin{equation}
\phi({\bf r}', t)={\gamma Z_1 e\over r'}+
\gamma \ \sqrt{\pi \over 5}\  {1\over r'^3} \ Y_{20} (\theta')\  Q^{(c)}_0
\label{phi_2}
\end{equation}
where $r'=\sqrt{b^2+\gamma^2v^2t^2}$, with $b$ equal to the impact
parameter, $v$ the projectile velocity, and $\gamma=(1-v^2/c)^{-1/2}$.

The first term in the above equation is the well-known Li\'enard-Wiechert
potential of a relativistic charge. It gives rise to monopole-multipole
excitations of the target, which we have discussed so far.
The second term accounts for
quadrupole-multipole excitations of the target and is a correction due
to the deformation of the projectile. This field will depend on the
orientation of the projectile with respect to its trajectory (see Fig.
\ref{a-f18}).
We can separate the orientation angles from the angular position of the
projectile (along its trajectory) with respect to the target by using the
identity
\begin{equation}
Y_{20}(\theta')=\sqrt{4 \pi \over 5} \sum_m Y_{2m}(\theta, \phi) \
Y_{2m}(\chi',0)
\label{Y} \ ,
\end{equation}
where $(\theta, \phi)$ denotes the orientation of the projectile symmetry
axis with respect to the bombarding axis and
$\chi' = \gamma vt / r'(t) $.

The dipole excitation of the target is the most relevant and we shall restrict
ourselves to this case only \cite{BB88}. At a point  ${\bf r}\equiv (x,y,z)$
from the center of mass of the target the field is obtained by replacing
${\bf r}'=(b,0,\gamma vt)$ by $[b-x,y,\gamma (vt-z)]$ in Eq. (\ref{phi_2}).
The excitation amplitude to first order
is given by Eq. (\ref{3.13}).
Using the continuity  equation and
expanding (\ref{3.13}) to lowest order in ${\bf r}$ we find
\begin{equation}
a_{fi}=a_{fi}^{(1)}+a_{fi}^{(2)}
\label{afi2}
\end{equation}
where
\begin{equation}
a_{fi}^{(1)}=-i\ {2Z_1e\over \hbar v}\  {\xi\over b}\
\biggl\{ K_1(\xi) D_{fi}^{(x)} +i \ {1\over \gamma} \ K_0(\xi) D_{fi}^{(z)}
\biggr\}
\label{ae1}
\end{equation}
and
\begin{equation}
a_{fi}^{(2)}=-i \ {4\pi \over 5} {Q^{(c)}_0\over \hbar v} \
{\xi^2\over b^3} \
\biggl\{ K_2(\xi) D_{fi}^{(x)} +i\ {1\over \gamma} \ K_1(\xi) D_{fi}^{(z)}
\biggr\}\
\sum_m Y_{2m}(\theta,
\phi)\  Y_{2m}({\pi\over 2},0)
\label{ae2}
\end{equation}
where $\xi=\omega b/\gamma v$ and $K_i$ is the modified Bessel function of
order $i$.
To simplify the notation, we have used the Cartesian definition of the
matrix elements.
The dipole matrix elements for the nuclear excitation are
given by
\begin{equation}
D^{x(z)}_{fi}=<f|x(z)|i> \ .
\label{Dxz}
\end{equation}
In terms of 
spherical coordinates, $D_{fi}^{(x)}={\cal M}_{fi}(E1,
1)+{\cal M}_{fi} (E1, -1)$
and $D_{fi}^{(z)} = {\cal M}_{fi} (E1, 0)$.
Thus, Eq. (\ref{ae1}) is equal to Eq. (\ref{a1st}).

In the expression (\ref{ae2}) we have used the approximation
$Y_{2m}(\chi,0)\simeq Y_{2m}({\pi \over 2}, 0)$ which is valid for high energy
collisions since the quadrupole field is strongly peaked at $t=0$,
corresponding to the distance
of closet approach.
Eqs.  (\ref{ae1})  and  (\ref{ae2})  allow  us  to  calculate  the   dipole
excitation cross
section by integrating their absolute squares over impact parameter, starting
from a minimum impact parameter for which the strong interaction sets in.
Neglecting the diffuseness of the matter distribution of the nuclei we can
write (see Fig. \ref{a-f18})
\begin{equation}
b_{min}(\theta)\simeq R_1+R_2\Bigl[1+\beta \ Y_{20}({\pi\over 2}+\theta)\Bigr]
\label{bmin}
\end{equation}
with the nuclear radii given by $R_i=1.2 \cdot A_i^{1/3}$. The total
cross section is
\begin{equation}
\sigma=2\pi \int_{{b_{min}}(\theta)} db \ b\ < | a_{fi}(b, \Omega)|^2 >
\label{sigma}
\end{equation}
where the $<\cdots>$ sign means that an average over all the possible
orientations of the
projectile, i.e., over all angles $\Omega=(\theta,\ \phi)$, is done.

We will apply the above formalism to the Coulomb excitation of
$^{208}$Pb by $^{238}$U  projectiles. We will give the $^{238}$U an
artificial deformation in the range $\beta=0-1$ to check the
dependence of the cross sections with this parameter. The cross section
given above contains three terms: $ \sigma=\sigma_1+\sigma_2+\sigma_{12}$.
$\sigma_1$ is due to the monopole-dipole excitation amplitude, $\sigma_2$ is
due to
the quadrupole-dipole excitation amplitude, and $\sigma_{12}$ is the
interference between them.

In Fig. {\ref{a-f19} we present the results for the numerical
calculation  of the quantity
\begin{equation}
\Delta=100 \times {\sigma_1-\sigma^{\beta=0}_1\over \sigma^{\beta=0}_1}
\label{delta}
\end{equation}
which is the percent correction
of dipole excitations  in $^{208}$Pb by a uranium projectile
due the average over the orientation of the projectile.
$\sigma^{\beta=0}_1$ is the cross section for $\beta=0$.
We present results for three bombarding energies, 10A GeV,
1A GeV  and 100A MeV, and as a function
of $\beta$. The quantity defined by Eq.  (\ref{delta}) is independent of the
nature of the state excited, since the dipole matrix elements cancel out.
They depend on the energy of the state.
In order to see how the effect depends qualitatively on the
energy of the state we used three different excitation energies
$E_{fi}=1$, 10 and 25 MeV, respectively. These correspond to the
dotted, dashed and solid lines in Fig. \ref{a-f19}, respectively.

One observes from Fig. \ref{a-f19} that the deformation effect
accounted for by an average of the
minimum impact parameter which enters Eq. (\ref{sigma}) increases
the magnitude of the cross section. Thus
the average is equivalent to a smaller "effective" impact
parameter, since the cross sections increase with decreasing
values of $b_{min}$.
The effect
is larger
the greater the excitation energy is.  This effect
also decreases with the bombarding energy. For very high
bombarding energies
it is very small even for the largest deformation. These
results can be explained as
follows. The Coulomb excitation cross section at very high bombarding
energies, or very small excitation energies,
is proportional to  $\ln \Bigl[\omega b_{min}(\theta)/\gamma v)
\Bigr]$. Averaging
over orientation of the projectile means an  average of $\ln(b_{min})$ due to
the additivity law of the logarithm. One can easily do this average  and the
net result is a rescaling of $b_{min}$ as $fb_{min}$, with f smaller,
but very close to one.

For high excitation energies,  or small bombarding energies, the cross section
is proportional to $\exp\Bigl\{-2\omega b_{min}(\theta)/\gamma v\Bigr\}$ due
to
the adiabaticity condition \cite{WA79}. Thus, in these situations, the cross
section
is strongly dependent on  the average over orientation due to the strong
variation of the exponential function with the  argument.

Now we consider the effect of the second term of Eq. (\ref{phi_2}), namely of
the
quadrupole-dipole excitations. In Fig. \ref{a-f20}
we show the  excitation of
a giant resonance dipole state in lead ($E_{fi}=13.5$ MeV)
due to the second term Eq. (\ref{phi_2}), as a function the deformation
parameter $\beta$ and for a bombarding energy of 100A MeV.
We assume that the giant dipole state exhausts fully the TRK sum-rule,
Eq. (\ref{5.4a}), in lead.
Now the average over orientation also includes
the dependence of the quadrupole-dipole interaction on $\Omega=(\theta,
\phi)$.
As expected the cross section increases with $\beta$. But it is small as
compared to the monopole-dipole excitations even for a large deformation. At
this beam energy the monopole-dipole excitation is of order of 1 barn.

The total cross section contains an interference between the amplitudes
$a_{fi}^{(1)}$ and $a_{fi}^{(2)}$. This is shown in Table \ref{a-t2}
for 100A MeV for which the effect is larger. The second
column gives the cross sections for monopole-dipole excitations
of a giant resonance dipole state in lead. The
effect of the orientation average can be seen as an
increase of the cross section as compared to the value in the
first row (zero deformation). For $\beta =0.3$ which is approximately the
deformation parameter for $^{238}$U the correction to the cross
section is negligible. In the third column the cross section
for quadrupole-dipole excitation are given. They are also much smaller
than those for the monopole-dipole excitations. The total
cross sections, given in the last column, are also little dependent
on the effect of the deformation. For $\beta=0.3$ it corresponds
to  an increase of 3\%  of the value of the original cross section (first
row ). This effect also decreases with the bombarding
energy. For 1A GeV, $\sigma^{\beta=0}=5922$ mb, while
$\sigma=5932$ mb for $\beta=0.3$,
with all effects included.

\begin{table}[tb]
\caption[ ]
{Cross sections (in mb) for Coulomb excitation
of the giant dipole resonance in
$^{208}$Pb by  $^{238}$U projectiles at  100A MeV.
In the second (third) column the cross sections are due to
the monopole (quadrupole)- dipole interaction.  The last column
is the total cross section. An average over the orientation of
the projectile was  done. A
realistic value of the deformation of $^{238}$U corresponds to
$\beta\simeq 0.3$. But, a variation of $\beta$ is used to obtain an insight
of the
magnitude of the effect.
\label{a-t2}}
\begin{center}
\begin{tabular}{ l l l l l r }
\hline
$\beta$&$\sigma_1$ [mb]&$\sigma_2$ [mb]&$\sigma$ [mb] \\ \hline
0&1171&0&1171 \\ \hline
0.1&1173&0.179&1174 \\
0.2&1179&0.748&1184 \\
0.3&1189&1.773&1200 \\
0.4&1202&3.34&1224 \\
0.5&1220&5.57&1242 \\
0.6&1241&8.61&1291 \\
0.7&1265&12.6&1335 \\
0.8&1294&17.9&1389 \\
0.9&1326&24.7&1446 \\
1&1362&33.3&1522 \\ \hline
\end{tabular}
\end{center}
\end{table}

In conclusion, the effect of excitation by a deformed projectile, which can
be studied  by averaging over the
projectile orientation, is to
increase slightly the cross sections. The inclusion of the
quadrupole-dipole interaction increases the cross
section, too. However, these corrections are small for realistic deformations.
They cannot be responsible for the large deviations of the experimental values
of
the Coulomb fragmentation
cross sections from the standard theory \cite{BB88,WA79}, as has been
observed \cite{Au93,Ju93} for deformed projectiles.

\section{\bf Heavy ion excitation of multiphonon resonances}
\subsection{Introduction}

Much of the interest on multiphonon resonances relies on the
possibility of looking at exotic particle decay of these states. For
example, in Ref. \cite{Va89} a hydrodynamical model was used to predict the
proton and neutron dynamical densities in a multiphonon state of a nucleus.
Large proton and neutron excesses at the surface are developed in a
multiphonon state. Thus, the emission of exotic clusters from the decay of
these states are a natural possibility. A more classical point of view is
that the Lorentz contracted Coulomb field in a peripheral relativistic heavy
ion collision acts as a hammer on the protons of the nuclei \cite{BB88}.
This (collective) motion of the protons seem only to be probed in
relativistic Coulomb excitation. It is not well known how this classical
view can be related to microscopic properties of the nuclei in a multiphonon
state.

Since there is more energy deposit in the nuclei, other decay channels are
open for
the multiphonon states. Generally, the GR's in heavy nuclei decay by
neutron emission. One expects that
the double, or triple, GDR decays mainly in the 2n and 3n decay channel. In
fact,
such a picture has been adopted by \cite{abs,Au93} with success to
explain the
total cross sections for the neutron removal  in  peripheral  collisions.
The method is the same
that we used to explain the one-neutron removal cross sections, i.e., by
replacing $f_n$ by $f_{2n}$, and $f_{3n}$, in Eq. (\ref{sigrem}).

Although the perspectives for an experimental evidence of the DGDR via
relativistic Cou\-lomb excitation were good, on the basis of the large
theoretical cross sections, it was first found in pion scattering at the Los
Alamos Pion Facility \cite{Mo88}. In pion scattering off nuclei the DGDR can
be described as a two-step mechanism induced by the pion-nucleus
interaction. Using the Axel-Brink hypotheses, the cross sections for
the excitation of the DGDR with pions were shown to be well within the
experimental possibilities \cite{Mo88}. Only about 5 years later, the first
Coulomb excitation experiments for the excitation of the DGDR were performed
at the GSI facility in Darmstadt/Germany \cite{Ri93,Sc93}. In Fig.
\ref{a-f21} we
show the result of one of these experiments, which looked for the neutron
decay channels of giant resonances excited in relativistic projectiles. The
excitation spectrum of relativistic $^{136}$Xe projectiles incident on Pb
are compared with the spectrum obtained in C targets. A comparison of the
two spectra immediately proofs that nuclear contribution to the excitation
is very small. Another experiment \cite{Ri93} dealt with the photon decay of
the double giant resonance. A clear bump in the spectra of coincident photon
pairs was observed around the energy of two times the GDR centroid energy in
$^{208}$Pb targets excited with relativistic $^{209}$Bi projectiles.

The advantages of relativistic Coulomb excitation of heavy ions over other
probes (pions, nuclear excitation, etc.) was clearly demonstrated in several
GSI experiments \cite{Ri93,Sc93,Au93,Bo96}.

A collection of the experimental data on the energy and width of the DGDR is
shown in Fig. \ref{a-f22}. The data points are from a compilation from
pion (open
symbols), and Coulomb excitation and nuclear excitation (full symbols)
experiments \cite{CF95}.

The dashed lines are guide to the eyes. We see from Fig.
\ref{a-f22}(a) that the
energy of the DGDR agrees reasonably with the expected harmonic prediction
that the energy should be about twice the energy of the GDR, although small
departures from this prediction are seen, especially in pion and nuclear
excitation experiments. The width of the DGDR seems to agree with an average
value of $\sqrt{2}$ times that of the GDR, although a factor 2 seems also to
be possible, as we see from Fig. \ref{a-f22}(b). Fig. \ref{a-f22}(c)
shows the ratio between
the experimentally determined cross sections and the calculated ones. Here
is where the data appear to be more dispersed. The largest values of $\sigma
_{\exp }/\sigma _{_{th}}$ come from pion experiments, yielding up to a value
of 5 for this quantity.

We now discuss many features of
the double GDR excitation theoretically and some attempts to solve the
discrepancies between theory and experiment observed in Fig.
\ref{a-f22}.

\subsection{Perturbation theory and harmonic models}
\subsubsection{Sum rules for single and double resonances}

The simplest way to determine the matrix elements of excitation of
giant resonances is by means of sum rules under the assumption that those
sum rules are exhausted by collective states. We have done this when we used
the sum rules (\ref{5.4a},\ref{5.4b}). Let us look at these with more details,
since they will be useful for the determination of the matrix elements
for multiphonon excitations.
The
conventional sum rules for the dipole and quadrupole transitions,
derived without exchange and velocity-dependent corrections, are $(\hbar =1)$
\begin{equation}
\sum_f \omega_{fi} \Big\vert D_{fi}^{(m)}\Big\vert^2= {3\over 4 \pi}
\ {1\over
2 m_N}\ {NZ \over  A} \ e^2;
\label{13}
\end{equation}
\begin{equation}
\sum_f \omega_{fi} \Big\vert Q_{fi}^{(m)}\Big\vert^2= 2\ {1 \over
2 m_N}\ {3R^2 \over 4 \pi}\ e^2 \times
\left\{ \begin{array}{lll}
           \mbox{$Z^2/ A,$}    & \mbox{\rm isoscalar excitations,}\\
           \mbox{$NZ/ A,$}    & \mbox{\rm isovector excitations.}
\end{array}
\right. \ ,
\label{14}
\end{equation}
where $D^{(m)}\equiv {\cal M}(E1m)$ and $Q^{(m)}\equiv{\cal M}(E2m)$.

We explain our procedure on the example of the dipole sum rule (\ref{13}).
The right hand side $S_{D}$ of (\ref{13}) being calculated
for the fixed initial state $|i>$
in fact does not depend on the choice of $|i>$. (This dependence
is rather weak even if the exchange terms are taken into account).
Since $S_{D}$ does not depend on the projection $m$ of the dipole
operator $D_{1}^{(m)}$ as well, it is convenient to introduce
in usual way the reduced matrix
elements of multipole operators,
\begin{equation}
<f;I_{f}M_{f}|{\cal O}_{l}^{(m)}|i;I_{i}M_{i}> = <I_{f}M_{f}|I_{i}lM_{i}m>
(f;I_{f}||{\cal O}_{l}||i;I_{i}),
\label{15}
\end{equation}
where $f$ stands now for all quantum numbers except angular momentum
ones, $I$ and $M$, and to perform the additional summation of Eq.
(\ref{13}) over $m$. In such a way one obtains
\begin{equation}
\sum_{f,I_{f}}\omega_{fi} (2I_{f}+1) (f;I_{f}||D||i;I_{i})^{2} =
3 (2I_{i}+1) S_{D}.                              \label{16}
\end{equation}

Now let us take the ground state $|0>$ of an even-even nucleus with
angular momentum $I_{0}=0$ as an initial one $|i;I_{i}>$.
If we assume that the single GDR $|1> \equiv |1;1>$ is an isolated state
saturating the corresponding sum rule, we just divide
the right hand side of (\ref{16})
by the excitation energy $\omega_{10}$ to
obtain the reduced matrix element
\begin{equation}
(1||D||0)^{2} = \frac {S_{D}}{\omega_{10}}.                \label{17}
\end{equation}

In order to be able to
calculate the cross section of excitation of the double GDR,
we have to take the single GDR state $|1>$ as an initial one. The
corresponding sum in Eq. (\ref{16}), according to our assumption, is
saturated
by (i) "down" transition to the ground state $|0>$, which has negative
transition energy $-\omega_{10}$ and, due to the symmetry properties of the
Clebsch-Gordan coefficients, the strength which is
3 times larger than that of Eq. (\ref{17}),
and (ii) "up" transitions to the double GDR states $|2;I_{2}=L>$
where $L$ can be equal to 0 and 2.
The resulting sum rule for the up transitions is
\begin{equation}
\sum_{L=0,2}(2L+1)\omega_{21}^{(L)}(2;L||D||1)^{2} = 12 S_{D}.    \label{18}
\end{equation}
where $\omega_{21}^{(L)} \equiv E_{2;L} - E_{1}$
is the energy of the second excitation.
Actually, considering, instead of the sum over $m$,
the original dipole sum rule (\ref{13}) for fixed $m$, one can
separate the two contributions to the sum (\ref{18}) and find
\begin{equation}
(2;L||D||1)^{2} = 2 \frac{S_{D}}{\omega_{21}^{L}}.    \label{19}
\end{equation}
Obviously, it is consistent with the sum rule (\ref{18}).

Eqs. (\ref{17}) and (\ref{19}) imply the relation between the strengths of
sequential
excitation processes,
\begin{equation}
(2;L||D||1)^{2} = 2 \frac
{\omega_{10}}{\omega_{21}^{(L)}} (1||D||0)^{2}.  \label{20}
\end{equation}
For the equidistant vibrational spectrum this result is nothing but
the standard Bose factor of stimulated radiation; our result is valid
under more broad assumptions. The resulting
enhancement factor includes, in addition, the ratio of transition
frequencies which, according to the data, is slightly larger than 1.
The generalization for the third and higher order excitation processes
is straightforward.

\subsubsection{Spreading widths of single and double resonances}

The above assumption of saturation certainly does not account for
the fact that the resonances are
wide. In fact, this might be also relevant for the calculation of total cross
sections since the Coulomb excitation amplitudes given by
may vary strongly with the excitation energy.
Therefore they might be sensitive to the shape of the strength function.
The widths of the resonances can be taken into account in a simplified
approach, as we describe next.

In a microscopic approach, the GDR is described by a coherent superposition
of one-particle one-hole states. One of the many such states is pushed up by
the residual interaction to the experimentally observed position of the GDR.
This state carries practically all the $E1$ strength. This situation is
simply
realized in a model with a separable residual interaction. We write the GDR
state as (one phonon with angular momentum 1M) $\left| 1,\;1M\right\rangle
=A_{1M}^{\dagger }\left| 0\right\rangle $ where $A_{1M}^{\dagger }$ is a
proper superposition of particle-hole creation operators. Applying the
quasi-boson approximation we can use the boson commutation relations and
construct the multiphonon states (N-phonon states). A N-phonon state will be
a coherent superposition of N-particle N-hole states. The width of the GDR
in heavy nuclei is essentially due to the spreading width, i.e., to the
coupling to more
complex quasibound configurations. The escape width plays only a minor
role. We are not interested in a detailed microscopic description of these
states here. We use a simple model for the strength function \cite{BB86}.
We couple a state $\left| a\right\rangle $ (i.e. a GDR state) by some
mechanism to more complex states $\left| \alpha \right\rangle $, for
simplicity we assume a constant coupling matrix element $V_{a\alpha
}=\left\langle a\left| V\right| \alpha \right\rangle =\left\langle \alpha
\left| V\right| a\right\rangle ={\rm v}$. With an equal spacing of $D$ of
the levels $\left| \alpha \right\rangle $ one obtains a width
\begin{equation}
\Gamma =2\pi \frac{{\rm v}^2}D\;,
\label{GD}
\end{equation}
for the state $\left| \alpha \right\rangle .$ We assume the same mechanism
to be responsible for the width of the N-phonon state: one of the
N-independent phonons decays into the more complex states $\left| \alpha
\right\rangle $ while the other (N-1)-phonons remain spectators. We write
the coupling interaction in terms of creation (destruction) operators $%
c_a^{\dagger }\;(c_\alpha )$ of the complex states $\left| \alpha
\right\rangle $ as
\begin{equation}
V={\rm v}\left( A_{1M}^{\dagger }\;c_\alpha +A_{1M}\;c_\alpha ^{\dagger
}\right) \;.
\label{VA}
\end{equation}
For the coupling matrix elements v$_N$, which connects an N-phonon state $%
\left| N\right\rangle $ to the state $\left| N-1,\alpha \right\rangle $ (N-1
spectator phonons) one obtains
\begin{equation}
{\rm v}_N=\left\langle N-1,\alpha \left| V\right| N\right\rangle ={\rm v}%
\left\langle N-1\left| A_{1M}\right| N\right\rangle ={\rm v}\;.\;\sqrt{N}\;,
\end{equation}
i.e., one obtains for the width $\Gamma _N$ of the N-phonon state
\begin{equation}
\Gamma =2\pi N\frac{{\rm v}^2}D\;=N\Gamma \;,
\label{GN}
\end{equation}
where $\Gamma $ is given by Eq. (\ref{GD}).

Thus, the factor N in (\ref{GN}) arises naturally from the bosonic character
of the
collective states. For the DGDR this would mean $\Gamma _2=2\Gamma _1$. The
data points shown in Fig. \ref{a-f22}(b) seem to favor a lower
multiplicative factor.

We can also give a qualitative explanation for a smaller $\Gamma _2/\Gamma
_1 $ value.
We again assume that the damping of the collective modes is mostly due to the
coupling to the background of complex configurations in the vicinity
of the resonance energy. Then the resonance state $|\lambda>$ gets
fragmented acquiring the spreading width $\Gamma_{\lambda}$. The
stationary final states $|f>$ in the region of the GR
are superpositions (with the same exact quantum numbers as the
collective mode) of the form
\begin{equation}
|f> = C_\lambda^{(f)} \ |\lambda> + \sum_\nu C_\nu^{(f)} |\nu >
\ ,\label{21}
\end{equation}
where $|\lambda>$ is a pure GR state and
$|\nu>$ are complex many particle-many hole states. If the
resonance component dominates in the excitation process as it should be
for the one-body multipole operator, we find
the first order amplitude $a_{fi}^{(\lambda)}$ of the excitation of the
individual state $|f>$ in the fragmentation region
\begin{equation}
a_{fi}^{(\lambda)}\simeq \Bigl[ C_\lambda^{(f)}\Bigr]^* \
a^{1st}_{\lambda} (\omega_{fi})\ .
\label{22}
\end{equation}
Here $a^{1st}_{\lambda}$ stands for the original first order excitation
amplitude. As a function of the transition energy,
the probability for the one-phonon excitation is
\begin{equation}
P_{\lambda}^{1st}({\omega})=\sum_f \biggl[
\Big\vert C_\lambda^{(f)} \Big\vert^2
\ \delta(\omega-\omega_{fi}) \biggr]
\Big\vert
a^{1st}_{\lambda} (\omega_{fi})
\Big\vert^{2} \equiv {\cal F}_{\lambda}(\omega)\Big\vert
a^{1st}_{\lambda}(\omega)\Big\vert^{2}                \label{23}
\end{equation}
where we introduced the strength function ${\cal F}_{\lambda}(\omega)$.

The traditional derivation of the strength function
(see Ref. \cite{BM75}) is based on the rough assumptions
concerning mixing matrix elements and the equidistant spectrum of complex
states. The matrix elements $V_{\lambda \nu}$ which couple the collective
mode to the background states are assumed to be of the same average
magnitude for all remote states $|\nu>$ from both sides of the resonance.
Under those conditions the resulting strength function has the
Breit-Wigner (BW) shape
\begin{equation}
 {\cal F}_{\lambda}(\omega) = {1 \over 2 \pi} \
{\Gamma_\lambda \over {(\omega-\omega_\lambda)^2 +\Gamma_\lambda^2/4}}
\ ,
\label{24}
\end{equation}
where $\Gamma_\lambda$ is the spreading width of the collective resonance,
\begin{equation}
\Gamma_{\lambda} = 2\pi \frac{<V^{2}_{\lambda \nu}>_{\nu}}{d}, \label{25}
\end{equation}
$d$ is the mean level spacing of complex states, coupling matrix
elements are averaged over the states $|\nu>$ and
$\omega_\lambda$ is the energy centroid. We will  use in our
numerical calculations the BW strength function (\ref{24}) with the empirical
parameters $\omega_{\lambda}$ and $\Gamma_{\lambda}$. However, the same
procedure can be applied to any specific form of ${\cal F}_{\lambda}(\omega)$.
Later we come back to the
question of justification of the model leading to Eqs. (\ref{24}) and
(\ref{25}).

The multiphonon states could also be reached by a direct excitation. Quite
similarly, we can repeat the above arguments to calculate the probability
for the direct excitation of a multiphonon state, with the appropriate
spreading width and energy centroid of that state. The direct (or first-order)
probabilities are then given by
\begin{equation}
P^{1st}_2(\omega) = {\cal F}_2(\omega)
\ \Big\vert a^{1st}_2 (\omega) \Big\vert^2
\ .\label{26}
\end{equation}

Let us now treat the case of the two-step excitation of GR
(double-phonon). For simplicity, we denote the single-phonon state by
$|1>$ and the double-phonon state by $|2>$, the
corresponding centroids being at $\omega_{1}$ and
$\omega_{2}$ respectively. The total probability to
excite the double-phonon state is obtained by
\begin{eqnarray}
P(\omega) &=& \sum_f \Big\vert a^{1st}_{fi}
+ a^{2nd}_{fi} \Big\vert^2 \ \delta (\omega -
\omega_{fi}) \nonumber \\
&\equiv& P^{1st} (\omega) + P^{2nd} (\omega) +P^{int} (\omega)\ ,
\label{27}
\end{eqnarray}
where $P^{1st}$ is the direct (or first-order) excitation of the double-phonon
state, $P^{2nd}$ is the two-step (or second-order) excitation term, and the
last term in Eq. (\ref{27}) is the interference between the two.

\subsubsection{Second-order perturbation theory}

To second-order, the amplitude  for a two-step excitation to a state $|2>$
via intermediate states $|1>$ is given by
\begin{equation}
a_{20}^{2nd}=\sum_{1}\
{1\over (i\hbar)^2} \ \int_{-\infty}^\infty dt \
e^{i\omega_{21} t} \ V_{21}(t)\ \int_{-\infty}^t \ dt' \ e^{i \omega_{10} t'}
V_{10}(t') \, ,
\label{h}
\end{equation}
where $V_{21}(t)$ is a short
notation for the interaction potential inside brackets of the integral of
Eq. (\ref{h}) for the transition $\vert 1>
\rightarrow \vert 2>$.

Using the integral representation of the step function
\begin{equation}
\Theta (t-t') = - \lim_{\delta \rightarrow 0^+}
\ {1\over 2 \pi i} \int_{-\infty}^\infty {e^{-iq(t-t')}\over q+i\delta} dq = \
\left\{ \begin{array}{ll}
           1,               & \mbox{if $t>t'$ \ , }\\
           0,               & \mbox{if $t<t'$},
\end{array}
\right.
\label{i}
\end{equation}
one finds \cite{AW66}
\begin{eqnarray}
a^{2nd}_{20}&=& {1\over 2} \ \sum_{1} a_{21}^{1st} (\omega_{21})
\ a^{1st}_{10}(\omega_{10}) \nonumber \\
&+& {i\over 2 \pi} \ \sum_{1} {\cal P} \int_{-\infty}^\infty
\ {dq \over q}\ a^{1st}_{21}(\omega_{21}-q)\
a_{10}^{1st}(\omega_{10}+q) \ ,
\label{j}
\end{eqnarray}
where ${\cal P}$ stands for the principal value of the integral.
For numerical evaluation it is more appropriate to rewrite the
principal value integral in Eq. (\ref{j}) as
\begin{eqnarray}
&&{\cal P} \int_{-\infty}^\infty
\ {dq \over q}\ a^{1st}_{21} (\omega_{21}-q)\
a_{10}^{1st}(\omega_{10}+q) =
\nonumber\\
& &\int_0^\infty
\ {dq \over q}\ \biggl[ a^{1st}_{21}
(\omega_{21}-q)\
a_{10}^{1st}(\omega_{10}+q)
-a^{1st}_{21} (\omega_{21}+q)\
a_{10}^{1st}(\omega_{10}-q)
\biggr] \ . \nonumber \\
& & \  \
\label{k}
\end{eqnarray}
To calculate $a^{1st}(\omega)$ for negative values of $\omega$, we note that
the interaction potential can be written as a sum of an even and an odd
part. This implies that $a^{1st}(-\omega)=
-\Bigl[a^{1st}(\omega)\Bigr]^*$.

For three-phonon excitation we use the third term of the time-dependent
perturbation expansion, and the same procedure as above (Eqs. (\ref{i} -
\ref{k})).

\subsubsection{Harmonic vibrator model}

A simplified model, often used in connection with multiphonon excitations,
is the harmonic vibrator model. In this model, the resonance widths are
neglected and the Coupled-Channel equations
can be solved exactly, in terms of the first-order excitation amplitudes
\cite{BB86}. The excitation amplitude of the $n$-th harmonic oscillator
state, for any time $t$, is given by
\begin{equation}
a_{h.o.}^{(n)} (t) ={\Big[ a_{1st} (t) \Big]^n \over \sqrt{n!}} \
\exp \Big\{ -|a_{1st} (t)|^2/2 \Big\}\ ,\label{aho}
\end{equation}
where $a_{1st} (t) $ is the excitation amplitude for the
$0 \ (g.s.)\longrightarrow 1\ (one \ phonon)$
calculated with the first-order perturbation
theory.

For the excitation of giant resonances, $n$ can be identified
with the state corresponding to a multiple $n$ of the single
giant resonance state.  This procedure has been often used in
order to calculate the cross sections for the excitation of
multiphonon giant resonances.  Since this result is exact in the
harmonic vibrator model, it accounts for all coupling between
the states. However, this result can be applied to studies of
giant resonance excitation only if the same class of multipole
states is involved.  I.e., if one considers only electric dipole
excitations, and use the harmonic oscillator model, one can
calculate the excitation probabilities, and cross sections, of
the GDR, double-GDR, triple-GDR, etc.  Eq. (\ref{aho}) is not
valid if the excitation of other multipolarities are involved,
e.g., if the excitation of dipole states  and quadrupole states
are treated simultaneously.  In Ref. \cite{norbury} a hybrid
harmonic oscillator model has been used. In this work, it is assumed
that the difference between the amplitudes obtained with the
harmonic oscillator model and with $n$-th order perturbation theory
is due to the appearance of the exponential term on
the r.h.s. of Eq.~(\ref{aho}). This exponential takes care of
the decrease in the  occupation amplitude of the ground state as
a function of time. As argued in Ref. \cite{norbury}, the
presence of other multipole states, e.g., of quadrupole states,
together with dipole states, may be accounted for by adding the first
order excitation amplitudes for the quadrupole states to the
exponent in Eq.~(\ref{aho}). This would correct for the flux
from the ground state to the quadrupole states. In other words,
Eq.~(\ref{aho}) should be corrected to read
\begin{equation}
a_{h.o.}^{(n)} (\pi\lambda,\ t) ={\Big[ a_{1st} (\pi\lambda, \ t)
\Big]^n \over \sqrt{n!}} \
\exp \Big\{ - \sum_{\pi'\lambda'} \Big|
a_{1st} (\pi'\lambda', \ t)\Big|^2/2 \Big\}\ .\label{ahon}
\end{equation}

The harmonic oscillator model is not in complete agreement with
the experimental findings. The double-GDR and double-GQR
states do not have exactly twice the energy of the respective
GDR and GQR states \cite{Em94,CF95,Aum98}. Apparently, the matrix
elements for the transition from the GDR (GQR) to the double-GDR
(double-GQR)
state does not follow the boson-rule \cite{BZ93}.
This is borne out by the discrepancy between the experimental cross
sections for the excitation of the double-GDR and the double-GQR with the
perturbation theory, and with the harmonic oscillator model
\cite{Em94,CF95,Aum98}. Thus, a Coupled-Channels calculation is useful
to determine which matrix elements for the transitions among the
giant resonance states reproduce the experimental data.

Assuming that one has $\sigma _\gamma ^{\pi \lambda }(E)$ somehow (either
from experiments, or from theory), a simple harmonic model following
the discussion above  can be formulated to obtain the
include the widths of the states. As we have mentioned, in the harmonic
oscillator model the
inclusion of the coupling between all multiphonon states can be performed
analytically \cite{BB86}. One of the basic changes is that the excitation
probabilities calculated to first-order, $P_{\pi \lambda }^{1st}(E,b),$ are
modified to include the flux of probability to the other states. That is,
\begin{equation}
P_{\pi \lambda }(E,\;b)=P_{\pi \lambda }^{1st}(E,b)\;\exp \left\{ -P_{\pi
\lambda }^{1st}(b)\right\} \;,
\end{equation}
where $P_{\pi \lambda }^{1st}(b)$ is the integral of over the excitation
energy $E$. In general, the probability to reach a multiphonon state with
the energy $E^{(n)}$ from the ground state, with energy $E^{(0)},$ is
obtained by an integral over all intermediate energies
\begin{eqnarray}
&&P_{\pi ^{*}\lambda ^{*}}^{(n)}(E^{(n)},b)=\frac 1{n!}\;\exp \left\{
-P_{\pi \lambda }^{1st}(b)\right\} \int dE^{(n-1)}\;dE^{(n-2)}\;...\;dE^{(1)}
\\
&&\times P_{\pi \lambda }^{1st}(E^{(n)}-E^{(n-1)},b)\;P_{\pi \lambda
}^{1st}(E^{(n-1)}-E^{(n-2)},b)\;...\;P_{\pi \lambda
}^{1st}(E^{(1)}-E^{(0)},b)  \nonumber
\end{eqnarray}

\subsubsection{Comparison with experiments}

The reactions $^{136}$Xe + $^{208}$Pb at 0.69A GeV
and $^{209}$Bi + $^{208}$Pb
at 1A GeV have been
measured
at GSI \cite{Ri93,Sc93}.
We apply the formalism developed in the preceding sections to
calculate the excitation probabilities and cross sections for these systems.

Cross sections (in mb) for the Coulomb excitation
of the GDR$_{iv}$,
GQR$_{is}$
and GQR$_{iv}$ in $^{136}$Xe incident on Pb at 0.69A GeV are
given in
Table {\ref{a-t3}. We have assumed that the GDR$_{iv}$, GQR$_{is}$ and
the GQR$_{iv}$ are located at
15.3, 12.3 and 24 MeV, and that they exhaust 100\%, 70\% and 80\% of the
corresponding sum rules, respectively \cite{Ber75}.
We used $b_{min}=1.2 \cdot (A_1^{1/3}+A_2^{1/3})$ fm = 13.3 fm
as a lower limit guess and $b_{min}=15.6$ fm suggested by
the parameterization \cite{Ko87} as an upper limit (number inside
parentheses). The parameterization \cite{BCV89} yields an intermediate
value for this quantity.
The
contributions to various angular momentum projections of each state are
shown separately. In the last column the total cross sections are
calculated with the widths of the states taken into account.
We use for the GDR$_{iv}$, GQR$_{is}$ and GQR$_{iv}$ the BW strength
functions (\ref{24}) with
the resonance widths $\Gamma=$
4.8, 4 and 7 MeV, respectively \cite{Ber75}. We see that
states with higher angular momentum projections are more populated.
The inclusion of the widths of the resonances in the calculation
increases  the cross sections by about 10-20\%.
The experimental value \cite{Sc93} $1110\pm80$ mb for the GDR
is much smaller which made the authors of \cite{Sc93} to claim that
the GDR absorbs only 65\% of the sum rule
(this number apparently contradicts to the systematics of
data for real monochromatic photons \cite{Ber75}). Using this value, our
result reduces to 1613 (1183) mb which seems to prefer
the upper value of $b_{min}$. The numbers in
parentheses are also in rough agreement with the data \cite{Sc93} for the
GQR$_{is}$ and GQR$_{iv}$.

\begin{table}[tb]
\caption[ ]
{Cross sections (in mb) for the Coulomb excitation
of the GDR$_{iv}$, GQR$_{is}$
and GQR$_{iv}$ in $^{136}$Xe incident on $^{208}$Pb at 0.69A GeV.
The cross sections in the last column are
calculated with the widths of the states taken into account. The values
outside (inside) parentheses use $b_{min}$=13.3 (15.6) fm.
\label{a-t3}}
\begin{center}
\begin{tabular}{ l l l l l r } \hline
  &$m=\pm 2$&$m=\pm 1$&$m=0$&$\sigma_{total}$&$\sigma_{width}$ \\ \hline
GDR$_{iv}$&-- &949 (712)&264 (201)&2162 (1630) &2482 (1820) \\
GQR$_{is}$&90 (64)&8.4 (6.09)&14.3 (10.6)&211 (150)&241 (169) \\
GQR$_{iv}$&29.7 (25.6)&6.1 (5.46)&14 (12.4)&84.1 (74.5)&102 (93) \\
\hline \end{tabular}
\end{center}
\end{table}

Using the formalism developed in Sect. 2.8 we have also calculated the
cross sections for the nuclear excitation of the GQR$_{is}$ in the same
reaction.
The cross sections for the excitation of isovector modes are reduced by
a factor $\Bigl[(N-Z)/A\Bigr]^2$ since the isovector mode is excited
due to the difference in strength of the nuclear interaction between
the target and the protons and
neutrons of the projectile \cite{BM75}.
This implies that the isovector excitations
are strongly suppressed in nuclear excitations.
Therefore, we do not consider them here. For the
excitation of the GQR$_{is}$ we find $\sigma^N=5.3$ mb, if we use the
deformation
parameter $\beta R= 0.7$ fm for $^{136}$Xe. In the calculation of the
nuclear potential  we used Fermi density distributions
with parameters $\rho_0=0.17$ fm$^{-3}$ and $R=$5.6 (6.5) fm,
$a=$0.65 (0.65) fm
for Xe (Pb). The nucleon-nucleon cross section used was 40 mb.
Again we see that the nuclear contribution to the total cross section is
very small.

The double dipole phonon state can couple to total angular momentum 0 or 2.
As we mentioned in Sect. 2, for the state with $L=2$ there is
the possibility of a direct quadrupole Coulomb excitation
($L=0$ states cannot be Coulomb excited \cite{BB88}). For simplicity,
we do not consider here the physics of the isospin coupling of the two GDR.

We calculated the direct and the two-step
probabilities for the excitation of the
double-phonon state according to the approach discussed in the previous
sections.
The total cross sections obtained  are shown in Table \ref{a-t4}. We
found that
the principal value term in Eq. (\ref{j}) contributes very little (less than
1\%) to
the GDR $\times$ GDR cross section via a two-step process.

\begin{table}[tb]
\caption[ ]
{Excitation cross sections (in mb) of the GDR$_{iv}$,
and of the [GDR]$^n$ states in the
reaction $^{208}$Pb~+~$^{208}$Pb
at 640A MeV. A comparison with first order perturbation theory and the
harmonic oscillator is made.
\label{a-t4}}
\begin{center}
\begin{tabular}{ l l l r } \hline
State & 1st pert. th. & harm. osc. & c.c. \\ \hline
GDR$_{iv}$     & 3891 & 3235 & 3210 \\
$[$GDR$_{iv}]^2$ &  388 & 281  & 280 \\
$[$GDR$_{iv}]^3$ & 39.2 & 27.3 & 32.7 \\
$[$GDR$_{iv}]^4$ &  4.2 &  2.4 & 3.2 \\ \hline
\end{tabular}
\end{center}
\end{table}

 From Table \ref{a-t5}
we see that the inclusion of the widths of the final (GDR $\times$ GDR)
and the intermediate (GDR) state increase the cross sections by 10-20\%.
For the position and width of the GDR $\times$  GDR state we took
$E=28.3$ MeV and $\Gamma=7$ MeV, respectively \cite{Sc93}
which corresponds to $\omega_{10}=15.3$ MeV and $\omega_{21}
=13$ MeV,
both for $L=2$ and $L=0$.
For the
calculation of the direct excitation we assumed that the resonance
would exhaust 20\% of the GQR$_{is}$ sum rule. It is based on the
hypotheses
that the missing strength of the low-lying GQR$_{is}$ could be located
at the double dipole phonon state as a consequence of the
anharmonic phonon coupling of the $(QDD)$-type. Obviously, it should be
considered as highly overestimated upper boundary of the direct
excitation.
In Ref. \cite{Pon92} the reduced transition
probability for the excitation of double-phonon states within the
Quasiparticle-Phonon Model have been calculated. The value $B(2^+,E2)=4.2$
$e^2$ fm$^4$ has been obtained. Using this value we get
that the cross section for the direct excitation of the $L=2$
state is 12 $\mu$b, much smaller than what we quote above.
We conclude that even in the more optimistic cases the contribution of
the direct mechanism to the total cross section for Coulomb excitation
of the double-phonon state is much less than that of the two-step
process.

\begin{table}[tb]
\caption[ ]
{Cross sections (in mb) for the Coulomb excitation
of the double GDR
in $^{136}$Xe incident on Pb at 0.69A GeV.
The cross sections in the last column are
calculated with the widths of the states taken into account.
The values outside (inside) parentheses use $b_{min}$=13.3 (15.6) fm.
\label{a-t5}}
\begin{center}
\begin{tabular}{ l l l l l r }
\hline
DGDR state &$m=\pm 2$&$m=\pm 1$&$m=0$&$\sigma_{total}$&
$\sigma_{width}$ \\ \hline
L=0 (two-step)&-- &--&22.8 (10.7)&22.8 (10.7)&28.4 (13.3)\\
L=2 (two-step)&23.3 (11.2)&13.4 (6.6)&51.4 (26.8)&124.8 (62.4)&154 (77)\\
L=2 (direct   &&&&&\\
$-$ 20\% of SR) &3.27 (2.85)&0.86
(0.77)&2.12 (1.88)&10.3 (9.12)&11.8 (10.8)\\ \hline
\end{tabular}
\end{center}
\end{table}

Another conclusion drawn from the numbers of Table \ref{a-t5} is that
the excitation of the $L=2$ double-phonon state is much stronger than
for the $L=0$ state.
Adding the two contributions we find that the total cross section for
the excitation of the double-phonon state (excluding the direct mechanism)
in the reaction above is equal to 182 (101) mb.
The experimental value of
Ref. \cite{Sc93} is about $215\pm50$ mb. As stated above, the
nuclear contribution to the
(direct) excitation of the double-phonon state is not relevant.
If we assume again that about 20\% of the sum rule
strength is exhausted by this state (using e.g. $\beta$R=0.1 fm), we get
1.1 mb for the nuclear excitation of the $L=2$ double-phonon state.
Contrary to the single phonon case, the appropriate value of
$b_{min}$ for the double GDR experiment \cite{Sc93} is
$b_{min}$=13.3 fm.

We also compare our results with the experiment of
Ritman et al. \cite{Ri93}. They measured the excitation of a $^{208}$Pb
target by means of $^{209}$Bi projectiles at 1A GeV and
obtained $770\pm220$ mb for the excitation
cross section  of the double resonance. We calculate the cross sections
for the same
system, using $E_1=13.5$ MeV, $\Gamma_1=4$ MeV, $E_2=27$ MeV and
$\Gamma_2=6$ MeV
for the energy position and widths of the GDR and the
GDR $\times$ GDR in $^{208}$Pb,
respectively. Using the formalism developed in Sects. 3.2.2 and 3.2.3
and including the effects of the widths of the states, we find
$\sigma_1=5234$ b for the excitation of the GDR and
$\sigma_2=692$ mb for the excitation  of the GDR $\times$ GDR,
using $b_{min}=1.2(A_P^{1/3}+A_T^{1/3})$ fm = 14.2 fm.
Thus, while the
cross section for the excitation of single phonons is a factor 2.8 larger than
that of the experiment of Ref. \cite{Sc93},
the cross sections for the excitation of double phonons is larger
by a factor 3.8. This is due to the larger value for the excitation
probabilities caused by a larger $B(E1)$ value
for this reaction.
The parameterization \cite{Ko87} with $b_{min}=b_{min}$=16.97 fm would lead
to smaller cross sections $\sigma_{1}=4130$ mb and
$\sigma_{2}=319$ mb.

We found the ratio of $(P_{m=+1} + P_{m=-1})/P_{m=0} = 9.4$ for
the excitation of
the GDR in the experiment of Ref. \cite{Ri93}. They quote the value 28 in
their
calculations and fit the gamma-ray
angular distribution  according to this value. We think that this result could
somewhat change the extracted value of the GDR $\times$ GDR
cross section which is
quoted in Ref. \cite{Ri93}.

Using the formalism shown of section 3.2.4  we find
that the cross sections for the excitation of three-phonon states in the
experiment of Schmidt et al. \cite{Sc93} is equal to 19.2 mb (with
$b_{min}=13.3$ fm)
while it is equal to
117 mb (with $b_{min}=14.2$ fm) for the
experiment of Ritman et al. \cite{Ri93}.  The identification of
these resonances is therefore more difficult, but possible with the present
experimental techniques. Using the same arguments leading to Eq. (\ref{20})
we find for the reduced matrix elements, in obvious notations,
$|D_{32}|^2=3(\omega_{10}/\omega_{32})|D_{10}|^2$, which we used
in our calculation. We assumed that $\omega_{10}/\omega_{32}\simeq
\omega_{10}/\omega_{21}$. These enhancement factors
for the excitation of higher phonon states are very
important to explain the magnitude of the cross sections.
The anharmonic effects,
suggested in \cite{Sc93} to explain the large excitation of
double GDR, are expected to be small since the mixing of
single- and double- phonon states is forbidden by the
angular momentum and parity. The main anharmonic
effect, apart from the weak coupling of the double GDR with
$L=2$ to GQR, is the IBM-like scattering of dipole phonons
which splits $L=0$ and $L=2$ states but
hardly changes excitation and decay properties.

Another important question is related to the expected width of the
multiphonon states.
Early estimates \cite{BB86} presented in Sect. 3.2.2 indicated that
these widths should
scale as $\Gamma_n=n\Gamma_1$.
The experiments show however that a scaling as
$\Gamma_n=\sqrt{n} \Gamma_{1}$ is more
appropriate, at least for the double GDR. We next address
in detail different aspects of physics responsible for
the width of the double phonon state.

\subsection{General arguments on the width of the double phonon state}

Here we discuss in qualitative terms the problem of the width of a
collective state which can be thought of as being created by the excitation of
two quanta in a complex many-body system. We assume that the genuine decay
to continuum is of minor importance at the given excitation energy.
Therefore we focus on the damping width which comes from the fact that
the collective mode is a specific coherent superposition of simple
configurations (for instance, of a particle-hole character) rather than
a pure stationary state.

In the actual excitation process the predominant mechanism is that of the
sequential one-phonon excitation. Under our assumption that the sum rule
is saturated by the GR the intermediate states contribute to this process
as far as they contain a significant collective component. Therefore the
interference of many incoherent paths can be neglected so that we are
interested  in the shape $P(E)$ of the excitation function at a given energy
$E = E_{1} + E_{2}$ which can be obtained as a convolution of the single
phonon excitation functions,
\begin{equation}
P(E) = \int dE_{1} dE_{2} P_{1}(E_{1})P_{2}(E_{2})\delta (E - E_{1} -
E_{2}).                                                    \label{43}
\end{equation}
The same shape should be revealed in the deexcitation process.

In this formulation the problem is different from what is usually
looked at when one is interested, for example, in sound attenuation.
In such classical problems the conventional exponential decrease of the
wave intensity does not correspond to the decay of the state with a
certain initial number of quanta. Contrary to that, here we have to
compare the damping rates of individual quantum states with the fixed
number of quanta, single- and double-phonon states in particular.

We have to mention also that in the nuclear GR case
quantum effects are more pronounced since the temperature corresponding to the
relevant  excitation energy is less than $\hbar \omega$
whereas in the measurements of the attenuation of the zero and first sound
in the macroscopic Fermi-liquid \cite{He-3}  the situation is always inverse
and the quantum limit is hardly attainable. (In nuclear physics the
classical case can be studied with low-lying quadrupole vibrations).

Independently of specific features of nuclear structure (level density,
$A$-dependence, shell effects, finiteness of the system leading
to the linear momentum nonconservation and, therefore, to the estimate
of the available phase space which could be different from that for
infinite matter, and so on) we can try to make several comments of
general nature.

If the anharmonic effects could be considered to be small we can assume
that the phonons decay independently by what can be described, using the
language of stationary quantum mechanics, as mixing to complex
background states. The decay rate $\Gamma_{1}(e)$ of an individual
quasiparticle (elementary excitation) with energy $e$ depends on the
background level density and, whence, on the excitation energy. The decay of
a state with $n$ quasiparticles occurs as far as one of the constituents
decays. It implies the simple estimate of the width $\Gamma_{n}$ of the
$n$-quantum state, $\Gamma_{n} \simeq n\Gamma_{1}(E/n)$. For the decay of
typical many particle-many hole configurations \cite{Laur,Ber83,Bush}
one usually
takes the Fermi-liquid estimate $\Gamma_{1}(e) \propto e^{2}$ which leads
to $\Gamma_{n} \propto T^{3} \propto E^{3/2}$ since the average number of
quasiparticles in a typical thermal configuration at temperature $T$ is
$n \propto T$. This estimate agrees with data. In the case of the pure
$n$-phonon state $E/n = \hbar \omega$ which results in the ratio
$r_{n} \equiv \Gamma_{n} / \Gamma_{1} \simeq n$.

Thus, the simplest line of reasoning favors the width of the double GR
to be twice as big as the width of the single GR. At the first glance,
this estimate is especially reasonable for the giant dipole since here the
anharmonic effects, determining the whole pattern of low-lying vibrations,
are expected to be very weak.
Angular momentum and parity conservation forbids cubic
anharmonicity which would mix single- and double-quantum states and influence
both excitation cross sections and spreading widths. The main anharmonic
term, apart from the mentioned in Sect. 3.2.2 weak mixing of the giant
quadrupole
to the double dipole state with $L=2$,
probably corresponds to the phonon scattering similar to that in the
IBM. It results in the shift of the double-phonon state from $2\hbar\omega$
and splitting of $L=0$ and $L=2$ states hardly changing the decay properties.
Experimentally, the energy shift seems to be rather small.

There are also other arguments for the width ratio $r_{2}=2$. In our
calculation of cross sections we assumed the BW shape (\ref{24}) of
strength functions (\ref{23}). If the sequential excitation is described by
the
BW functions $P_{1}(E_{1})$ with the centroid at $e$ and the width
$\Gamma$, and $P_{2}(E_{2})$ with corresponding parameters $e'$ and
$\Gamma'$, the convolution (\ref{43}) restores the BW shape
with the centroid at $e+e'$ and the total width $\Gamma+\Gamma'$.
For identical phonons it means that the width ratio $r_{2}=2$.

As we mentioned in Sect. 3.2.2,
the BW shape of the strength function is derived analytically within the
simple model \cite{BM75} of coupling between a phonon and complex
background states. One diagonalizes first the Hamiltonian in the subspace of
those complex states and get their energies $\epsilon_{\nu}$. If the
underlying dynamics is nearly chaotic, the resulting spectrum will show
up level repulsion and rigid structure similar to that of the Gaussian
Orthogonal Ensemble (GOE), with the mean level spacing $d$.
Roughly speaking, one can assume the equidistant energy spectrum. The
collective phonon $|1>$ at energy $E_{1}$ is coupled to those states and
corresponding matrix elements $V_{1\nu}$ are assumed to be of the same order
of magnitude (much larger than the level spacing $d$) for all states $|\nu>$
in the large energy interval around the collective resonance. Then
the energies of the stationary states (final states $|f>$ in the notations
of previous sections) are the roots $E=E_{f}$ of the secular equation
\begin{equation}
F(E) \equiv E - E_{1} - \sum_{\nu}\frac{V_{1\nu}^{2}}{E - \epsilon_{\nu}} =0,
\label{44}
\end{equation}
and the distribution of the collective strength, Eq. (\ref{21}),
\begin{equation}
|C_{1}^{(f)}|^{2} = [dF/dE]^{-1}_{E=E_{f}} = [1 + \sum_{\nu} \frac
{V_{1\nu}^{2}}{(E_{f} - \epsilon_{\nu})^{2}}]^{-1}         \label{45}
\end{equation}
reveals the BW shape (\ref{24}) and the "golden rule" expression (\ref{25})
for the width $\Gamma_{1}$.

We can repeat the procedure for the double phonon state. Phonons of
different kind would couple to different background states with different
level spacing and coupling matrix elements. It corresponds to independent
decay leading as we discussed above to $\Gamma_{2} = \Gamma + \Gamma'$.
For the identical phonons, we should take into account that the double
phonon state $|2>$ is coupled to the states "single phonon + background"
and the background states here are the same as those determining the width
of the single phonon state $|1>$. This picture is in accordance with the
famous Axel-Brink hypotheses. Therefore the expression for the width,
Eq. (\ref{25}), contains the same level density whereas all coupling
matrix elements
for the transition to a complex state $|\nu>$ (plus a remaining phonon)
have to be multiplied by the Bose factor, $V_{2\nu} = \sqrt{2} V_{1\nu}$.
Thus, we come again to $r_{2}=2$.

The approach of the proceeding paragraph can be slightly modified
by introducing explicitly coupling via a doorway state \cite{Mord} or GOE
internal dynamics \cite{Zel}. In both cases the Bose factor $\sqrt{2}$ leads
to the same result $r_{2}=2$.

In addition, the collective resonance might be further broadened by the
coupling to low-lying collective vibrational or rotational modes. For example,
in the simplest model where the dipole phonon radiates and absorbs low
energy scalar quanta, it is easy to show that, in the stationary cloud
of scalar quanta, their average number, which determines the fragmentation
region of the dipole mode, is proportional to the squared number of
dipole phonons. Hence it gives a large width ratio $r_{2} = 4.$ For the
nuclei where actual data exist, this is not important since they are rather
rigid spherical nuclei with no adiabatic collective modes.

On the other hand, one can present some arguments in favor of the width ratio
$r_{2}=\sqrt{2}$ which apparently is preferred by the existing data.

First of all, this value follows from the convolution (\ref{43}) of Gaussian
distribution functions (instead of BW ones). Of course, this
is the inconsistent approach since the experimentalists use BW
or Lorentzian fit. But one can easily understand that the result $r_{2} =
\sqrt{2}$ is not restricted to Gaussian fit. For an arbitrary sequence
of two excitation processes we have $<E> = <E_{1} +E_{2}>$
and $<E^{2}> \ = \ <(E_{1} + E_{2})^{2}>$; for uncorrelated steps it results
in the addition of fluctuations in quadrature, $(\Delta E)^{2} =
(\Delta E_{1})^{2} + (\Delta E_{2})^{2}$. Identifying these fluctuations
with the widths up to a common factor, we get for the identical phonons
$\Gamma_{2}^{2} = 2\Gamma_{1}^{2}$, or $r_{2} = \sqrt{2}$.

The same conclusion will be valid for any distribution function which, as
the Gaussian one, has a finite second moment, contrary to the BW
or Lorentzian ones with the second moment diverging. In some sense we may
conclude that, in physical terms, the difference between $r_{2}=2$ and
$r_{2}=\sqrt{2}$ is due to the different treatment of the wings of the
distribution functions which reflect small admixtures of far remote states.

In the standard model of the strength function \cite{BM75} all remote states
are coupled to the collective mode equally strong. This is obviously an
unrealistic assumption. The shell model (more generally, mean field) basis
is the "natural" one \cite{MF} for estimating a degree of complexity of
various
states in a Fermi system at not very high excitation energy. In this
representation matrix elements of residual interaction couple the
collective state (coherent superposition of particle-hole excitations
found for example in the framework of the RPA) only to the states of the
next level of complexity (exciton class). Those states, in turn, become mixed
with more complex configurations. This process proliferates and each
simple state acquires its spreading, or fragmentation, width $2a=Nd$ where $N$
stands for a typical number of stationary states carrying the noticeable
weight
of the ancestor state and the level spacing $d$ is basically the same
as in the mean field approximation. Inversely, $N$ can be viewed as the
localization length of a stationary complex state in the mean field
basis.

In the stochastic limit the local background dynamical properties
can be modeled by those of the GOE with the semicircle radius $a$.
This intrinsic spreading width $a$, which is expected to be of the order of
magnitude of typical matrix elements of the original residual interaction
between simple configurations, is the dynamical scale missed in the
standard model which corresponds to the limit $a\rightarrow \infty$. The
existence of this intrinsic scale can be associated  with the
saturation \cite{Bra} of the width of a single GR at high temperature.

The standard model supposedly is valid for the spreading width $\Gamma$
small in comparison with $a$. Because of the relatively weak interaction
leading to the isospin impurity, this is the case for the isobaric analog
states (IAS) \cite{Har,PvB} where typical spreading widths are less than
100 keV. This approach allows one to explain, at least qualitatively,
small variations of the spreading widths of the IAS. The tunneling
mixing of superdeformed states with the normal deformed
background presents an extreme example of the small spreading width.
However, in the case
of GR the situation might be different.

To illustrate the new behavior in the opposite case of $\Gamma\geq a$, we can
imagine the limit of the almost degenerate intrinsic states with very
strong coupling to a collective mode. (The actual situation presumably is
intermediate). Assuming that the unperturbed phonon state has an energy
in the same region, one can easily see from Eqs. (\ref{44}) and (\ref{45})
that the
coupling results in the appearance of the two collective states sharing
evenly the collective strength and shifted symmetrically from the
unperturbed region by $\Delta E = \pm \sqrt{\sum_{\nu}V_{\nu}^{2}}$.
The physical reason is evident: the interaction of the background states
through the collective mode creates a specific coherent superposition which
is hybridized with and repelled from the original collective state.
The similar effect was discussed in different context in \cite{Ann}
and observed in numerical simulations \cite{Herz}. The well known doubling
of the resonance peak at the passage of a laser beam through a cavity
containing a two-level atom is the simplest prototype of such a phenomenon.

In this limit one gets the effective width of collective response
$2\Delta E = 2\sqrt{N<V_{\nu}^{2}>} =
2\sqrt{a\Gamma_{s}/\pi}$ where $\Gamma_{s}$ is the
standard spreading width (\ref{25}). This effective width is linearly
proportional to the average coupling matrix element. Therefore it should
increase by factor $\sqrt{n}$ when applied to a $n-$phonon collective
state. Thus, we anticipate in this limit $r_{2} = \sqrt{2}$.
One may say that the phonons do not decay independently
being correlated via common decay channels. In the
literature the similar result, due to apparently the same physical reasons,
was mentioned in \cite{PGR} referring to the
unpublished calculations in the framework of the second RPA.

\bigskip

\subsection{Coupled-channels calculations with inclusion of the GR widths}

We have seen that the excitation probabilities of excitation
of single and double giant resonances are quite large. It is
worthwhile to study the excitation process with a coupled channels
calculation and compare to the other approximations. We will now
study this effect by using the Coupled-Channels Born
approximation. This approximation was used in Ref. \cite{Ca94} to
describe the excitation of the double giant resonance in
relativistic heavy ion collisions.  It is based on the idea that
in such cases only the coupling between the ground state and the
dominant giant dipole state has to be treated  exactly.  The
reason is that the transitions to giant quadrupole and to the
double-phonon states have low probability amplitudes, even for
small impact parameters.  However, an exact treatment of the
back-and-forth transitions between the ground state and the
giant dipole state is necessary. This leads to
modifications of the transitions amplitudes to the remaining
resonances, which are populated by the ground state and the GDR.
In Ref. \cite{Ca94} the application of the method
was limited to the use of an schematic interaction, and the
magnetic substates were neglected.  These deficiencies are
corrected here. The method allows the inclusion of the width of
the giant resonances in a very simple and straightforward way.
It will be useful for us to compare with the Coupled-Channels
calculations with isolated states, as we described in the
previous sections.  Fig. \ref{a-f23} represents the procedure. The GDR
is coupled to the ground state while the remaining resonances are
fed by these two states according to first order perturbation
theory. The coupling matrix elements involves the ground state and
a set of doorway states $|{\cal D}^{\scriptstyle  (n)}_{\scriptstyle
\lambda\mu}>$, where $n$
specifies the kind of resonance and $\lambda\mu$ are angular momentum
quantum numbers. The amplitudes of these resonances in real continuum
states are
\begin{equation}
\alpha^{\scriptstyle  (n)}(\epsilon)=<\phi(\epsilon)\Big|{\cal
D}^{\scriptstyle  (n)}_{\scriptstyle  \lambda\mu}>
\label{aeps}
\ ,
\end{equation}
where $\phi(\epsilon)$ denotes the wavefunction of one
of the numerous states which are responsible for the broad
structure of the resonance.  In this equation
$\epsilon=E_x-E_n$, where $E_x$ is the excitation energy and
$E_n$ is the centroid of the resonance considered.

As we have stated above, in this approach we use the
Coupled-Channels equations for the coupling between the
ground state and the GDR. This results in the following
Coupled-Channels equations:
\begin{eqnarray}
i\hbar \ {\dot a}_{\scriptstyle  0}(t) &=&
\sum_\mu \ \int d\epsilon <\phi(\epsilon) |{\cal
D}^{\scriptstyle  (1)}_{\scriptstyle  1\mu}>
\ <{\cal D}^{\scriptstyle  (1)}_{\scriptstyle  1\mu}|V_{E1,\mu}(t) |0> \
\exp\Big\{ -
{i\over \hbar} (E_{\scriptstyle  1}+\epsilon)t\Big\}
\ a^{\scriptstyle  (1)}_{\scriptstyle  \epsilon,1\mu}(t)\nonumber \\
&=&
\sum_\mu \ \int d\epsilon \ \alpha^{\scriptstyle  (1)}(\epsilon)
\ V^{\scriptstyle  (01)}_{\scriptstyle  \mu}(t) \ \exp\Big\{ -
{i\over \hbar} (E_{\scriptstyle  1}+\epsilon)t \Big\}
\ a^{\scriptstyle  (1)}_{\scriptstyle  \epsilon,1\mu}(t)\ ,\label{da0}
\end{eqnarray}
and
\begin{equation}
i\hbar \ {\dot a}^{\scriptstyle  (1)}_{\scriptstyle  \epsilon,1\mu}(t) =
\Big[ (\alpha^{\scriptstyle  (1)}(\epsilon)\ V^{\scriptstyle
(01)}_{\scriptstyle  \mu}(t)\Big]^* \
\exp\Big\{ i (E_1+\epsilon)t/\hbar \Big\}
\ a_{\scriptstyle  0}(t)\ .\label{da1}
\end{equation}
Above, $(n=1)$ stands for the GDR, $a_{\scriptstyle  0}$ denotes the
occupation
amplitude of the ground state and $a^{\scriptstyle  (1)}_{\scriptstyle
\epsilon,1\mu}$
the occupation amplitude of a state located at an
energy $\epsilon$ away from the GDR centroid, and with
magnetic quantum number $\mu$ ($\mu = -1,0,1$).
We used the short hand notation $V^{\scriptstyle  (01)}_{\scriptstyle
\mu}(t)=
<{\cal D}^{\scriptstyle
(1)}_{\scriptstyle  1\mu}|V_{E1,\mu}(t) |0>$.

Integrating Eq.~(\ref{da1}) and inserting the result in
Eq.~(\ref{da0}), we get the integro-differential equation for the
ground state occupation amplitude
\begin{eqnarray}
{\dot a}_{\scriptstyle  0}(t)&=& - \frac{1}{\hbar^2}\ \sum_\mu \
V^{\scriptstyle  (01)}_{\scriptstyle  \mu}(t)\
\int
d\epsilon\  |\alpha^{\scriptstyle  (1)}(\epsilon)|^2 \nonumber\\
& \times &
\int_{-\infty}^t dt'\
\left[ V^{\scriptstyle  (01)}_{\scriptstyle  \mu}(t')\right]^* \ \exp
\Big\{ -i (E_{\scriptstyle  1}+\epsilon)
(t-t')/\hbar\Big\} \ a_{\scriptstyle  0}(t')\ ,\label{da01}
\end{eqnarray}
where we used that $a^{\scriptstyle  (1)}_{\scriptstyle
\epsilon,1\mu}(t=-\infty)=0$.
To carry out the integration over $\epsilon$, we should use
an appropriate parametrization for the doorway amplitude
$\alpha^{\scriptstyle  (1)}(\epsilon)$. A convenient choice is the
Breit-Wigner (BW) form
\begin{equation}
|\alpha^{\scriptstyle  (1)}(\epsilon)|^2 = { 1 \over 2 \pi}\
\left[ {\Gamma_{\scriptstyle  1} \over \epsilon^2+\Gamma_{\scriptstyle  1}^2/4
}\right]
\ ,\label{aeps2}
\end{equation}
where $\Gamma_1$ is chosen to fit the experimental width.
In this case, this integral will be the simple exponential
\begin{equation}
\int d\epsilon \ |\alpha^{\scriptstyle  (1)}(\epsilon)|^2 \
\exp\Big\{
-i {(E_1+\epsilon)t\over \hbar}\Big\}
=\exp\Big\{ -i{(E_1-i\Gamma_1/2) t\over\hbar}\Big\}
\ .
\end{equation}

A better agreement with the experimental line shapes of the
giant resonances is obtained by using a Lorentzian (L)
parametrization for $|\alpha^{\scriptstyle  (1)}(\epsilon)|^2$, i.e.,
\begin{equation}
|\alpha^{\scriptstyle  (1)}(\epsilon)|^2 = {2 \over \pi}
\ \left[ {\Gamma_{\scriptstyle  1}\ E_x^2 \over (E_x^2-E_{\scriptstyle
1}^2)^2+
\Gamma_{\scriptstyle  1}^2 E_x^2 }\right]
\ ,\label{L}
\end{equation}
where $E_x=E_{\scriptstyle  1}+\epsilon$.
The energy integral can still be performed exactly~\cite{Pato} but now
it leads to the more complicated result
\begin{eqnarray}
\int d\epsilon \ |\alpha^{\scriptstyle  (1)}(\epsilon)|^2\ \exp\Big\{
-i {(E_{\scriptstyle  1}+\epsilon)t\over \hbar} \Big\}
&=&\Big(1-i{\Gamma_{\scriptstyle  1}\over 2E_{\scriptstyle  1}} \Big) \
\exp\Big\{ -i{(E_{\scriptstyle  1}-i\Gamma_{\scriptstyle  1}/2)
t\over\hbar}\Big\}\
\nonumber \\
&+&\ \Delta C(t)\ ,
\end{eqnarray}
where $\Delta C(t)$ is a non-exponential correction to the decay.
For the energies and widths involved in the excitation of giant
resonances, this correction can be shown numerically to be negligible.
It will therefore be ignored in our subsequent calculations.
After integration over $\epsilon$, Eq.~(\ref{da01}) reduces to
\begin{eqnarray}
{\ddot a}_{\scriptstyle  0}(t)\ =-\ {\cal S}_1\
\sum_\mu \ V^{\scriptstyle  (01)}_{\scriptstyle  \mu} (t)  \int_{-\infty}^t
dt'\
\left[ V^{\scriptstyle  (01)}_{\scriptstyle  \mu}(t')\right]^*  \exp \Big\{ -i
{(E_{\scriptstyle  1}-i\Gamma_{\scriptstyle  1}/2)
(t-t') \over \hbar}\Big\} \ a_{\scriptstyle  0}(t')\ ,\nonumber \\
\label{da02}
\end{eqnarray}
where the factor ${\cal S}_1$ is ${\cal S}_1 = 1$ for BW-shape and
${\cal S}_1 =  1 - i \Gamma_{\scriptstyle  1}/2 E_{\scriptstyle  1}$ for
L-shape.

We can take advantage of the exponential time-dependence in the integral
of the above equation, to reduce it to a set of second order differential
equations. Introducing the auxiliary amplitudes $A_{\scriptstyle  \mu}(t)$,
given
by the relation
\begin{equation}
a_{\scriptstyle  0}(t)= 1\ +\ \sum_\mu A_{\scriptstyle  \mu} (t) \ ,\label{A}
\end{equation}
with initial conditions $A_{\scriptstyle  \mu} (t=-\infty) = 0$, and
taking the derivative of Eq.~(\ref{da02}), we get
\begin{eqnarray}
{\ddot A}_{\scriptstyle  \mu}(t) - \bigg[ {{\dot V}^{\scriptstyle
(01)}_{\scriptstyle  \mu} (t) \over
V^{\scriptstyle  (01)}_{\scriptstyle  \mu} (t)}\  -\
{i\over \hbar} \Big( E_{\scriptstyle  1} -i {\Gamma_{\scriptstyle  1} \over 2}
\Big)\bigg] \ {\dot A}_{\scriptstyle  \mu}(t)\
+\ {\cal S}_1\ {|V^{\scriptstyle  (01)}_{\scriptstyle  \mu}(t)|^2 \over
\hbar^2} \ \Big[
1+\sum_{\mu'}A_{\mu'}(t)\Big] =0
\ .
\nonumber \\
\end{eqnarray}

Solving the above equation, we get $a_{\scriptstyle  0}(t)$. Using this
amplitude
and integrating Eq.~(\ref{da1}), one can evaluate $a^{\scriptstyle
(1)}_{\scriptstyle
\epsilon,1\mu}(t)$. The probability density for the population of a
GDR continuum state with energy $E_x$ in a collision with impact parameter
$b$,
$P_1(b,E_x)$, is obtained trough the summation over
the asymptotic ($t\rightarrow \infty$) contribution from each magnetic
substate. We get
\begin{equation}
P_1(b,\ E_x) = |\alpha^{\scriptstyle  (1)}(E_x-E_{\scriptstyle  1})|^2
\ \sum_\mu \bigg| \int_{-\infty}^\infty dt'\
\exp\Big\{iE_xt'\Big\} \ \left[ V^{\scriptstyle  (01)}_{\scriptstyle
\mu}(t')\right]^* \ a_0(t')
\bigg|^2\ ,\label{pro1}
\end{equation}
where $|\alpha^{\scriptstyle  (1)}(E_x-E_{\scriptstyle  1})|^2$ is given by
Eq.~(\ref{aeps2}) or
by Eq.~(\ref{L}), depending on the choice of the resonance shape.

To first order, DGDR continuum states can be populated through $E2$-coupling
from the ground state or through $E1$-coupling from GDR states.
The probability density arising from the former is given by Eq.~(\ref{pro1}),
with the replacement of the line shape $|\alpha^{\scriptstyle (1)}|^2$ by its
DGDR
counterpart $|\alpha^{\scriptstyle (2)}|^2$
(defined in terms of parameters $E_{\scriptstyle  2}$ and
$\Gamma_{\scriptstyle  2}$)
and the use of the appropriate coupling-matrix elements $V^{\scriptstyle
(02)}_{\scriptstyle  \mu}(t)$ with the $E2$ time dependence given by
(\ref{Vfi6}). On the other hand, the contribution from the latter
process is
\begin{eqnarray}
P_2(b,E_x)& =& |\alpha^{\scriptstyle  (2)}(E_x-E_{\scriptstyle  2})|^2\, {\cal
S}_1\
\ \sum_\nu \bigg| \int_{-\infty}^\infty dt' \
\exp\Big\{iE_xt'\Big\} \ \Big\{\sum_\mu \ \left(V^{\scriptstyle
(12)}_{\scriptstyle  \nu\mu}
(t')\right)^*
\nonumber \\
&\times& \int_{-\infty}^{t'} dt''\
\ \left( V^{\scriptstyle  (01)}_{\scriptstyle  \mu}(t'')\right) \
\exp \Big\{ -i {(E_{\scriptstyle  1}- i \Gamma_{\scriptstyle  1}/2)
(t-t')\over \hbar} \Big\} \ a_{\scriptstyle  0}(t'')
\bigg|^2 \ . \label{P2}
\end{eqnarray}
We should point out that Eq.~(\ref{P2}) is {\bf not} equivalent to
second-order perturbation theory. This would be  true only in the limit
$a_0(t) \longrightarrow 1$. In this approach, $a_0(t)\ne 1$, since it
is modified by the time-dependent coupling to the GDR state.
This coupling is treated exactly by means of the
Coupled-Channels equations. We consider that this is the main
effect on the calculation of the DGDR excitation probability.
This approach is justified due to the small
excitation amplitude for the transition $1 \longrightarrow 2$,
since $a_1(t) \ll a_0(t)$.

Equations  similar to (\ref{pro1}) can also be used to calculate the
GQR$_{is}$ and GQR$_{iv}$ excitation probabilities, with the proper
choice
of energies, widths, and transition potentials (e.g., $V_{E2}(t)$, or
$V_{N2}(t)$, or both).

In the next section we will apply the results of this section
to analyze the effect of the widths of the GR's in a Coupled-Channels
approach to relativistic Coulomb excitation.

\bigskip

\subsubsection{Zero-width calculations}

We consider the excitation of giant resonances in $^{208}$Pb
projectiles, incident on $^{208}$Pb targets at 640A MeV, which has
been studied at the GSI/SIS,
Darmstadt~\cite{Em94,Aum98}. For this system the excitation
probabilities of the isovector giant dipole (GDR$_{iv}$) at 13.5 MeV
are large and, consequently, high order effects of channel coupling
should be relevant. To assess the importance of these
effects, we assume that the GDR state  depletes 100\% of
the energy-weighted sum-rule and neglect the resonance width.

As a first step, we study the time evolution of the excitation
process, solving the Coupled-Channels equations for a reduced
set of states.  We consider only the ground state (g.s.) and the
GDR. The excitation probability is then compared with that
obtained with first order perturbation theory. This is done in
Fig. \ref{a-f24}, where we plot the occupation probabilities of the
g.s., $|a_0(t)|^2$, and of the GDR, $|a_1(t)|^2$, as functions
of time, for a collision with impact parameter $b= 15$ fm. As
discussed earlier, the Coulomb interaction is strongly peaked
around $t=0$, with a width of the order $\Delta t \simeq
b/\gamma v$. Accordingly, the amplitudes are rapidly varying in
this time range. A comparison between the CC-calculation (solid
line) and first order perturbation theory (dashed line) shows
that the the high order processes contained in the former lead
to an appreciable reduction of the GDR excitation probability.
 From this figure we can also conclude that our numerical
calculations can be restricted to the interval $-10 < \tau <10,$
where $\tau = (\gamma v/b)\ t$ is the time variable measured in
natural units. Outside this range, the amplitudes reach
asymptotic values.

It is worthwhile to compare the predictions of first order
perturbation theory with those of the harmonic oscillator model
and the CC calculations.  In addition to the GDR, we include the
following multiphonon states: a double giant dipole state
([GDR$_{iv}]^2$) at 27 MeV, a triple giant dipole state
([GDR$_{iv}]^3$) at 40.5 MeV, and a quadruple giant dipole
state ([GDR$_{iv}]^4$) at 54 MeV. The coupling between the
multiphonon states are determined by  boson factors, i.e.,
for $0 \rightarrow 1$ and $n-1 \rightarrow n$ \cite{BZ93}:
\begin{equation}
\vert <n-1||V_{E/N,1}||n>\vert^2=n\ \vert <0||V_{E/N,1}||1>\vert^2 \ .
\label{VE_N}
\end{equation}
Direct excitations of the
multiphonon states from the g.s.  are not considered. The
angular momentum addition rules for bosons yields the following
angular momentum states: $L=0$ and $2$, for the [GDR]$^2$
state; $L=1$, 2, and 3, for the [GDR]$^3$ state; and $L=0$,
1, 2, 3, and 4, for the [GDR]$^4$ state. We assume that
states with the same number of phonons are degenerate. In Table
\ref{a-t5}, we show the resulting cross sections. The excitation
probabilities and the cross section were calculated with the
formalism of section 3.4. The integration over impact parameter
was carried out in the interval $b_{\scriptstyle  min} < b < \infty$.  As we
discuss below, the low-$b$ cut-off value~\cite{BZ93} $b_{\scriptstyle
min} = 14.3$~fm mocks up absorption effects. We have checked that
the CC results are not significantly affected by the unknown
phases of the transition matrix elements. Since the multiphonon
spectrum is equally spaced, and the coupling matrix-elements are
related through boson factors (as in Eq. (\ref{VE_N})), the
harmonic oscillator and the CC cross sections should be equal.
In fact the numerical results of these calculations given in the
table are very close.  We also see that the excitation cross
sections of triple- and quartic-phonon states are much smaller
than that for the [GDR]$^2$.  Therefore, we shall
concentrate our studies on the [GDR]$^2$, neglecting other
multiphonon states.

Next, we include the remaining important giant resonances in
$^{208}$Pb.  Namely, the isoscalar giant quadrupole (GQR$_{is}$) at
10.9~MeV and the isovector giant quadrupole (GDR$_{iv}$) at 22~MeV.
Also in this case, we use 100 \% of the energy-weighted sum
rules to deduce the strength matrix elements.  In Table \ref{a-t6}, we
show the excitation probabilities in a grazing collision, with
$b=14.3$~fm. We see that first order perturbation theory yields
a very large excitation probability for the GDR$_{iv}$ state. This
is strongly reduced in a c.c.  calculation, as we have already
discussed in connection with Fig. \ref{a-f24}. The excitations of
the remaining states are also influenced. They are reduced due
to the lowering of the occupation probabilities of the g.s. and
of the GDR$_{iv}$ state in the c.c.  calculation. As expected,
perturbation theory  and  c.c. calculations agree at large
impact parameters, when the transition probabilities are small.
For the excitation of the [GDR$_{iv}]^2$ state we used
second-order perturbation theory to obtain the value in the
second column. The presence of the GQR$_{is}$ and the GQR$_{iv}$
influence the c.c. probabilities for the excitation of the GDR and the
[GDR$_{iv}]^2$, respectively.

\begin{table}[tb]
\caption[ ]
{Transition probabilities at $b=14.3$ fm, for the
reaction $^{208}$Pb~+~$^{208}$Pb at 640A
MeV.
A comparison with first order perturbation theory is made.
\label{a-t6}}
\begin{center}
\begin{tabular}{ l l r } \hline
Trans.&1st pert. th.&c.c. \\ \hline
g.s.$\longrightarrow$ g.s.&---&0.515 \\
g.s.$\longrightarrow$ GDR$_{iv}$ &0.506&0.279 \\
g.s.$\longrightarrow$ GQR$_{is}$ &0.080&0.064 \\
g.s.$\longrightarrow$ GQR$_{iv}$&0.064&0.049 \\
g.s.$\longrightarrow$ [GDR$_{iv}]^2$ &0.128&0.092 \\ \hline
\end{tabular}
\end{center}
\end{table}

\subsubsection{Effect of resonance widths}

We now turn to the influence of the giant resonance widths on the
excitation dynamics. We had considered this in Sect. 3.2. But, now we
show that the coupled-channels effects lead to important quantitative
modifications of the results.
We use the CCBA
formalism developed in Sect. 3.4. Schematically, the CC problem is
that represented Fig. \ref{a-f23}. As we have seen above, the strongest
coupling occurs
between the g.s. and the GDR.

In Fig. \ref{a-f25}, we show the excitation energy spectrum for the GDR,
the DGDR
(a notation for the [GDR$_{iv}]^2$), GQR$_{is}$ and GQR$_{iv}$. The
centroid energies and the widths of these resonances are listed in Table
\ref{a-t7}. The
figure show excitation spectra obtained with both Breit-Wigner (BW) and
Lorentzian (L) line shapes. One observes that the BW and L spectra have
similar
strengths at the resonance maxima. However, the low energy parts (one or two
widths below the centroid) of the spectra are more than one order of magnitude
higher in the BW calculation. The reason for this behavior is that Coulomb
excitation favors low energy transitions and the BW has a larger low energy
tail as compared with the Lorentzian line shape. The contribution from the
DGDR leads to a pronounced bump in the total energy spectrum. This bump
depends
on the relative strength of the DGDR with respect to the GDR. In Fig.
\ref{a-f26}, we
show the ratio $\sigma_{\mbox{\tiny DGDR}}/\sigma_{\mbox{\tiny GDR}}$
as a function of the
bombarding energy. We observe that this ratio is roughly constant in the
energy range $E_{\scriptstyle  lab}/A = 200\ -\ 1000$~MeV and it falls beyond
these
limits. This range corresponds to the SIS-energies at the GSI-Darmstadt
facility.

\begin{table}[tb]
\caption[ ]
{Centroid energies and widths of the main Giant Resonances in
           $^{208}$Pb.
\label{a-t7}}
\begin{center}
\begin{tabular}{ l l l l l r } \hline
&GDR&DGDR&GQR$_{is}$&GQR$_{iv}$ \\ \hline
$E$ [MeV]&13.5&27.0&10.9&20.2 \\
$\Gamma$ [MeV]&4.0&5.7&4.8&5.5 \\ \hline
\end{tabular}
\end{center}
\end{table}

We now study the influence of the resonance widths and shapes on the GDR
and DGDR cross sections. This study is similar to that presented in
Ref.~\cite{Ca94}, except that we now have a realistic three dimensional
treatment of the states and consider different line shapes. In the upper part
of Fig. \ref{a-f27}, denoted by (a), we show $\sigma_{\mbox{\tiny GDR}}$
as a function of
$\Gamma_{\mbox{\tiny GDR}}$, treated as a free parameter. We note that
the BW and L
parameterizations lead to different trends. In the BW case the cross section
grows with $\Gamma_{\mbox{\tiny GDR}}$ while in the L case it decreases.
The growing
trend is also found in Ref.~\cite{Ca94}, which uses the BW line shape. The
reason for this trend in the BW case is that an increase in the GDR width
enhances the low energy tail of the line shape, picking up more
contributions from the low energy transitions, favored in Coulomb excitation.
On the other hand, an increase of the GDR width enhances the doorway amplitude
to higher energies where Coulomb excitation is weaker. In Fig.
\ref{a-f27} (b) and (c),
we study the dependence of $\sigma_{\mbox{\tiny GDR}}$ on
$\Gamma_{\mbox{\tiny GDR}}$. In (b),
the DGDR width is kept fixed at the value 5.7~MeV while in (c) it is kept
proportional to $\sigma_{\mbox{\tiny GDR}}$, fixing the ratio
$\Gamma_{\mbox{\tiny DGDR}} /
\Gamma_{\mbox{\tiny GDR}} = \sqrt{2}$. The first point to be noticed is
that the BW results are systematically higher than the L ones. This is a
consequence of the different low energy tails of these functions, as
discussed above. One notices also that $\sigma_{\mbox{\tiny DGDR}}$
decreases
with $\Gamma_{\mbox{\tiny GDR}}$ both in the BW and L cases. This trend
can be
understood in terms of the uncertainty principle. If the GDR width is
increased, its life-time is reduced. Since the DGDR is dominantly populated
from the GDR, its short life-time leads to decay before the transition to the
DGDR.

To assess the sensitivity of the DGDR cross section on the strength of the
matrix elements and on the energy position of the resonance, we
present in Table \ref{a-t8} the cross sections for the excitation of the
GDR,
DGDR, GQR$_{is}$ and GQR$_{iv}$, obtained with the CCBA approximation
and 100\% of the sum-rules for the respective modes.
In this calculation we have included
the strong absorption, as explained in Sect. 2.8.
For comparison, the values inside
parenthesis (and brackets) of the DGDR excitation cross section include a
direct excitation of the L=2 DGDR state. We assumed that 20\% of the
$E2$ sum rule could be allocated for this excitation mode of the DGDR.
The cross sections increase by less than 10\% in this case.
The value inside parenthesis (brackets) assume a positive (negative)
sign of the matrix element for the direct excitation.

\begin{table}[tb]
\caption[ ]
{Cross sections (in mb) for the excitation of giant resonances
in lead, for the
reaction $^{208}$Pb~+~$^{208}$Pb at
640A MeV. See text for details.\label{a-t8}}
\begin{center}
\begin{tabular}{ l l l l r } \hline
GDR&DGDR&GQR$_{is}$&GQR$_{iv}$ \\ \hline
2704&184 (199) [198]&347&186 \\ \hline
\end{tabular}
\end{center}
\end{table}

Since the excitation of the DGDR is weak, it is very well described
by Eq.~(\ref{P2}) and the DGDR population is approximately
proportional to the squared strength of $V^{\scriptstyle  (12)}$.
Therefore, to increase the DGDR cross section by a factor of 2,
it is necessary to violate the relation $E_{\mbox{\tiny
DGDR}}=2E_{\mbox{\tiny GDR}}$ by
the same factor.  This would require a strongly anharmonic Hamiltonian
for the nuclear collective modes, which would not be supported
by traditional nuclear models~\cite{BZ93}. Arguments supporting
such anharmonicities have recently been presented in
Ref.~\cite{Vo95,Lan97,Bor97}. Another effect arising from anharmonicity
would be the spin or isospin splitting of the DGDR. Since the
Coulomb interaction favors lower energy excitations, it is clear
that a decrease of the DGDR centroid  would increase its cross
section. A similar effect would occur if a strongly populated
substate is splitted to lower energies.  To study this point, we
have varied the energy of the DGDR centroid in the range $20 \
{\mbox MeV}\le E_{\mbox{\tiny DGDR}} \le 27 \ {\mbox MeV}$. The obtained
DGDR cross sections (including direct excitations) are equal to 620
mb, 299 mb and 199 mb, for the centroid energies of 20 MeV, 24
MeV, and 27 MeV, respectively. Although the experimental data on the
DGDR excitation \cite{Em94,CF95,Aum98} seem to indicate that $E_{\mbox{\tiny
DGDR}} \sim 2 E_{\mbox{\tiny GDR}}$, a small  deviation (in the range of
10\%-15\%) of the centroid energy from this value might be possible.
However, the data are not
conclusive, and more experiments are clearly necessary. We
conclude, that from the arguments analyzed here, the
magnitude of the DGDR cross section is more sensitive to the
energy position of this state. The magnitude of the DGDR cross
section would increase by a factor 2 if the energy position of
the DGDR decreases by 20\%, as found in Ref. \cite{Vo95,Lan97,Bor97},
due to  anharmonic effects. In Ref. \cite{BCH96} one
obtained
$\sigma _{\mbox{\tiny DGDR}}=$ 620 mb, 299 mb, and 199 mb for the
centroid energies of $%
E_{\mbox{\tiny DGDR}}=\;$20 MeV, 24 MeV and 27 MeV, respectively. This
shows that
anharmonic effects can play a big role in the Coulomb excitation cross
sections of the DGDR, depending on the size of the shift of
$E_{\mbox{\tiny DGDR}}$.
However, in Ref. \cite{BZ93} the source for anharmonic effects were
discussed and it was suggested that it should be very small, i.e., $\Delta
^{(2)}E$ $=E_{\mbox{\tiny DGDR}}-2E_{\mbox{\tiny GDR}}\simeq 0.$

The anharmonic behavior of the giant resonances as a possibility to explain
the increase of the Coulomb excitation cross sections has been studied by
several authors (see also Ref. \cite{BF97}, and references
therein). It was found that the effect is indeed negligible and it could be
estimated \cite{BF97} as $\Delta ^{(2)}E$ $<E_{\mbox{\tiny
GDR}}/(50.A)\sim A^{-4/3}$ MeV.

One attempt to explain the larger experimental cross section is to
include contributions from the excitation of a single coherent phonon
on "hot" fine structure states (Brink-Axel mechanism). Recently
\cite{Hus3,Hus4} this has been done through two different approaches.
In the first one \cite{Hus3}, the nucleus is described as a collective
harmonic oscillator interacting with a set of oscillators representing 
statistical degrees of freedom. In the second \cite{Hus4}, a statistical approach along the lines proposed by Ko \cite{Ko78} (see also \cite{BB86})
is used. These works indicate that the Brink-Axel mechanism should play an important role, being able to explain, in part, some discrepancies between
theory and experimental cross sections. Further work along these lines
were published by, Hussein and collaborators \cite{Hus1,Hus2,Hus5}. In
a recent publication \cite{Hus6},  the influence of the isospin 
structure of the double  giant resonance was studied in details. It was
shown that this structure also leads to an enhancement of the 
calculated cross sections.

The calculations discussed so far are based on macroscopic properties of the
nuclei, sum rules, etc. Now we show that, in order to obtain a better
quantitative description of double giant resonances it is necessary to
include the internal degrees of freedom of the nuclei appropriately.
We will discuss this next. But, we first describe the formalism that
we will use for this purpose.

\section{\bf Description of one- and multi-phonon excited states within the
Quasipar\-ticle-Phonon Model}

\subsection{The model Hamiltonian and phonons}

The Hamiltonian, $H$, of the Quasiparticle-Phonon Model (QPM)
(see Refs. \cite{Sol92,Vdo83,Vor83} for more details) 
is introduced on basis of physical ideas of
nucleons moving in an average field and interacted among each other by
means of a residual interaction.
Schematically it can be written in the form:
\begin{equation}
H = H_{s.p.} + H_{pair} + H_{r.i.}~.
\label{h1}
\end{equation}
We limit ourselves here only by the formalism for even-even spherical
nuclei.
The first term of Eq.~(\ref{h1}), $H_{s.p.}$, corresponds to the average
field for neutrons (n) and protons (p).
In the second-quantized representation it can be written in terms
of creation (annihilation) $a^{+}_{jm}$ ($a_{jm}$) operators of
particles on the level of the average field with quantum numbers
$j \equiv [n,l,j]$ and $m$ as following
\begin{equation}
H_{s.p.} = \sum_{\tau}^{n,p} \sum_{j,m} E_{j} a^{+}_{jm} \: a_{jm}~,
\label{h1a}
\end{equation}
where $E_{j}$ is the energy of the single particle level degenerated in
spherical nuclei by magnetic quantum number $m$.
The second term of Eq.~(\ref{h1}), $H_{pair}$, corresponds to residual
interaction responsible for pairing in non-magic nuclei.
In the QPM this interaction is described by monopole pairing with a
constant matrix element $G_{\tau}^{(0)}$
\begin{equation}
H_{pair} = \sum_{\tau}^{n,p} G_{\tau}^{(0)} \sum_{j,j'}
\sqrt{(2j+1) \cdot (2j'+1)}
[a^{+}_{jm} a^{+}_{j-m}]_{00}
[a_{j'-m'} a_{j'm'}]_{00}~,
\label{h1b}
\end{equation}
\begin{equation}
[a^{+}_{j} a^{+}_{j'}]_{\lambda \mu} = \sum_{m,m'}
C_{j m j' m'}^{\lambda \mu}
a^{+}_{jm} a^{+}_{j' m'}
\nonumber
\end{equation}
where $C_{j m j' m'}^{\lambda \mu}$ is the Clebsch-Gordan coefficient.
Since the QPM is usually applied for a description of properties of
medium and heavy nuclei with a filling of different subshells for neutrons
and protons, the neutron-proton monopole pairing is neglected.
The residual interaction, $H_{r.i.}$, is taken in the QPM in a separable
form as a multipole decomposition.
Its part in the particle-hole channel can be written as
\begin{equation}
H_{r.i.}^{(p-h)} = \sum_{\lambda \mu} \sum_{\tau \rho}^{\pm 1}
(\ae_{0}^{(\lambda)} + \rho \ae_{1}^{(\lambda)})
M^{+}_{\lambda \mu}(\tau) M_{\lambda \mu}(\rho \tau)~,
\label{h1c}
\end{equation}
where $\ae_{0(1)}^{(\lambda)}$ are the model parameters which determine
the strength of isoscalar (isovector) residual interaction.
The multipole operator $M^{+}_{\lambda \mu}(\tau)$ has the form
\begin{equation}
M^{+}_{\lambda \mu}(\tau) = \sum_{j,m,j',m'}
\langle j m \mid i^{\lambda} f_{\lambda}^{\tau}(r) Y_{\lambda
\mu}(\Omega) \mid j' m' \rangle a^{+}_{jm} \: a_{j'm'}
\label{m}
\end{equation}
for the natural parity states and the form
\begin{equation}
M^{+}_{\lambda \mu}(\tau) = \sum_{j,m,j',m',lm_1}
\langle j m \mid i^{l} f_{l}^{\tau}(r)
[\vec{\sigma} \cdot \vec{Y}_{lm_1}(\Omega)]_{\lambda\mu}
\mid j' m' \rangle a^{+}_{jm} \: a_{j'm'}
\label{mm}
\end{equation}
for the unnatural parity states.
The function $f_{\lambda}^{\tau}(r)$ is a radial formfactor which in
actual calculations is taken either as $r^{\lambda}$ or as a derivative
of the central part of the average field:
$f_{\lambda}^{\tau}(r) = dU^{\tau}(r)/dr$.
The value $\tau = -1 (+1) $ corresponds to neutrons (protons).
We will not consider here the residual interaction in the particle-particle
channel which is the most important for the description of
two-nucleon transfer reactions.

The basic QPM equations are obtained by means of step-by-step
diagonalization of the model Hamiltonian (\ref{h1}).
On the first step its first two terms (\ref{h1a}) and (\ref{h1b})
are diagonalized.
For that the Bogoliubov's canonical transformation from particle creation
(annihilation) operators to quasiparticle creation (annihilation) operators
$\alpha_{jm}^{+}$~($\alpha_{jm}$) is applied:
\begin{equation}
a_{jm}^{+}\,=\,u_{j}\alpha_{jm}^{+}\,+\,(-1)^{j-m}v_{j}\alpha_{j-m}~.
\label{qua}
\end{equation}
The ground state of even-even nucleus, $| \rangle _{q}$, is assumed as a
quasiparticle vacuum:
$\alpha_{jm} | \rangle _{q} \equiv 0$.
Then the energy of the ground state is minimized:
\begin{equation}
\delta\{\langle \mid H_{s.p.} + H_{pair} \mid \rangle _{q} +
\sum_{j} \mu_{j} (u_{j}^{2} + v_{j}^{2} - 1) \} = 0~,
\label{mi1}
\end{equation}
where $\mu_{j}$ are Lagrange coefficients.
The result of this minimization are the well-known BCS equations solving
which one obtains correlation functions
$C_{\tau} = G_{\tau}^{(0)} \sum_{j} u_{j} v_{j}$ and chemical
potentials $\lambda_{\tau}$ for neutron and proton systems.
The coefficients of the Bogoliubov transformation $u_{j}$ and $v_{j}$
can be calculated from these values as following:
\begin{equation}
v_j^2 = \frac{1}{2} \left \{ 1 - \frac{E_j - \lambda_{\tau}}
{\varepsilon_{j}} \right \}~,~~~~~u_j^2 = 1 - v_j^2
\label{uv}
\end{equation}
where $\varepsilon_{j}$ is the quasiparticle energy:
\begin{equation}
\varepsilon_{j} = \sqrt{C^{2}_{\tau} + [E_{j} - \lambda_{\tau}]^{2} }~.
\label{eps}
\end{equation}

In magic nuclei the BCS equations yield a zero value for the
correlation function and the position of the chemical potential in the
gap between particles and hole is uncertain.
This results in vanishing of monopole pairing correlations and the
Bogoliubov's coefficients $u_{j}(v_{j}$) equal to
0(1) for holes and to 1(0) for particles, respectively.

After diagonalization of the first two terms of the model Hamiltonian
(\ref{h1}) they can be written as following:
\begin{equation}
H_{s.p.} + H_{pair} = \sum_{\tau}^{n,p} \sum_{j,m} \varepsilon_{j}
\alpha^{+}_{jm} \alpha_{jm}
\label{h1ab}
\end{equation}
and the multiple operator (\ref{m}) in terms of quasiparticle operators
has the form
\begin{equation}
M^{+}_{\lambda \mu}(\tau) =
\sum_{jj'}^{\tau} \frac{f^{(\lambda)}_{jj'}}{\sqrt{2 \lambda +1}}
\left \{
\frac{u^{(+)}_{jj'}}{2}
\left (
[\alpha^{+}_{j} \alpha^{+}_{j'}]_{\lambda \mu}  \right . \right .
+ 
\left . (-1) ^{\lambda
- \mu} [\alpha_{j'} \alpha_{j}]_{\lambda - \mu} \right )
- 
\left . v^{(-)}_{jj'}  B_{\tau}(j j'; \lambda \mu)  \right \}~;
\label{m1}
\end{equation}
\begin{equation}
B_{\tau}(j j'; \lambda \mu) = \sum_{m m'} (-1)^{j'+m'}
C_{j m j' m'}^{\lambda \mu}
\alpha^{+}_{jm} \alpha_{j'-m'}~,
\label{bb}
\end{equation}
where $f^{(\lambda)}_{jj'} = < j || i^{\lambda} f_{\lambda}^{\tau}(r)
Y_{\lambda}(\Omega) ||j'>$ is the reduced matrix element of the
multipole operator.
We also introduced the following combinations of the Bogoliubov's
coefficients:
$ u^{(\pm)}_{jj'} = u_{j} v_{j'} \pm u_{j'} v_{j}$
and
$ v^{(\mp)}_{jj'} = u_{j} u_{j'} \mp v_{j} v_{j'}$ to be used below.

We have determined the ground state of even-even nuclei as the
quasiparticle vacuum.
In this case, the simplest excited states of
nucleus are two-quasiparticle states,
$\alpha^{+}_{jm} \alpha^{+}_{j'm'}| \rangle _{q}$,
which correspond to particle-hole transitions if monopole pairing
vanishes.
Two fermion quasiparticle operators couple to the total integer angular
momentum corresponding to the Bose statistics.
Thus, it is convenient to project the bi-fermion terms
$[\alpha^{+}_{j} \alpha^{+}_{j'}]_{\lambda \mu}$ and
$[\alpha_{j'} \alpha_{j}]_{\lambda - \mu}$
in Eq. (\ref{m1}) into the space of quasi-boson operators.
Following this boson mapping procedure we introduce the phonon
operators of the multipolarity $\lambda$ and projection $\mu$ as
following:
\begin{equation}
Q^{+}_{\lambda \mu i}\,=\,\frac{1}{2}
\sum_{\tau}^{n,p}\sum_{jj'} \left \{ \psi_{jj'}^{\lambda i}
[\alpha^{+}_{j} \alpha^{+}_{j'}]_{\lambda \mu}
-(-1)^{\lambda - \mu} \varphi_{jj'}^{\lambda i}
[\alpha_{j'}\alpha_{j}]_{\lambda -\mu} \right \}~.
\label{ph}
\end{equation}
The total number of different phonons for the given multipolarity
$\lambda$ should be equal to the sum of neutron and proton
two-quasiparticle states coupled to the same angular momentum.
The index $i$ is used to number these different phonons.

The coefficients $\psi_{jj'}^{\lambda i}$ and $\varphi_{jj'}^{\lambda i}$
of the linear transformation (\ref{ph}) one obtains by
diagonalization of the model Hamiltonian in the space of one-phonon
states, $Q^{+}_{\lambda \mu i} | \rangle _{ph}$.
This can be done for example by applying again the variation procedure:
\begin{equation}
\delta\{\langle \mid Q_{\lambda \mu i} H
Q^{+}_{\lambda \mu i}  \mid \rangle _{ph} -
\frac{\omega_{\lambda i}}{2} [ \sum_{jj'}
\{ (\psi^{\lambda i}_{jj'})^{2} -
(\varphi^{\lambda i}_{jj'})^{2} \} - 2] \} = 0~.
\label{mi2}
\end{equation}

It yields the well-known equations of the random-phase approximation
(RPA) which for the case of the separable form of the residual
interaction in $ph$--channel may be written as
\begin{equation}
\begin{tabular}{||ll||}
$(\ae^{(\lambda)}_{0} + \ae^{(\lambda)}_{1})
X^{\lambda}_n(\omega) - 1 $ &
$(\ae^{(\lambda)}_{0} - \ae^{(\lambda)}_{1})
X^{\lambda}_n(\omega) $
\\
$(\ae^{(\lambda)}_{0} - \ae^{(\lambda)}_{1})
X^{\lambda}_p(\omega)  $ &
$(\ae^{(\lambda)}_{0} + \ae^{(\lambda)}_{1})
X^{\lambda}_p(\omega) - 1 $
\end{tabular}
= \: 0
\label{det-rpa}
\end{equation}
where the following notation have been used
\begin{equation}
X^{\lambda}_{\tau}(\omega) =
\frac{1}{2 \lambda +1} \sum_{j j'}^{\tau}
\frac{(f_{j j'}^{\lambda} u_{j j'}^{(+)})^2
(\varepsilon_{j} + \varepsilon_{j'})}
{(\varepsilon_{j} + \varepsilon_{j'})^2 - \omega^2}~.
\end{equation}
The determinant equation (\ref{det-rpa}) is a function of the nucleus
excitation energy $\omega$. Solving this equation for each
multipolarity $\lambda^{\pi}$ one obtains the spectrum of nuclei one-phonon
excitation $\omega_{\lambda i}$. The index $i$ in the definition of
the phonon operator (\ref{ph}) gets the meaning of the order number
of the solution of Eq.~(\ref{det-rpa}).
The fermion structure of phonon excitation, i.e. the amplitudes
$\psi$ and $\varphi$, corresponding to the contribution of different
two-quasiparticle components to the phonon operator, are obtained from
the following equation
\begin{equation}
{\psi \choose \varphi}^{\lambda i}_{j j'} (\tau) =
\frac{1} {\sqrt{2 \cal{Y} _{\tau} ^{\lambda \mit{i}} }}
\cdot
\frac{f^{\lambda}_{j j'}(\tau) u_{j j'}^{(+)}}
{\varepsilon_{j} + \varepsilon_{j'} \mp \omega_{\lambda i}}
\label{psiphi}
\end{equation}
where the value $\cal{Y} _{\tau} ^{\lambda \mit{i}}$ is determined from
normalization condition for phonon operators:
\begin{equation}
\langle | Q_{\lambda \mu i} Q_{\lambda \mu i}^+ | \rangle_{ph} =
 \sum_{\tau}^{n,p} \sum_{jj'} \left \{ ( \psi^{\lambda i}_{jj'})^{2} -
(\varphi^{\lambda i}_{jj'} \right )^{2} \}  = 2
\label{nor}
\end{equation}
and one obtains
\begin{eqnarray}
\cal{Y} _{\tau} ^{\lambda \mit{i}} &=&
Y _{\tau} ^{\lambda \mit{i}} + Y _{-\tau} ^{\lambda \mit{i}}
\left\{ \frac{1- (\ae^{(\lambda)}_{0} + \ae^{(\lambda)}_{1})
X^{\lambda}_{\tau} (\omega_{\lambda i})}
{(\ae^{(\lambda)}_{0} - \ae^{(\lambda)}_{1})
X^{\lambda}_{- \tau} (\omega_{\lambda i})}
\right\} ^2;
\label{yy}
 \\
Y _{\tau} ^{\lambda \mit{i}} &=& \frac{1}{2 \lambda + 1} \cdot
\sum_{j j'}^{\tau}
\frac{(f_{j j'}^{\lambda} u_{j j'}^{(+)})^2
(\varepsilon_{j} + \varepsilon_{j'}) \omega_{\lambda i}}
{\left [ (\varepsilon_{j} + \varepsilon_{j'})^2 - \omega^2
\right ] ^2}~.
\nonumber
\end{eqnarray}

Equations (\ref{det-rpa}), (\ref{psiphi}) and (\ref{yy}) correspond
to natural parity
phonons. Similar equations are valid for unnatural parity phonons
by substituting the reduced spin-multipole matrix element
$f_{j j'}^{[\sigma l]_{\lambda}}$ and combination of coefficients of
Bogoliubov transformation $u_{j j'}^{(-)}$ for
$f_{j j'}^{\lambda}$ and $u_{j j'}^{(+)}$, respectively.
Also, amplitude $\varphi^{\lambda i}_{jj'}$ changes the sign in
Eq.~(\ref{psiphi}) for unnatural parity phonons.

The RPA equations have been obtained under the assumption that the
nucleus ground state is the phonon vacuum,
$Q_{\lambda \mu i} | \rangle _{ph} \equiv 0$.
This means that the ground state correlations due to the last term
of the model Hamiltonian, $H_{r.i}$, are taken into account. If they are
not accounted for and the ground state is still considered as a
quasiparticle vacuum $| \rangle _{q}$, one obtains the so-called
Tamm-Dankov approximation (TDA). The TDA equations can be easily
obtained from the RPA ones by neglecting backward going amplitudes in
the definition of the phonon operator (\ref{ph}), i.e. applying
$\varphi^{\lambda i}_{jj'} \equiv 0$.

The relation between the wave functions of the phonon and quasiparticle
vacuums is the following \cite{Sol92}:
\begin{equation}
| \rangle _{ph} = \frac{1}{\cal N} \prod_{\lambda} \exp \left \{
-\frac{1}{4} \sum_{i \mu} \sum_{j_1 j_2 \atop j_3 j_4}
(\psi^{\lambda i}_{j_3 j_4})^{-1} \varphi^{\lambda i}_{j_1 j_2}
(-1)^{\lambda - \mu}
[\alpha^+_{j_1} \alpha^+_{j_2}]_{\lambda \mu}
[\alpha^+_{j_3} \alpha^+_{j_4}]_{\lambda -\mu}
\right \} | \rangle _{q}
\label{gswf}
\end{equation}
where ${\cal N}$ is a normalization factor.

For actual numerical calculations one needs to determine the model
parameters. The average field for neutrons and protons is described in
the QPM by phenomenologic Woods-Saxon potential:
\begin{equation}
U^{\tau}(r) = \frac{V_0^{\tau}}{1+e^{(r-R_0^{\tau})/a_0^{\tau}}} -
\frac{\hbar^2}{\mu^2 c^2} ~\frac{1}{r} ~\frac{d}{dr}
\left( \frac{V_{ls}^{\tau}}{1+e^{(r-R_{ls}^{\tau})/a_{ls}^{\tau}}}
\vec{l} \cdot \vec{s} \right)
+V_C (r)                  \ .
\label{wse}
\end{equation}

The parameters of this potential for different A-mass
regions are listed in Table~\ref{ws} (see, also Ref. \cite{Pon79}).
We usually use $R_{ls}^{\tau}=R_0^{\tau}$,
$a_{ls}^{\tau}=a_0^{\tau}$, and $R_C=R_0^p$.
All single particle levels from the bottom are included in
calculation. The single particle continuum is approximated by narrow
quasibound states. This approximation gives a good description of the
exhaust of the
energy weighted sum rules (EWSR) for low values of $\lambda$ in medium
and heavy nuclei. For the lead region we use the single particle spectrum
near the Fermi surface from Ref.~\cite{Vor84} which was adjusted to
achieve a correct description of low-lying states in neighboring odd
nuclei. The parameters of the monopole $G^{(0)}_{\tau}$ have been
fitted to reproduce the pairing energies.

\begin{table}[tb]
\caption[ ]
{Parameters of Woods-Saxon potential Eq.~(\protect\ref{wse})
for different A-mass regions.
\label{ws}}
\begin{center}
\begin{tabular}{rcccccccc} \hline
A
&\multicolumn{4}{c}{\underline{neutrons}}
&\multicolumn{4}{c}{\underline{protons}} \\
     &       $V_0^n$   &  $R_0^n$  & $a_0^n$ &  $V_{ls}^n$
     &       $V_0^p$   &  $R_0^p$  & $a_0^p$ &  $V_{ls}^p$ \\
      &      [MeV]  &   [fm]  &  [fm] &   [MeV]
      &      [MeV]  &   [fm]  &  [fm] &   [MeV] \\
\hline
   49 &     -41.35  &   4.852  &   0.6200  &   -9.655
      &     -58.68  &   4.538  &   0.6301  &   -9.506 \\
   59 &     -46.20  &   5.100  &   0.6200  &   -9.540
      &     -53.70  &   4.827  &   0.6301  &   -8.270 \\
   91 &     -44.70  &   5.802  &   0.6200  &   -9.231
      &     -56.86  &   5.577  &   0.6301  &   -9.609 \\
  121 &     -43.20  &   6.331  &   0.6200  &   -8.921
      &     -59.90  &   6.133  &   0.6301  &  -10.363 \\
  141 &     -45.95  &   6.610  &   0.6200  &   -9.489
      &     -57.70  &   6.454  &   0.6301  &  -10.069 \\
  209 &     -44.83  &   7.477  &   0.6301  &   -8.428
      &     -60.30  &   7.359  &   0.6301  &  -11.186 \\
\hline
\end{tabular}
\end{center}
\end{table}

The parameters of the residual interaction are obtained
the following way. The strength of the residual interaction for
$\lambda^\pi $ = 2$^{+}$ and 3$^{-}$ is
adjusted to reproduce the properties (excitation energy and
$B(E \lambda )$ value, known from experiment) of the 2$_1^{+}$ and
3$_1^{-}$ states.
Usually it is not possible within one-phonon approximation,
discussed in this subsection, if sufficiently large single particle
spectrum is used. When the energy of the lowest excitation is adjusted
to the experimental value, the RPA equation yield an overestimated
collectivity, $B(E \lambda )$ value, for this state. And vice versa, if
the collectivity of this state is reproduced, the excitation energy is
too high as compared to the experimental value. The situation
sufficiently improves when the coupling of one-phonon states to more
complex configurations is taken into account as will be discussed in
the next subsection. For the lowest excited state the coupling to
complex configurations results in the energy shift downwards. Thus,
for nuclei not very far from  a  closed  shell  it  becomes  possible  to
achieve a
good description of both, the excitation energy and the $B(E \lambda )$
value. The ratio between isoscalar and isovector strength of the
residual interaction is usually fixed as
$\ae^{(\lambda)}_{1} / \ae^{(\lambda)}_{0} = -1.2$ in calculation with
the radial formfactor of the multipole operator as a derivative of the
average field. With this ratio the best description of isovector
multipole resonances with $\lambda > 1$ is achieved although the
experimental information on these resonances is still sparse.
For the dipole-dipole residual interaction the strength parameter are
adjusted to exclude the spurious center of mass motion and to obtain a
correct position of the GDR centroid.
For the phonons with the multipolarity $\lambda \ge 4$ the same
procedure of adjusting the strength parameters as for $\lambda^{\pi} =
2^+$ and $3^-$ cannot be applied. First, it is because the lowest
states of high multipolarity are much less collective and their
properties are more sensitive to description of single particle
levels near the Fermi surface than to the strength of the residual
interaction. Second, in many cases the lowest states with
$\lambda \ge 4$ are either two-phonon states or the states with a
large admixture of two-phonon configurations, thus, their properties
are determined by phonons of another multipolarity. For these reasons
we use $\ae^{(\lambda)}_{0,1} = \ae^{(2^+)}_{0,1}$ for even parity
phonons and $\ae^{(\lambda)}_{0,1} = \ae^{(3^-)}_{0,1}$ for odd parity
in calculation with $f_{\lambda}^{\tau}(r) = dU^{\tau}(r)/dr$.
In fact, the difference between
$\ae^{(2^+)}_{0,1}$ and $\ae^{(3^-)}_{0,1}$ does not exceed a few
percent with this radial formfactor of residual force.

\subsection{Mixing between simple and complex configurations in
wave fun\-ctions of excited states}

Diagonalization of the model Hamiltonian in the space of one-phonon states
allows us to write it in the form
\begin{equation}
H = \sum_{\lambda \mu i} \omega_{\lambda i} Q^{+}_{\lambda \mu i}
Q_{\lambda \mu i} + H_{int.}~,
\label{h3}
\end{equation}
\begin{equation}
H_{int.} = - \frac{1}{2} \sum_{\lambda \mu i}
\left \{ \left [ (-1)^{\lambda - \mu} Q^{+}_{\lambda \mu i}
+ Q_{\lambda -\mu i} \right ]
\sum_{j j' \tau} \frac{f^{\lambda}_{j j'} v_{j j'}^{(-)}}
{\sqrt{2 \cal{Y} _{\tau} ^{\lambda \mit{i}} }} B_{\tau}(j j'; \lambda -
\mu) + h. c. \right \}~,
\label{hint}
\end{equation}
where the origin of the second term in Eq.~(\ref{h3}) can be traced back to
the last term of multipole operator (\ref{m1}) which cannot be projected
onto the space of the phonon operators.  On the other hand, applying
Marumori expansion technique \cite{Mar64}, one may expand the operator
$B_{\tau}(j j'; \lambda - \mu) \sim \alpha^+ \alpha$ in
an infinite sum of even-number phonon operators.
Keeping only the first term of this expansion, the non-diagonal term of the
model Hamiltonian, $H_{int.}$, in the space of phonon operators may be
re-written as
\begin{equation}
H_{int.} = \sum_{\lambda \mu i \atop {\scriptstyle \lambda_1 \mu_1 i_1 \atop
\scriptstyle \lambda_2 \mu_2 i_2}}
U^{\lambda_1 i_1}_{\lambda_2 i_2}(\lambda i) Q^+_{\lambda \mu i}
[Q_{\lambda_1 \mu_1 i_1} Q_{\lambda_2 \mu_2 i_2}]_{\lambda \mu} +h.c.
\label{hint1}
\end{equation}
where the matrix element of interaction between one- and two-phonon
configurations,
$U^{\lambda_1 i_1}_{\lambda_2 i_2}(\lambda i)$,
can be calculated by making use of the internal fermion structure 
of phonons, i.e.
$\psi$ and $\varphi$ coefficients, and reduced matrix elements of the
separable force formfactor, $f^{\lambda}_{j_1j_2}$. It has the form
\begin{eqnarray}
U^{\lambda_1 i_1}_{\lambda_2 i_2}(\lambda i) &=&
<Q_{\lambda i} | H | [Q^+_{\lambda_1 i_1}
Q^+_{\lambda_2 i_2}]_{\lambda} > =
(-1)^{\lambda_1+\lambda_2-\lambda}\sqrt{\frac{(2\lambda_1+1)(2\lambda_2+1)}{2}
}
\nonumber \\
&\times& \sum_{\tau}^{n,p} \sum_{j_1 j_2 j_3}  \left [
\frac{f^{\lambda}_{j_1j_2}v^{(\mp)}_{j_1j_2}}
{\sqrt{{\cal{Y}}^{\lambda i}_{\tau}}}
\left\{
\begin{array}{ccc}
\lambda_1 & \lambda_2 & \lambda \\
j_2  & j_1  & j_3 \end{array}
\right\}
\left( \psi^{\lambda_1 i_1}_{j_3 j_1}  \:
\varphi^{\lambda_2 i_2}_{j_2 j_3} \pm
\psi^{\lambda_2 i_2}_{j_2 j_3}   \:
\varphi^{\lambda_1 i_1}_{j_3 j_1}
\right)
\right .
\nonumber \\
&&\hspace*{12.5mm}
+\left .
\frac{f^{\lambda_1}_{j_1j_2}v^{(\mp)}_{j_1j_2}}
{\sqrt{{\cal{Y}}^{\lambda_1 i_1}_{\tau}}}
\left\{
\begin{array}{ccc}
\lambda_1 & \lambda_2 & \lambda \\
j_3  & j_2  & j_1 \end{array}
\right\}
\left( \varphi^{\lambda i}_{j_2 j_3}  \:
\varphi^{\lambda_2 i_2}_{j_3 j_1} \pm
\psi^{\lambda i}_{j_2 j_3}   \:
\psi^{\lambda_2 i_2}_{j_3 j_1}
\right) \right .
\nonumber \\
&&\hspace*{12.5mm}
+\left .
\frac{f^{\lambda_2}_{j_1j_2}v^{(\mp)}_{j_1j_2}}
{\sqrt{{\cal{Y}}^{\lambda_2 i_2}_{\tau}}}
\left\{
\begin{array}{ccc}
\lambda_1 & \lambda_2 & \lambda \\
j_1  & j_3  & j_2 \end{array}
\right\}
\left( \psi^{\lambda i}_{j_3 j_1}  \:
\psi^{\lambda_1 i_1}_{j_2 j_3} \pm
\varphi^{\lambda i}_{j_3 j_1}   \:
\varphi^{\lambda_1 i_1}_{j_2 j_3}
\right)
\right ]~.
\label{U}
\end{eqnarray}
The upper (lower) sign in each of three terms in Eq.~(\ref{U}) correspond to
multipole (spin-multipole) matrix element $f^{\lambda}_{j_1j_2}$,
$f^{\lambda_1}_{j_1j_2}$ or $f^{\lambda_2}_{j_1j_2}$, respectively.

Thus, we have completed a projection of the nuclear Hamiltonian into the
space of phonon operators. Now we may assume that phonons obey boson
statistics and work in the space of boson operators only.
The presence of the term of interaction, $H_{int}$, in the model Hamiltonian
means that the approximation, in which excited states of the nucleus are
considered as pure one-, two-, multi-phonon states, is not sufficient.
In fact, we have already mentioned above that it is not possible to describe
the properties of the lowest collective vibrations in spherical nuclei in
one-phonon approximation. It is also well-known that the coupling between
one- and two-phonon configurations is the main mechanism for the damping of
giant resonances. All this means that one needs to go beyond the
approximation of independent phonons and take into account a coupling
between them.
To accomplish this task we write the wave function of excited states
with angular momentum $J$ and projection $M$ in
even-even nuclei in the most general form as a mixture of one-, two-,
$\cdots$ phonon configurations:
\begin{eqnarray}
\Psi^{\nu}(JM) &=&
\left \{ \sum_{\alpha_1} S^{\nu}_{\alpha_1}(J) Q^{+}_{\alpha_1} \right.
~+~
\sum_{\alpha_2 \beta_2} \frac{D^{\nu}_{\alpha_2 \beta_2}(J)}
{\sqrt{1+\delta_{\alpha_2, \beta_2}}}
[Q^{+}_{\alpha_2}  Q^{+}_{\beta_2}]_{JM}
\label{wf1}
   \\[3mm]
 &+& \left.
\sum_{\alpha_3 \beta_3 \gamma_3} \frac{T^{\nu}_{\alpha_3 \beta_3
\gamma_3}(J) }
{\sqrt{1+\delta_{\alpha_3, \beta_3, \gamma_3}}}
[Q^{+}_{\alpha_3}  Q^{+}_{\beta_3} Q^{+}_{\gamma_3}]_{JM}
+ \cdots
 \right \}
\mid \rangle _{ph}~,    \nonumber
\end{eqnarray}
\begin{equation}
\delta_{\alpha_3, \beta_3, \gamma_3}  =
\delta_{\alpha_3, \beta_3} + \delta_{\alpha_3, \gamma_3}
+ \delta_{\beta_3, \gamma_3}
+ 2 \delta_{\alpha_3, \beta_3}  \delta_{\alpha_3, \gamma_3}~.
\end{equation}
By greek characters we mean the phonon's identity, i.e. its multipolarity and
order number, $\alpha \equiv \lambda^{\pi} i$, the index
$\nu$ (= 1, 2, 3 $\ldots$) labels
whether a state $J$ is the first, second, etc., state in the total energy
spectrum of the system.
It is assumed that any combination $\alpha, \beta, \gamma$ of phonons
appears only once.
The second and the third terms in Eq. (\ref{wf1}) include phonons of different
multipolarities and parities, they only must couple to the same total
angular momentum $J$ as the one-phonon term.

Let us limit the wave function of excited states by three-phonon terms and
diagonalize the model Hamiltonian of Eqs. (\ref{h3},\ref{hint1}) in the
space of these states. We use for that a minimization procedure
\begin{equation}
\delta \left\{ <\Psi^{\nu}(JM) \mid  H  \mid  \Psi^{\nu}(JM)>
- E_x^{J}<\Psi^{\nu}(JM) \mid\Psi^{\nu}(JM)>
\right\}  =  0~,
\label{min3}
\end{equation}
which yields a set of linear equations over unknown
wave function coefficients
$S^{\nu}_{\alpha_1}(J)$, $D^{\nu}_{\alpha_2 \beta_2}(J)$ and
$T^{\nu}_{\alpha_3 \beta_3 \gamma_3}(J)$:
\begin{eqnarray}
&&(\omega_{\alpha_1}-E_x^{J})
S^{\nu}_{\alpha_1}(J)+
\sum_{\alpha_2, \beta_2}
D^{\nu}_{\alpha_2 \beta_2}(J)
\tilde{U}_{\alpha_2 \beta_2}^{\alpha_1}
=  0~,
\label{se1}
 \\[3mm]
&&\sum_{\alpha_1}
S^{\nu}_{\alpha_1}(J)
\tilde{U}_{\alpha_2 \beta_2}^{\alpha_1}
+
(\omega_{\alpha_2} + \omega_{\beta_2}
- E_x^{J})
D^{\nu}_{\alpha_2 \beta_2}(J)
+
\sum_{\alpha_3 \beta_3 \gamma_3}
T^{\nu}_{\alpha_3 \beta_3 \gamma_3}(J)
\tilde{U}^{\alpha_2 \beta_2}_{\alpha_3 \beta_3 \gamma_3}
=  0~,
\nonumber  \\[3mm]
&&\sum_{\alpha_2 \beta_2}
D^{\nu}_{\alpha_2 \beta_2}(J)
\tilde{U}^{\alpha_2 \beta_2}_{\alpha_3 \beta_3 \gamma_3}
+
(\omega_{\alpha_3} + \omega_{\beta_3} + \omega_{\gamma_3}
- E_x^{J})
T^{\nu}_{\alpha_3 \beta_3 \gamma_3}(J)
= 0~.
\nonumber
\end{eqnarray}
Applying boson commutation relations for phonons, the matrix element of
interaction between two- and three-phonon configurations,
\begin{equation}
\tilde{U}^{\alpha_2 \beta_2}_{\alpha_3 \beta_3 \gamma_3} =
{\sqrt{1+\delta_{\alpha_2, \beta_2}}}
{\sqrt{1+ \delta_{\alpha_3, \beta_3, \gamma_3}}}
\langle
[Q_{\alpha_2}  Q_{\beta_2}]_{JM}
    \mid H_{int.}    \mid
[Q^{+}_{\alpha_3}  Q^{+}_{\beta_3} Q^{+}_{\gamma_3}]_{JM}
 \rangle~,
\label{u-c1}
\end{equation}
can be expressed as a function of matrix elements of interaction between
one- and two-phonon configurations,
\begin{equation}
\tilde{U}_{\alpha_2 \beta_2}^{\alpha_1} =
\sqrt{1+\delta_{\alpha_2, \beta_2}}
\langle  Q_{\alpha_1}
    \mid H_{int.}    \mid
[Q^{+}_{\alpha_2}  Q^{+}_{\beta_2}]_{JM}
 \rangle =
\sqrt{1+\delta_{\alpha_2, \beta_2}}
U_{\alpha_2}^{\beta_2}(\alpha_1)~,
\label{u-c}
\end{equation}
as following
\begin{eqnarray}
\tilde{U}^{\alpha_2 \beta_2}_{\alpha_3 \beta_3 \gamma_3}  &=&
\sqrt{1+\delta_{\beta_3, \gamma_3}}
[\tilde{U}_{ \beta_3 \gamma_3}^{\alpha_2} \delta_{ \beta_2, \alpha_3}
+     \tilde{U}_{ \beta_3 \gamma_3}^{ \beta_2} \delta_{\alpha_2, \alpha_3}]
\label{U-U}  \\[3mm]
&+&\sqrt{1+\delta_{\alpha_3, \gamma_3}}
[\tilde{U}_{\alpha_3 \gamma_3}^{\alpha_2} \delta_{ \beta_2,  \beta_3}
+\tilde{U}_{\alpha_3 \gamma_3}^{ \beta_2} \delta_{\alpha_2,  \beta_3}]
+  \sqrt{1+\delta_{\alpha_3, \beta_3}}
[\tilde{U}_{\alpha_3 \beta_3 }^{\alpha_2} \delta_{ \beta_2, \gamma_3}
+\tilde{U}_{\alpha_3 \beta_3 }^{ \beta_2} \delta_{\alpha_2, \gamma_3}]
\nonumber
\end{eqnarray}
and the value $\tilde{U}_{\alpha_2}^{\beta_2}(\alpha_1)$
 is calculated according to
Eq.~(\ref{U}). Since we have used pure boson commutation relations for phonons
the two-phonon configuration $[\alpha_2 \beta_2]_J$ couples only to those
three-phonon configurations $[\alpha_3 \beta_3 \gamma_3]_J$ where either
$\alpha_3$, $\beta_3$ or $\gamma_3$ are equal to $\alpha_2$ or $\beta_2$.
This is governed by $\delta$-functions in Eq.~(\ref{U-U}).

The number of linear equations (\ref{se1}) equals to the number of one-,
two- and three-phonon configurations included in the wave function
(\ref{wf1}). Solving these equations we obtain the energy spectrum
$E^J_{\nu}$ of excited states described by wave function (\ref{wf1})
and the coefficients of wave function (\ref{se1}), $S$, $D$ and $T$.

It should be pointed out that within this approximation, in which
phonons are considered as ideal bosons and nuclear Hamiltonian includes
one-phonon exchange term,
multi-phonon configurations of course posses no anharmonicity features.
The strength of any one- or many-phonon configuration included
in the wave function (\ref{wf1}) fragments over some energy interval due to
the interaction with other configurations. But the centroid of the strength
distribution remains at the unperturbed energy. Thus, the energy centroid of
two-phonon configuration $[\alpha_2 \beta_2]_J$ equals exactly to the sum of
energies of $\alpha_2$ and $\beta_2$ phonons for all values of $J$.
To consider anharmonic properties of multi-phonon states one needs to go
beyond pure boson features of excitations in even-even nuclei and take into
account their internal fermion structure.

Another reason to return back to the fermion origin of phonon excitations
is two main problems in considering multi-phonon states associated with
the boson mapping procedure.
The first problem is an admixture of
spurious $npnh$ configurations which violate Pauli principle in the
wave function of $n$-phonon state.
The second one is related to the
fact that the set of pure n-phonon states is mathematically non-orthonormal
if the internal fermion structure of phonons is taken into account (see
Refs.~\cite{Pie95,Gri96} for more details).
To overcome these problems we will keep
on using a phonon's imaging of nuclear excitation and use the same expression
for the wave function of excited states (\ref{wf1}) but in calculation of
the norm of this wave function, $<\Psi^{\nu}(JM) \mid\Psi^{\nu}(JM)>$,
and the energy of this state,
$<\Psi^{\nu}(JM) \mid  H  \mid  \Psi^{\nu}(JM)>$,
we will use exact commutation relations between phonon operators:
\begin{eqnarray}
[Q_{\lambda \mu i}, Q^{+}_{\lambda' \mu' i'} ]_{\_}
&=& 
\delta_{\lambda, \lambda'} \delta_{\mu, \mu'}
\delta_{i,i'} 
- \sum_{\scriptstyle jj'j_{2} \atop \scriptstyle m m'm_{2}}
\alpha^{+}_{jm} \alpha_{j'm'}
\label{q-q-c}
\\
&\times&
\left \{
\psi^{\lambda i}_{j'j_{2}} \psi^{\lambda'i'}_{jj_{2}}
C_{j'm' j_2m_2}^{\lambda \mu}
C_{j m  j_2m_2}^{\lambda' \mu'}
- (-)^{\lambda + \lambda'+\mu + \mu'}
\varphi^{\lambda i}_{jj_{2}} \varphi^{\lambda'i'}_{j'j_{2}}
C_{j m  j_2m_2}^{\lambda -\mu}
C_{j'm' j_2m_2}^{\lambda' -\mu'}
\right \}
\nonumber
\end{eqnarray}
and exact commutation relations between phonon and quasiparticle operators:
\begin{eqnarray}
[\alpha_{jm}, Q^+_{\lambda \mu i} ]_{\_} &=& \sum_{j'm'}
\psi^{\lambda i}_{jj'} C^{\lambda \mu}_{jm j'm'} \alpha^+_{j'm'}~,
\label{a-q-c}
  \\
\nonumber
[\alpha^+_{jm}, Q^+_{\lambda \mu i} ]_{\_} &=&
(-1)^{\lambda - \mu} \sum_{j'm'}
\varphi^{\lambda i}_{jj'} C^{\lambda -\mu}_{jm j'm'} \alpha_{j'm'}~.
\end{eqnarray}
Also we will not expand the operator $B_{\tau}(jj'; \lambda \mu)$ in
Eq.~(\ref{hint}) into a sum
of phonon operators but use its exact fermion structure.
The first term of Eq.~(\ref{q-q-c}) corresponds to the ideal boson
approximation while the second one is a correction due to the fermion
structure of phonon operators. The overlap matrix elements between different
two-phonon configurations modifies as
\begin{equation}
\langle [Q_{\beta'} Q_{\alpha'}]_J | [Q^+_{\alpha} Q^+_{\beta}]_J \rangle =
\langle [b_{\beta'} b_{\alpha'}]_J | [b^+_{\alpha} b^+_{\beta}]_J \rangle +
 K^J(\beta' \alpha'  | \alpha \beta)
\label{k2}
\end{equation}
where $b^+_{\alpha}$ is the ideal boson operator and the
quantity $K$,
\begin{eqnarray}
&&K^J(\beta' \alpha'  | \alpha \beta) =
 K^J(\lambda_4 i_4 \lambda_3 i_3   | \lambda_1 i_1 \lambda_2 i_2 )  =
\sqrt{(2 \lambda_1 +1)(2 \lambda_2 +1)(2 \lambda_3 +1)(2 \lambda_4 +1) }
\nonumber \\
&&(-1)^{\lambda_2 + \lambda_4}
\sum_{j_1 j_2 \atop j_3 j_4} (-1)^{j_2+j_4}
\left\{
\begin{array}{ccc}
j_1 & j_2 & \lambda_4 \\
j_4 & j_3 & \lambda_3  \\
\lambda_1 & \lambda_2 & J
\end{array}
\right\}
\left(
\psi^{\lambda_3 i_3}_{j_3 j_4}   \:
\psi^{\lambda_1 i_1}_{j_1 j_4}   \:
\psi^{\lambda_2 i_2}_{j_3 j_2}   \:
\psi^{\lambda_4 i_4}_{j_1 j_2}   -
\varphi^{\lambda_3 i_3}_{j_3 j_4}   \:
\varphi^{\lambda_1 i_1}_{j_1 j_4}   \:
\varphi^{\lambda_2 i_2}_{j_3 j_2}   \:
\varphi^{\lambda_4 i_4}_{j_1 j_2}
\right),
\nonumber \\
&&
\label{k-coef}
\end{eqnarray}
is the Pauli principle correction coefficient.
The experience of realistic calculations shows that usually
$|K^J(\beta \alpha  | \alpha \beta)| \gg
|K^J(\beta' \alpha'  | \alpha \beta)| $ (where $\alpha \neq \alpha'$
and/or $\beta \neq \beta'$)
and that the so-called diagonal Pauli principle
approximation,
$K^J(\beta' \alpha'  | \alpha \beta) =
K^J(\alpha \beta) \delta_{\alpha, \alpha'}
\delta_{\beta, \beta'}$, provides rather good accuracy and
sufficiently simplifies the calculation.
For these reasons we will use this diagonal Pauli principle approximation in
what follows.

The similar expression, as (\ref{k2}), is valid for the overlap matrix
elements between different three-phonon configurations. It can be used as a
definition of the Pauli principle correction quantity
$K^J_I(\gamma' \beta' \alpha'  | \alpha \beta \gamma)$ which we will also
keep in diagonal approximation only.
The relation between $K^J(\alpha \beta)$ and $K^J_I(\alpha \beta \gamma)$
quantities is the following \cite{Din92}:
\begin{equation}
K^J_I(\alpha \beta \gamma) =
 K^I(\alpha \beta) \left (3+
\sum_{I '} \overline{U^2}(\alpha \beta J \gamma ;I, I')
K^{I'}(\beta \gamma)\right )
\end{equation}
where $\overline{U^2}$ stands for the Jahn coefficients \cite{Var88}.

When internal fermion structure of phonons is taken into account and exact
commutation relations (\ref{q-q-c},\ref{a-q-c}) are applied the secular
equation (\ref{se1}) transforms into
\begin{eqnarray}
&&(\omega_{\alpha_1}-E_x^{J})
S^{\nu}_{\alpha_1}(J)+
\sum_{\alpha_2, \beta_2}
D^{\nu}_{\alpha_2 \beta_2}(J)
\tilde{U}_{\alpha_2 \beta_2}^{\alpha_1}
=  0~,
\label{se2}
 \\[3mm]
&&\sum_{\alpha_1}
S^{\nu}_{\alpha_1}(J)
\tilde{U}_{\alpha_2 \beta_2}^{\alpha_1}
+
(\omega_{\alpha_2} + \omega_{\beta_2}
+ \Delta \omega_{\alpha_2 \beta_2}^J
- E_x^{J})
D^{\nu}_{\alpha_2 \beta_2}(J)
+
\sum_{\alpha_3 \beta_3 \gamma_3 I}
T^{\nu}_{\alpha_3 \beta_3 \gamma_3}(J)
\tilde{U}^{\alpha_2 \beta_2}_{\alpha_3 \beta_3 \gamma_3}
=  0~,
\nonumber  \\[3mm]
&&\sum_{\alpha_2 \beta_2}
D^{\nu}_{\alpha_2 \beta_2}(J)
\tilde{U}^{\alpha_2 \beta_2}_{\alpha_3 \beta_3 \gamma_3}
+
(\omega_{\alpha_3} + \omega_{\beta_3} + \omega_{\gamma_3}
+ \Delta \omega_{\alpha_3 \beta_3 \gamma_3}^J
- E_x^{J})
T^{\nu}_{\alpha_3 \beta_3 \gamma_3}(J)
= 0~.
\nonumber
\end{eqnarray}

The values $\Delta \omega_{\alpha_2 \beta_2}^J$ and
$\Delta \omega_{\alpha_3 \beta_3 \gamma_3}^J =
 \Delta \omega_{\alpha_3 \beta_3}^J+
 \Delta \omega_{\beta_3 \gamma_3}^J+
 \Delta \omega_{\alpha_3 \gamma_3}^J$
are anharmonicity shifts of two- and three-phonon configurations,
respectively, due to the Pauli principle corrections.
In diagonal approximation they can be calculated according to:
\begin{equation}
\Delta \omega_{\alpha_2 \beta_2}^J = -\frac{K^J(\alpha_2 \beta_2)}{4}
\sum_{\tau}^{n,p} \left [
\frac{X^{\alpha_2}_{\tau}}{{\cal{Y}} _{\tau} ^{\alpha_2}} +
\frac{X^{\beta_2}_{\tau}}{{\cal{Y}} _{\tau} ^{\beta_2}}
\right ]~.
\label{delt}
\end{equation}

Another role of Pauli principle corrections is a somewhat renormalization of
the interaction between $n$- and ($n+1$)-phonon configurations. We have
used the same notations for these matrix elements
$\tilde{U}_{\alpha_2 \beta_2}^{\alpha_1}$ as in the case of the 'ideal boson
approximation' (see,  Eqs.~(\ref{u-c1},\ref{u-c})).
But calculating  the matrix elements
$\langle  Q_{\alpha_1}    \mid H_{int.}    \mid
[Q^{+}_{\alpha_2}  Q^{+}_{\beta_2}]_{JM} \rangle$
we take into account the fermion structure of phonons and nuclear
Hamiltonian and obtain
\begin{equation}
\tilde{U}_{\alpha_2 \beta_2}^{\alpha_1} =
\sqrt{1+\delta_{\alpha_2, \beta_2}}
U_{\alpha_2}^{\beta_2}(\alpha_1) \times
\left[  1+\frac{1}{2}
K^J(\alpha_2 \beta_2) \right]
\label{u-c2}
\end{equation}
where the value $U_{\alpha_2}^{\beta_2}(\alpha_1)$ is calculated again
according to Eq.~(\ref{U}).
A similar additional factor
$[1+1/2 \times K^J_I(\alpha_3 \beta_3 \gamma_3)]$ receives
the matrix element of interaction between two- and three-phonon
configurations.

The minimal value of the quantity $K^J(\alpha \beta)$ equals to $-2$.
It corresponds to the case of the maximal Pauli principle violation,
i.e. to a spurious multi-phonon configuration. It happens only when
$\alpha_2$ and $\beta_2$ phonons are purely two-quasiparticle states.
In such a case the matrix element of interaction
$\tilde{U}_{\alpha_2 \beta_2}^{\alpha_1} \equiv 0$
(see, Eq.~(\ref{u-c2})) and the spurious state is completely separated
from other states.
While dealing with collective $\alpha_2$ and $\beta_2$ phonons, when a
possible admixture of the spurious four-quasiparticle configurations is
small, the value of $K^J(\alpha \beta)$ is close to 0. Nevertheless the
value of the anharmonicity shift $\Delta \omega_{\alpha_2 \beta_2}^J$
is not vanishingly small for the later because of the relatively large
value of the ratio $X^{\alpha}_{\tau}/{\cal{Y}} _{\tau} ^{\alpha}$ in
Eq.~(\ref{delt}). This shift is the largest one for the collective
low-lying multi-phonon configurations. For non-collective multi-phonon
states the shift is small because of the small value of the above
mentioned ratio.

Equations (\ref{se2}) have been obtained under two main assumptions. The
first one is the already discussed diagonal Pauli principle approximation.
The second assumption is the neglecting of the higher 
order terms of the interaction
part of the nuclear Hamiltonian as compared to the one
in Eq.~(\ref{hint1}) which couples $n$- and $(n\pm 1)$-phonon configurations.
For Eqs.~(\ref{se2}) it means that a direct coupling
between one- and three-phonon configurations of the wave
function (\ref{wf1}) which is possible due to non-zero matrix element
$\langle  Q_{\alpha_1}    \mid \alpha^+_{jm} \alpha_{jm} \mid
[Q^{+}_{\alpha_3}  Q^{+}_{\beta_3} Q^+_{\gamma_3}]_{JM} \rangle$,
is neglected.
In realistic calculation we will also use a selection of three-phonon
configurations provided by Eq.~(\ref{U-U}) although now the matrix element
$\tilde{U}^{\alpha_2 \beta_2}_{\alpha_3 \beta_3 \gamma_3} \neq 0$ even if
one of $\alpha_3$, $\beta_3$ or $\gamma_3$ is not necessarily equal to
$\alpha_2$ or $\beta_2$. These omitted matrix elements are orders of
magnitude smaller as compared to the accounted for ones.

Solving the system of linear equations (\ref{se2}) we obtain the spectrum of
excited states, $E_{\nu}^J$, described by the wave function (\ref{wf1}) and
coefficients $S^{\nu}_{\alpha_1}(J)$,  $D^{\nu}_{\alpha_2 \beta_2}(J)$
and $T^{\nu}_{\alpha_3 \beta_3 \gamma_3}(J)$ reflecting the phonon structure
of excited states.
Usually, in calculation of the properties of single giant resonances
the three-phonon terms of the wave function (\ref{wf1}) are omitted.
Then it is possible to solve the system of linear equations (\ref{se2})
with the rank of the $10^3-10^4$ order by a direct diagonalization.
But while considering the damping properties of two-phonon resonances,
three-phonon configurations cannot be omitted. For this case instead of
the diagonalization of the linear matrices of very high orders, an
alternative solution is possible.
We may substitute the first and last equations of (\ref{se2}) into the
second equation and obtain the system of non-linear equations
\begin{eqnarray}
&&det
\left | \left | \right.\right.  (\omega_{\alpha_2} + \omega_{\beta_2}
+\Delta\omega^J_{\alpha_2 \beta_2}
 - E_x^J)
\delta_{\alpha_2 \beta_2,\alpha_2' \beta_2'}
-
\sum_{\alpha_1} \frac{
\tilde{U}^{\alpha_1}_{\alpha_2 \beta_2} ~
\tilde{U}^{\alpha_1}_{\alpha_2' \beta_2'}   }
{\omega_{\alpha_1} - E_x^J} \nonumber \\[3mm]
&& \left. \left.
~-~ \sum_{\alpha_3 \beta_3 \gamma_3} \frac{
\tilde{U}_{\alpha_3 \beta_3 \gamma_3}^{\alpha_2 \beta_2} ~
\tilde{U}_{\alpha_3 \beta_3 \gamma_3}^{\alpha_2' \beta_2'}
}
{\omega_{\alpha_3} + \omega_{\beta_3} + \omega_{\gamma_3}
+\Delta\omega^J_{\alpha_3 \beta_3 \gamma_3}
- E_x^J}
\right | \right |
= 0~,
\label{det}
\end{eqnarray}
the rank of which equals to the number of two-phonon configurations included
in the wave function (\ref{wf1}). The solution of the system (\ref{det})
by some iterative method yield again the spectrum of excited states
$E_{\nu}^J$ and coefficients $D^{\nu}_{\alpha_2 \beta_2}(J)$.
Other coefficients of the wave function (\ref{wf1}) are related to these
coefficients as following
\begin{eqnarray}
S^{\nu}_{\alpha_1} (J) &=&
~-~
\frac{\sum_{\alpha_2 \beta_2}  D^{\nu}_{\alpha_2 \beta_2} (J)~
\tilde{U}^{\alpha_1}_{\alpha_2 \beta_2}  }
{\omega_{\alpha_1} - E_x^{\nu}}~,
\nonumber \\[3mm]
T^{\nu}_{\alpha_3 \beta_3 \gamma_3} (J) &=&
~-~
\frac{\sum_{\alpha_2 \beta_2} D^{\nu}_{\alpha_2 \beta_2} (J)~
\tilde{U}_{\alpha_3 \beta_3 \gamma_3}^{\alpha_2 \beta_2}  }
{\omega_{\alpha_3} + \omega_{\beta_3} + \omega_{\gamma_3}
+\Delta\omega^J_{\alpha_3 \beta_3 \gamma_3}
- E^J_{\nu}}~.
\label{st}
\end{eqnarray}

It may be argued that the boson mapping with keeping the fermion information
of the phonons' images at all stages of transformations gives no advantage as
compared to $npnh$ approach since, mathematically, a direct
correspondence
between two methods can be established only if the full basis of
$n$-phonon
states is used. However many $npnh$ configurations interact very weakly
with
other ones and as a result practically do not mix with them. It allows a
sufficient truncation of multi-phonon configurations in the wave function
(\ref{wf1}) based on their physical properties with keeping a good accuracy
for the components important for the subject of research. From the point
of view of the Pauli principle violation the most dangerous
multi-phonon configurations are the ones made of non-collective RPA
states. On the other hand, these configurations interact with the other
ones much weaker than the multi-phonon configurations including at
least one collective phonon. For these reasons the first are not
accounted for in the wave function (\ref{wf1}) in realistic calculation.
As the criteria ``collective/non-collective" we take the contribution of
the main two-quasiparticle component to the wave function of the phonon
operator. If the contribution exceeds 50-60\% we will call the phonon
non-collective.

Let us consider now the electromagnetic excitation of pure one- and
multi-phonon states from the ground state. The one-body operator of
electromagnetic transition has the form
\begin{equation}
{\cal M}(E \lambda \mu) = \sum_{\tau}^{n,p} e^{(\lambda)}_{\tau}
\sum_{\scriptstyle j j' \atop \scriptstyle m m'} (-1)^{j'+m'}
\frac{<j||E \lambda||j'>}{\sqrt{2 \lambda +1}}
C_{jm j'm'}^{\lambda \mu}
a^+_{jm} a_{j' -m'}
\label{tr1}
\end{equation}
where the single particle transition matrix element $<j||\mbox{E}
\lambda||j'> \equiv <j||i^{\lambda} Y_{{\lambda}} r^{\lambda}||j'>$
and $e^{(\lambda)}_{\tau}$ are effective charges for neutrons and
protons. In calculations we use the following values of effective
charges: $e^{(1)}_{n} = -Z/A$ and $e^{(1)}_{p} = N/A$ to separate the
center of mass motion and $e^{(\lambda \ne 1)}_{n} = 0$ and
$e^{(\lambda \ne 1)}_{p} = 1$. Performing the transformation from
particle operators to quasiparticle and phonon ones in Eq.~(\ref{tr1}),
this equation transforms into
\begin{eqnarray}
{\cal M}(E \lambda \mu ) &=&
\sum_{\tau}^{n,p} e^{(\lambda)}_{\tau}
\sum_{j j'} \frac{<j||E \lambda||j'>}
{\sqrt{2 \lambda +1}}  \left \{
\frac{u^{(+)}_{j j'}}{2} \sum_{i}
(\psi^{\lambda i}_{j j'} + \varphi^{\lambda i}_{j j'})
(Q^{+}_{\lambda \mu i}+(-)^{\lambda -\mu }Q_{\lambda -\mu i})
\right .     \nonumber
\\
&&\left .
+ v^{(-)}_{j j'}
\sum_{mm'}C_{jmj'm'}^{\lambda \mu }(-)^{j'+m'}
\alpha_{j' m'}^{+}\alpha _{j'-m'}
\right \}
\label{tr2}
\end{eqnarray}
where the first term corresponds to one-phonon exchange between initial
and final states and the second one is responsible for ``boson
forbidden"  electromagnetic transitions (see, for details
Ref.~\cite{Pon98}).
Then the reduced matrix element of the electromagnetic excitation of the
one-phonon state $\lambda i$ from the ground state $0^{+}_{g.s.}$ in
even-even nuclei may be calculated according to
\begin{equation}
<Q_{\lambda i}||{\cal M}(E \lambda)||0^{+}_{g.s.}> =
 \sum_{\tau}^{n,p} e^{(\lambda)}_{\tau} \sum_{j_1 j_2} {1\over 2}
<j_1 ||E \lambda || j_2> u^{(+)}_{j_1 j_2}
\left ( \psi^{\lambda i}_{j_1 j_2} + \varphi^{\lambda i}_{j_1 j_2}
\right )~.
\label{gs-1ph}
\end{equation}

Due to the ground state correlations the direct excitation of pure
two-phonon states $[Q^+_{\lambda_1 {i_1}} \times
Q^+_{\lambda_2 {i_2}}]_{\lambda}$ from the ground state is also
possible when we are dealing with the RPA phonons.
The physical reason for that becomes clear if we remember that the
ground state wave function includes a small admixture of four-, eight-,
$\cdots$ quasiparticle configurations (see, Eq. (\ref{gswf})).
The second term of Eq.~(\ref{tr2}) is responsible for these transitions
and the reduced matrix element can be obtained by applying the
commutation relations (\ref{a-q-c}). It has the form
\begin{eqnarray}
<[Q_{\lambda_2 {i_2}} \times
 Q_{\lambda_1{i_1}}]_{\lambda}
&||&{\cal M} (E \lambda)||0^{+}_{g.s.}>
=  \sqrt{(2\lambda_{1}+1)(2\lambda_{2}+1)}
\sum_{\tau}^{n,p} e^{(\lambda)}_{\tau}
\sum_{j_{1} j_{2} j_{3}}  v^{(-)}_{j_1 j_2}
 \nonumber
\\
& \times&
<j_1 ||E \lambda || j_2>
\left\{
\begin{array}{ccc}
\lambda_{2} & \lambda_{1} & \lambda \\
j_{1}  & j_{2}  & j_{3} \end{array}
\right\}
\left( \psi^{\lambda_{2} i_{2}}_{j_{2} j_{3}}  \:
\varphi^{\lambda_{1} i_{1}}_{j_{3} j_{1}} +
\psi^{\lambda_{1} i_{1}}_{j_{3} j_{1}}   \:
\varphi^{\lambda_{2} i_{2}}_{j_{2} j_{3}}
\right)~.
\label{gs-2ph}
\end{eqnarray}
Another type of boson forbidden $\gamma$-transitions which take place
due to the internal fermion structure of phonons are the ones between
one-phonon initial, $Q^+_{\lambda_{1} i_{1}} | \rangle_{ph}$, and final,
$Q^+_{\lambda_{2} i_{2}} | \rangle_{ph}$, states. The reduced matrix
element  of such transitions can be calculated according to
\begin{eqnarray}
<Q_{\lambda_2 {i_2}} ||{\cal M}(E \lambda)||Q^+_{\lambda_1{i_1}}>
&=&
\sqrt{2\lambda_{2}+1}
\sum_{\tau}^{n,p} e^{(\lambda)}_{\tau}
\sum_{j_{1} j_{2} j_{3}}
v^{(-)}_{j_1 j_2}
<j_1 ||E \lambda || j_2>  \nonumber \\
&\times&\left\{
\begin{array}{ccc}
\lambda_{1} & \lambda_{2} & \lambda \\
j_{1}  & j_{2}  & j_{3} \end{array} \right\}
\left( \psi^{\lambda_{1} i_{1}}_{j_{2} j_{3}}  \:
\psi^{\lambda_{2} i_{2}}_{j_{3} j_{1}}   +
\varphi^{\lambda_{1} i_{1}}_{j_{2} j_{3}}  \:
\varphi^{\lambda_{2} i_{2}}_{j_{3} j_{1}}
\right)~.
\label{q-q}
\end{eqnarray}
The matrix element for transitions between the two-phonon states
$[Q^+_{\lambda_1 i_1 }\times Q^+_{\lambda _2 i_2}]_{\lambda'}
| \rangle_{ph}$ and
$[Q^+_{\lambda_3 i_3 }\times Q^+_{\lambda _4 i_4}]_{\lambda''}
| \rangle_{ph}$
is very complex and not presented here.
Its first order term is very similar to the one for transitions
between the one-phonon states
$Q^+_{\lambda_1 i_1} | \rangle_{ph}$ and
$Q^+_{\lambda_4 i_4} | \rangle_{ph}$ and may be obtained by
assuming that the
fermion structure of one phonon is ``frozen", i.e., assuming that
$\lambda_2 i_2  \equiv \lambda_3 i_3$.

When the coupling between one- and multi-phonon configurations is
accounted for in the wave function of excited states,
the reduced matrix element of the electromagnetic excitation of the
states of Eq.~(\ref{wf1}) may be written as
\begin{eqnarray}
<\Psi^{\nu}(J) ||{\cal M}(E \lambda)||0^{+}_{g.s.}>  &=&
\left \{ \sum_{\alpha_1} S^{\nu}_{\alpha_1}(J)
<Q_{\lambda i}||{\cal M}(E \lambda)||0^{+}_{g.s.}>
\right.   \label{gs-wf1}   \\
&+&   \left .
\sum_{\alpha_2 \beta_2} \frac{D^{\nu}_{\alpha_2 \beta_2}(J)}
{\sqrt{1+\delta_{\alpha_2, \beta_2}}}
<[Q_{\lambda_2 {i_2}} \times Q_{\lambda_1{i_1}}]_{\lambda}
||{\cal M}(E \lambda)||0^{+}_{g.s.}>       \right \}
\nonumber
\end{eqnarray}
where we have neglected the direct excitation of three-phonon
configurations from the ground state.
Since an admixture of multi-quasiparticle configurations in the
ground state wave function is very small, the reduced matrix element
Eq.~(\ref{gs-2ph}) is typically about two orders of magnitude smaller as
compared to the reduced matrix element Eq.~(\ref{gs-1ph}).
For this reason, in most of the cases keeping only the first term in
Eq.~(\ref{gs-wf1}) and neglecting the second one together with 
interference effects provides very good accuracy in calculation.
Nevertheless, there are a few exceptional cases.

The first one is the excitation of the lowest $1^-$ state in spherical
nuclei.
It is well known that no collective one-phonon $1^-$ configurations
appear in the low energy region and the wave function of the $1^-_1$
state has the dominant two-phonon component $[2^+_1 \times 3^-_1]_{1^-}$.
There are three main mechanisms to explain the $E1$-excitation of this
state observed in the experiment \cite{Kne96}. The first is an influence
of the GDR. In microscopic theories it appears in a natural way due to
the coupling of one- and two-phonon configurations. Since the GDR is
located about 10~MeV higher, this
coupling yields only a very small portion of the observed strength.
The second mechanism is the excitation of non- and weakly-collective
one-phonon $1^-$ configurations which have relatively small $B(E1)$
values but are located in low energy region.  The last mechanism is
the direct excitation of two-phonon configurations from the ground
state. Although the direct excitation of two-phonon
configurations from the ground state is a second order effect,
excitation of collective two-phonon configurations
$[2^+_1 \times 3^-_1]_{1^-}$ play an essential role since the other
two mechanisms yield much weaker $E1$ strengths. In this case,
interference effects between the first and the third mechanisms are
also important \cite{Pon98}.

The second term of Eq.~(\ref{gs-wf1}), although very weak as compared to
the first one, may also play some role at the excitation energies above
20~MeV where the density two-phonon configurations is a few orders of
magnitude higher as compared to the density of one-phonon
configurations. It will be discussed below.

Considering the two-step mechanism of the DGDR excitation in  second
order perturbation theory we also need the reduced matrix element of
the electromagnetic excitation of the two-phonon DGDR state
$[1^-_{i} \times 1^-_{i'}]_J$
from the one-phonon GDR state $1^-_{i}$. In ideal boson approximation
this matrix element
\begin{equation}
< [1^-_{i'} \times 1^-_{i}]_J || {\cal M}(E 1) || 1^-_{i} > =
\sqrt{(1+\delta_{i, i'}) \frac{(2J+1)}{3}}
< 1^-_{i'} || {\cal M}(E 1) || g.s. > .
\label{eq:v3}
\end{equation}

\subsection{Comparison with other approaches}

The properties of the double giant resonances have been also
microscopically studied with the Skyrme forces \cite{Lan97,Lan98} and
within the second-RPA approach \cite{Nis95,Nis98}.

The most close to the QPM approach is the one of the first group of papers.
The main difference between these two approaches is that in calculations
with the Skyrme forces the properties of the ground and $1p1h$ excited
states are calculated self-consistently. As within the QPM, in
calculations with the Skyrme forces the $1p1h$ basis is mapped into the
phonon space. Multi-phonon states are obtained by folding of one-phonon
states. The phonon basis in Refs.~\cite{Lan97,Lan98} is restricted by
only a few, the most collective, phonons for each multipolarity.
Calculations are performed with the wave function including one- and
two-phonon terms. The main attention is paid to the effects of
anharmonicity and non-linearity. The later is an influence of taking
into account of boson forbidden transition matrix elements,
Eqs.~(\ref{gs-2ph},\ref{q-q}), on the absolute value of the DGDR
excitation in heavy ion collisions.

In the second-RPA approach \cite{Dro90}
the wave function of excited states is written as a mixture of 1p$-$1h
and 2p$-$2h configurations:
\begin{eqnarray}
Q^+_{\nu} |\rangle_{g.s.} &=&
\sum_{ph} (X^{\nu}_{ph} a^+_p a_h - Y^{\nu}_{ph} a^+_h a_p)
+\sum_{pp'hh'} (X^{\nu}_{pp'hh'} a^+_p a^+_{p'} a_{h'} a_h -
Y^{\nu}_{pp'hh'} a^+_h a^+_{h'} a_{p'} a_p) |\rangle_{g.s.}.
\nonumber \\
&&
\label{srpa}
\end{eqnarray}
The operators $Q^+_{\nu}$ are assumed as bosons and the energy
spectrum and coefficients $X$ and $Y$ are obtained by diagonalization
of the model Hamiltonian in the space of states described by the wave
functions of Eq.~(\ref{srpa}).

\section{\bf Physical properties of the double giant resonances}

In the present chapter we will consider the properties of the DGDR
as predicted by the QPM mainly in $^{136}$Xe and $^{208}$Pb for which
experimental data in relativistic heavy ion collision (RHIC) are
available.
Before proceeding with that let us briefly check an accuracy of the
description of the properties of low-lying states and single giant
resonances within this approach. It provides an estimate how good the
phonon basis, to be used in the forthcoming calculation of the DGDR 
properties, is
since no extra free parameters are used after this basis is fixed.
The results of our calculations of the position and exhaust of the energy
weighted sum rule (EWSR) of low-lying states and giant resonances as
well as the width of resonances in $^{136}$Xe and $^{208}$Pb are
presented in Table~\ref{vp-t1} in comparison with the experimental 
findings.
The comparison indicate a rather good correspondence between calculated
characteristics and experimental data.
The calculation somewhat underestimate the width of resonances,
especially of the isovector GQR.
The main reason is related to the necessity of truncating of complex
configurations included in the wave function of excited states in actual
calculation. The density of multi-phonon configurations is rapidly
increasing with the excitation energy. That is why the effect of the
basis truncation the most strongly influences on the width of the
GQR$_{iv}$ located at higher energies.

\begin{table}[tb]
\caption[ ]
{Integral characteristics (position, $E_x$, exhaust of the energy
weighted sum rule (EWSR) and width of resonances,
$\Gamma$) of low-lying excited states and one-phonon giant resonances
in $^{136}$Xe and $^{208}$Pb.
\label{vp-t1}}
\begin{center}
\begin{tabular}{ c c c c c c c c } \hline
 & & \multicolumn{3}{ c }{Calculation} &
\multicolumn{3}{c }{Experiment} \\ \cline{3-8}
Nucl. & $\lambda^{\pi}$ & $E_x$  &  $\Gamma$   &
EWSR, &  $E_x$  &  $\Gamma$   &
EWSR,  \\
&      &  [MeV]  & [MeV] & \% &  [MeV]&   [MeV] & \%\\
\hline
& $2^{+}_{1}$  &   1.4   &        &     2.6
            &      1.31   &          &   2.4    \\
& $3^{-}_{1}$  &   3.3   &        &      5.6
            &      3.28   &          &   5.2    \\
$^{136}$Xe & GDR$_{iv}$ & 15.1 & 4.0 & 107  & 15.2$^{a)}$
&4.8$^{a)}$ &   80--120 \\[2mm]
&  GQR$_{is}$  & 12.5 & 3.2 & 75 & 12.3$^{b)}$ &4.0$^{b)}$ &
70$^{b)}$ \\[2mm]
& GQR$_{iv}$  & 23.1 & 3.6& 80  & 22.1$\pm$0.7 &  $\leq$5.4& 93$\pm$45
\\ \hline
& $2^{+}_{1}$  & 4.2    &        & 16.4
            &  4.09      &          &  16.9   \\
& $3^{-}_{1}$  &  2.4   &        & 21.3
            &  2.61      &          &  20     \\
$^{208}$Pb & GDR$_{iv}$ & 13.35  &   3.5  & 94
                & 13.4  &   4.0  & 89--122  \\
& GQR$_{is}$  & 10.6   &   3.1  & 67
             &  10.5--10.9 & 2.4--3.0  &  60--80  \\
& GQR$_{iv}$  & 21.9   &   5.0  & 81
             &  22.6$\pm$0.4 & 6$\pm$2  & $\approx$50 \\
             \hline
\end{tabular}
\end{center}
\noindent
$^{a)}$Interpolation of experimental data \cite{Ber75}. \\
$^{b)}$Interpolation of experimental data \cite{Ber76}. 
\end{table}

\subsection{One-step excitation of two-phonon states in the energy
region of giant resonances}

Let us consider a direct photoexcitation of the two-phonon states
in the energy region of giant resonances from the ground state of
even-even nuclei (see, Refs.~\cite{Pon92,Cat92} for more details).
Since in RHIC experiments the Coulomb mechanism of excitation plays the
most essential role, the cross sections of photoexcitation can be easily
recalculated into RHIC cross sections for different energies and
Z-values of target and projectile nuclei.
In calculation of the $B(E \lambda)$ values we use only the terms
proportional to $\psi\,\varphi$ (see, Eq.~(\ref{gs-2ph})).
The complete set of diagrams corresponding to a direct transition to
two-phonon states from the ground state is presented in
Ref.~\cite{Bes86}.
As one can see from the analytical expressions 
the main part of the contributions from different terms disappears due
to the cancellation between particles and holes.

The cross sections of the direct photoexcitation of the groups of
two-phonon states made of phonons of definite multipolarities in
$^{136}$Xe and $^{208}$Pb are presented in Fig.~\ref{vp-f1}.
$E2$-excitation of $[1^{-} \otimes 1^{-}]_{2^{+}}$ states is plotted in
the top part of the figure.
$E1$-excitation of the two-phonon states
$[1^{-} \otimes 2^{+}]_{1^{-}}$ and
$[2^{+} \otimes 3^{-}]_{1^{-}}$ is shown in the middle and the
bottom parts, respectively.
The integral characteristics of two-phonon states which are a single
giant resonances built on top of either a low-lying state or another
single resonance in the same nuclei are given in Table~\ref{vp-t2}.

\begin{table}[tb]
\caption[ ]
{Integral characteristics (energy centroid, width and cross section of direct
photoexcitation from the ground state) of some groups of two-phonon
states which are a giant resonance built on top of either a low-lying
state or another single giant resonance
in $^{136}$Xe and $^{208}$Pb.
\label{vp-t2}}
\begin{center}
\begin{tabular}{lcccc} \hline
Nucl. & Configuration & Centroid &  Width &
 $\sigma_{\gamma}$\\
         &        &   [MeV]   &  [MeV]    & [mb] \\ \hline
&   $[1^{-}_{\mbox{\tiny GDR}_{iv}} \otimes
2^{+}_{\mbox{\tiny GQR}_{is}}]_{1^{-}}$ &  24.0   &   2.9   & 4.3
\\
$^{136}$Xe&   $[1^{-}_{\mbox{\tiny GDR}_{iv}} \otimes 1^{-}_{
\mbox{\tiny GDR}_{iv}}]_{2^{+}}$ &  30.2   &   4.0   & 0.33 \\
&   $[2^{+}_{\mbox{\tiny GQR}_{is}}  \otimes 2^{+}_{\mbox{\tiny
GQR}_{is}}]_{2^{+}}$ &  21.3   &   0.5   & 0.1 \\
\hline
& $[1^{-}_{\mbox{\tiny GDR}_{iv}} \otimes 2^{+}_{1}]_{1^{-}}$ &
 17.4   &   2.2   & 1.7\\
&  $[1^{-}_{\mbox{\tiny GDR}_{iv}} \otimes 2^{+}_{1}]_{2^{-}}$ &
 17.2   &   2.1   & $8.7 \cdot 10^{-4}$\\
$^{208}$Pb &    $[1^{-}_{\mbox{\tiny GDR}_{iv}} \otimes
2^{+}_{1}]_{3^{-}}$ &  17.7   &   3.4   & $4.9 \cdot 10^{-5}$ \\
&  $[1^{-}_{\mbox{\tiny GDR}_{iv}} \otimes 3^{-}_{1}]_{2^{+}}$ &
 15.3   &   3.9   & $5.2 \cdot 10^{-2}$ \\
&   $[1^{-}_{\mbox{\tiny GDR}_{iv}}  \otimes
2^{+}_{\mbox{\tiny GQR}_{is}}]_{1^{-}}$ &  25.1   &   3.8   & 9.6
\\
&   $[1^{-}_{\mbox{\tiny GDR}_{iv}}  \otimes
1^{-}_{\mbox{\tiny GDR}_{iv}}]_{2^{+}}$ &  25.5   &   4.4   & 0.22
\\ \hline
\end{tabular}
\end{center}
\end{table}

The main feature of the top part of Fig.~\ref{vp-f1} is that just all
two-phonon states which form this double-phonon resonance are
constructed of the one-phonon $1^{-}_i$ states belonging to the GDR in
the one-phonon approximation.
The structure of the $[1^{-} \otimes 2^{+}]_{1^{-}}$ and
$[2^{+} \otimes 3^{-}]_{1^{-}}$ states is more complex.
For example, among $[1^{-} \otimes 2^{+}]_{1^{-}}$ states
the substructure in the energy range from 15 to 20~MeV in $^{208}$Pb
(right middle part of Fig.~\ref{vp-f1}) is
formed mainly by $1^{-}_i$ phonons from the GDR region
coupled to the $2^{+}_{1}$ state. The small substructure above 32~MeV is
due to the GDR $1^{-}_i$ phonons coupled to the $2^{+}_{i'}$ phonons
of the isovector GQR. As for the broad structure between 20 and
30~MeV not only $[\mbox{GDR} \otimes \mbox{GQR}_{is}]_{1^-}$
states but many other two-phonon states built of less collective
$1^{-}_i$ and $2^{+}_{i'}$ phonons, the role of which is marginal for
properties of single resonances, play an essential role.
The same conclusions are valid for the direct photoexcitation of
$[1^{-} \otimes 2^{+}]_{1^{-}}$ states in $^{136}$Xe.
The cross section of the direct photoexcitation of the two-phonon
$1^-$ states built of phonons of the higher multipolarities yield
non-resonance feature.
It is already seen for the case of $[2^{+} \otimes 3^{-}]_{1^{-}}$
states (bottom part of Fig.~\ref{vp-f1}), especially in $^{208}$Pb.

While dealing with electromagnetic, or with Coulomb excitation from
a $0^+$ ground
state, the priority attention has to be paid to the final states with the
total angular momentum and parity $J^{\pi} = 1^-$.
For that we have calculated the cross section
for the photoexcitation of two-phonon states $[\lambda_1^{\pi_1} \otimes
\lambda_2^{\pi_2}]_{1^-}$, where $\lambda_1^{\pi_1}$ and $\lambda_2^{\pi_2}$
are both natural $\lambda^{\pi^n}$ ($\pi^n = (-1)^{\lambda}$) and unnatural
$\lambda^{\pi^u}$ ($\pi^u = (-1)^{\lambda + 1}$) parity phonons with
multipolarity $\lambda$ from 0 to 9.

\begin{table}[tb]
\caption[ ]
{Cross sections for the direct photoexcitation of different
two-phonon configurations from the ground state integrated over the energy
interval from 20 to 35~MeV in $^{136}$Xe and $^{208}$Pb. The GDR cross
section integrated over the energy of its location is presented in the last
line for a comparison.
\label{vp-t3}}
\begin{center}
\begin{tabular}{rcc}   \hline
& \multicolumn{2}{c}{$\sigma_{\gamma}$ [mb]} \\ \cline{2-3}
Configuration & $^{136}$Xe & $^{208}$Pb \\ \hline
$[0^+ \otimes 1^-]_{1^-}$ & 4.4 & 3.9 \\
$[1^- \otimes 2^+]_{1^-}$ & 36.6 & 44.8 \\
$[2^+ \otimes 3^-]_{1^-}$ & 82.8 & 33.1 \\
$[3^- \otimes 4^+]_{1^-}$ & 101.0 & 56.7 \\
$[4^+ \otimes 5^-]_{1^-}$ & 68.9 & 37.3 \\
$[5^- \otimes 6^+]_{1^-}$ & 49.2 & 46.2 \\
$[6^+ \otimes 7^-]_{1^-}$ & 31.9 & 49.8 \\
$[7^- \otimes 8^+]_{1^-}$ & 13.6 & 12.5 \\
$[8^+ \otimes 9^-]_{1^-}$ & 4.9 & 9.0 \\ \hline
$
\begin{array}{c}
^9 \\[-3.5mm]
\sum \\[-3.0mm]
_{ \lambda_1,\lambda_2 = 1}
\end{array}
\hspace*{-2mm} [\lambda_1^{\pi_1^n} \otimes \lambda_2^{\pi_2^u}]_{1^-}$ &
71.4 & 58.5 \\
$
\begin{array}{c}
^9 \\[-3.5mm]
\sum \\[-3.0mm]
_{ \lambda_1,\lambda_2 = 1}
\end{array}
\hspace*{-2mm} [\lambda_1^{\pi_1^u} \otimes \lambda_2^{\pi_2^u}]_{1^-}$ &
46.7 & 71.1 \\ \hline
$
\begin{array}{c}
^9 \\[-3.5mm]
\sum \\[-3.0mm]
_{ \lambda_1,\lambda_2 = 0}
\end{array}
\hspace*{-2mm} [\lambda_1^{\pi_1^{n,u}} \otimes
\lambda_2^{\pi_2^{n,u}}]_{1^-}$ & 511.4 & 422.9 \\ \hline
$[\mbox{GDR} \otimes \mbox{GDR}]_{2^+}$ & 0.33 & 0.22 \\
$
\begin{array}{c}
^9 \\[-3.5mm]
\sum \\[-3.0mm]
_{ \lambda_1,\lambda_2 = 1}
\end{array}
\hspace*{-2mm} [\lambda_1^{\pi_1^n} \otimes \lambda_2^{\pi_2^n}]_{2^+}$ &
38.1 & 21.7 \\ \hline
GDR\hspace*{10mm} & 2006 & 2790  \\ \hline
\end{tabular}
\end{center}
\end{table}

The results of the calculation for $^{136}$Xe and $^{208}$Pb integrated over
the energy interval from 20 to 35~MeV are presented in
Table~\ref{vp-t3}.
Each configuration $[\lambda_1^{\pi_1} \otimes \lambda_2^{\pi_2}]$ in the
table means a sum over a plenty of two-phonon states made of phonons with a
given spin and parity $\lambda_1^{\pi_1}$,
$\lambda_2^{\pi_2}$, but different RPA root numbers $i_1$,
$i_2$ of its constituents
\begin{equation}
\sigma([\lambda_1^{\pi_1} \otimes \lambda_2^{\pi_2}]) = \sum_{i_1,i_2}
\sigma([\lambda_1^{\pi_1}(i_1) \otimes \lambda_2^{\pi_2}(i_2)])~.
\end{equation}
The total number of two-phonon $1^-$ states included in this calculation for
each nucleus is about $10^5$ and they exhaust 25\% and 15\%
of the EWSR in $^{136}$Xe and $^{208}$Pb, respectively.
The absolute value of the photoexcitation of
any two-phonon state under consideration is negligibly small but altogether
they produce a sizable cross section. Table~\ref{vp-t3} demonstrates
that
different two-phonon configurations give comparable contributions to the
total cross section which decreases only for very high spins because of the
lower densities of such states. As a rule, unnatural parity phonons play
a less important role than natural parity ones. For these reasons we presented
in the table only the sums for [natural$\otimes$unnatural] and
[unnatural$ \otimes$unnatural] two-phonon configurations.

The cross section for the photoexcitation of all two-phonon $1^-$ states in
the energy region 20-35~MeV from the ground state equals in our calculation
to 511~mb and 423~mb for $^{136}$Xe and $^{208}$Pb, respectively. It is not
surprising that we got a larger value for $^{136}$Xe than for $^{208}$Pb.
This is because the phonon states in Xe are composed of a larger number of
two-quasiparticle configurations due to the pairing. The same values for
two-phonon states with angular momentum and parity $J^{\pi} = 2^+$ are an
order of magnitude smaller. We point out that the direct excitation of
$[1^-\otimes 1^-]_{2^+}$ or $[\mbox{GDR} \otimes \mbox{GDR}]_{2+}$
configurations is negligibly weak (compare results in Tables~\ref{vp-t2}
and \ref{vp-t3}). The calculated values should be compared to the cross
section for the photoexcitation of the single-phonon GDR which in our
calculation equals to 2006~mb and 2790~mb, respectively. A contribution of
two-phonon $1^-$ states to the total cross section at GDR energies is weaker
than at higher energies because of the lower density of two-phonon states
and the lower excitation energy and can be neglected considering the GDR
itself. It is clearly demonstrated in Fig.~\ref{vp-f2}b.
In this figure the cross sections of the photoexcitation of $1^-$ states
in $^{136}$Xe and $^{208}$Pb are presented. The top part of the figure
corresponds to a calculation performed in one-phonon approximation. The
results of calculations with the wave function which includes a coupling
between one- and two-phonon $1^-$ configurations are plotted in the
bottom part of the figure. For a visuality the last calculations are
also presented as strength functions
\begin{equation}
b(\sigma,E) =\frac{1}{2 \pi}  \sum_{\nu} \sigma^J_{\nu} \cdot
\frac{\Delta}{(E - E^J_{\nu})^2 + \Delta^2 / 4}
\label{sf}
\end{equation}
with a smearing parameter $\Delta = 1$~MeV, where $\sigma^J_{\nu}$ is a
partial cross section for the state with the excitation energy $E^J_{\nu}$
plotted also by a vertical line.
The $E1$-transitions to one-phonon components of the wave function of
excited $1^-$ states are plotted by dashed curve. It should be compared
with the solid curve which is the sum of transitions to one- and
two-phonon $1^-$ configurations in the GDR energy region.

For $^{208}$Pb photoexcitation cross sections are known from experimental
studies in $(\gamma, n)$ reactions up to the excitation energy about 25~MeV
\cite{Bel92,Bel95}. It was shown that QPM provides a very good
description
of the experimental data in the GDR region \cite{Bel92} (see,
Fig.~\ref{vp-f3}), while theoretical
calculations at higher excitation energies which account for contributions
 from the single-phonon GDR and GQR$_{iv}$ essentially underestimated the
experimental cross section \cite{Bel95}. The experimental cross sections
above 17~MeV are shown in Fig.~\ref{vp-f4} together with theoretical
predictions. The results of the calculations are presented as strength
functions obtained with averaging parameter equal to 1~MeV. The contribution
to the total cross section of the GQR$_{iv}$ (short-dashed curve), the high
energy tail of GDR (long-dashed curve), and their sum (squared curve), are
taken from Ref.~\cite{Bel95}. The curve with triangles represents the
contribution of the direct excitation of the two-phonon states
 from our present studies. The two-phonon states form practically a flat
background in the whole energy region under consideration. Summing together
the photoexcitation cross sections of all one- and two-phonon states we get
a solid curve which is in a very good agreement with the experimental data.

Thus, from our investigation of photoexcitation cross sections we
conclude that in this reaction very many different two-phonon states
above the GDR contribute
on a comparable level, forming altogether a flat physical background which
should be taken into account in the description of experimental data.
On the
other hand, Coulomb excitation in relativistic heavy ion collisions provides
a unique opportunity to excite a very selected number of two-phonon states
by the absorption of two virtual  $\gamma$'s in a single process of
projectile-target interaction \cite{BB88}. Theoretically this process
is
described using a the second order perturbation theory of the semi-classical
approach of A. Winther and K. Alder \cite{BB88,WA79} and discussed in
Sect. 3.2.3. Since excitation cross
sections to second order are much weaker than to first order of the theory,
two-phonon states connected to the ground states by two $E1$-transitions
are predominantly excited. These two-phonon states have the structure
$[1^-(i) \otimes 1^-(i^{\prime})]_{J^+}$ and form the DGDR.

\subsection{$1^+$ component of the DGDR}

According to the rules of angular momentum coupling two one-phonon
states with the spin and parity equal to $1^-$ may couple to the total
angular momentum $J^{\pi} = 0^+, 1^+$ and $2^+$. Thus, in principle,
three components of the DGDR with these quantum numbers should exist.
In phenomenological approaches describing the single GDR as one collective
state, the $[1^- \otimes 1^-]_{1^+}$ component the DGDR is forbidden by
symmetry properties.
Taking into account the Landau damping this collective state splits
into a set of different $1^-_i$ states distributed over an energy
interval.
In microscopic studies the Landau damping is taken into account by
solving the RPA equations.
Again, the diagonal components $[1^-_i \otimes 1^-_i]_{1^+}$ are forbidden
by the same symmetry properties but nondiagonal ones
$[1^-_i \otimes 1^-_{i'}]_{1^+}$ exist and should be taken into
consideration.
Consequently, the role of these nondiagonal components depends on
how strong is the Landau damping.

We produce here two-phonon DGDR states with quantum numbers
$J^{\pi} =$~$0^+$, $1^+$ and $2^+$ by coupling one-phonon RPA states
with the wave function $|1^-_i>_{m}$, to each other.
The index $m$ stands for different magnetic substates.
The wave function of the two-phonon states has the form:
\begin{equation}
| [1^-_i \otimes 1^-_{i}]_{J^{\pi}= 0^+, 2^+}>_{M} =
\frac{1}{\sqrt{2}} \sum_{m,m'} (1m 1 m'|JM)
|1^-_i>_{m} |1^-_i>_{m'} \ ,
\label{1p1}
\end{equation}
for two-phonon states made of two identical phonons while for other
DGDR states it is:
\begin{equation}
| [1^-_i \otimes 1^-_{i'}]_{J^{\pi}= 0^+, 1^+, 2^+}>_{M} =
\sum_{m,m'} (1m 1 m'|JM)
|1^-_i>_{m} |1^-_{i'}>_{m'} \ .
\label{1p2}
\end{equation}

In the present calculation we do not include the interaction between DGDR
states, of Eqs. (\ref{1p1},\ref{1p2}), and we do not couple them
to
states with different than two number of phonons (it will be considered
below).
Thus, our two-phonon states $| [1^-_i \otimes 1^-_{i'}]_{J^{\pi}}>_{M}$
have excitation energy equal to the sum of one-phonon energies
$\omega_i + \omega_{i'}$
and are degenerated for different values of the total spin $J^{\pi}$
and its projection $M$.

Since the main mechanism of excitation in projectile
ions at relativistic energies is the Coulomb part of interaction with a
target, the nuclear part of interaction has been neglected in the
present analysis.
In a semi-classical approach \cite{BB88} the two-phonon DGDR states can
be excited in second-order perturbation theory via the two-step
process g.s. $\rightarrow$ GDR $\rightarrow$ DGDR.
The second order amplitude can be written as
\begin{equation}
a^{(2)}_{[1^-_i \otimes 1^-_{i'}],M} =
\frac{1}{2} \sum_m
\
a^{(1)}_{1^-_i,m  \ \rightarrow  \ [1^-_i \otimes 1^-_{i'}],M}
\
a^{(1)}_{g.s. \ \rightarrow  \ 1^-_i,m}
\label{eq:n}
\end{equation}
where assuming the Coulomb mechanism of excitation the first order
amplitude
$a^{(1)}_{J_i \ \rightarrow J_f} $ is proportional to the reduced matrix
element of  $< J_f || E1 || J_i >$.
The reduced matrix element
$< [1^-_{i'} \otimes 1^-_{i}]_{J^{\pi}} || E1 || 1^-_{i} >$
of electromagnetic excitation of two-phonon
states, Eqs. (\ref{1p1},\ref{1p2}), from the one-phonon state
$| 1^-_{i} >_{m}$ is related,
in the boson picture of nuclear excitation, to the
excitation of $| 1^-_{i} >_{m}$ from the ground state according to
(\ref{eq:v3}).
It should be noted that although for the two-phonon states, Eq.
(\ref{1p1}),
we have an extra factor $\sqrt{2}$, the states of Eq. (\ref{1p2}) play a
more important role in two-step excitations since
they can be reached by two different possibilities:
$g.s. \rightarrow 1^-_{i} \rightarrow [ 1^-_{i} \otimes 1^-_{i'}]$ and
$g.s. \rightarrow 1^-_{i'} \rightarrow [ 1^-_{i} \otimes 1^-_{i'}]$.

First of all, we point out that in second-order
perturbation theory the amplitude for this process is identically zero
in a semi-classical approach.
This can be understood by looking at Fig.~\ref{vp-f5}.
The time-dependent
field $V_{E1}$ carries angular momentum with projections $m=0,\ \pm 1$.
Thus, to reach the $1^+$ DGDR magnetic substates, many routes are possible.
The lines represent transitions caused by the different projections of
$V_{E1}$: (a) dashed lines are for $m=0$, (b) dashed-dotted lines are for
$m=-1$, and (c) solid-lines are for $m=+1$.
The relation $V_{E1,m=0}\ne V_{E1,m=\pm 1}$ holds, so that not all routes
yield the same excitation amplitude.
Since the phases of the
wave functions of each set of magnetic substates are equal, the difference
between the transition amplitudes to a final $M$,
can also arise from different values
of the Clebsch-Gordan coefficients $(1m 1 m'|1M)$.
It is easy to see that, for any route to a final $M$, the
second-order amplitude will be proportional to $(001m|1m)
(1m 1m'|1M)\ V_{E1,m'} \ V_{E1,m} + (m \leftrightarrow m'$).
The two amplitudes carry opposite signs from the value of the
Clebsch-Gordan coefficients.
Since $(001m|1m) \equiv 1$, the identically
zero result for the excitation amplitude of the
$1^+$ DGDR state is therefore a consequence of
\begin{equation}
\sum_{m m'} (1m 1 m'|1M)=0~~.
\end{equation}

We have also performed a coupled-channels calculation \cite{Ber96b}
following the theory described in Ref. \cite{BCH96}.
As shown in Ref. \cite{BCH96},  the coupling of the electric
quadrupole (isovector and isoscalar) and the electric dipole states is very
weak and can be neglected. We therefore include
in our space only one-phonon $1^-$ and two-phonon
$[1^-_i \otimes 1^-_{i'}]_{J^{\pi}}$
($J^{\pi} =$ $0^+$, $1^+$ and $2^+$) states.
In coupled-channels calculation we take into account interference
effects in the excitation of different GDR and DGDR states and
obtain the occupation amplitudes by solving the
coupled-channels equations.
By solving these equations we thus account for
unitarity and for multi-step excitations, beyond
the two-step processes of Eq. (\ref{eq:n}).
The time dependent electric dipole field is
that of a straight-line moving particle with charge $Ze$, and impact
parameter $b$ (we use Eqs. (25-26) of Ref. \cite{BCH96}).

Due to the large number of degenerate magnetic
substates, to make our coupled-channels calculation feasible,
we have chosen a limited set of GDR and DGDR states.
We have taken six $1^-$ states which have the largest value of
the reduced matrix element $<1^-_i || E1 || g.s.>$.
These six states exhaust 90.6\% of the classical
EWSR, while all $1^-$ states up to 25~MeV
in our RPA calculation exhaust 94.3\% of it.
This value is somewhat smaller than the 122\% reported in
Ref.~\cite{Ve70}.
It is because the continuum in our RPA calculation was approximated
by narrow quasibound states. From these six one-phonon $1^-$ states we
construct two-phonon
$[1^-_i \otimes 1^-_{i'}]_{J^{\pi}}$ states, Eqs.
(\ref{1p1},\ref{1p2}),
which also have the largest matrix element of excitation
$<[1^-_i \otimes 1^-_{i'}]_{J^{\pi}} || E1 || 1^-_i>$
for excitations starting from one-phonon states
\footnote{As demonstrated in the previous
subsection the direct excitation of
two-phonon configurations from the ground state is very weak.
It allows us to exclude in our calculation matrix elements of the form
$<[1^-_i \otimes 1^-_{i'}]_{2^+(1^+)} || E2(M1) || g.s.>$ which
correspond to direct transitions and produce higher order effects in
comparison with accounted ones. These matrix elements give rise
to DGDR excitation in first order perturbation theory.
Thus, to prove our approximation we have calculated such cross sections
and got total values equal to 0.11~mb and $<$0.01~mb for the twenty one
$2^+$ and the fifteen $1^+$ basic two-phonon states, respectively.
These values have to be compared to 244.9~mb for the total DGDR cross
section in the second order perturbation theory.}.
The number of two-phonon states equals to twenty one for $J^{\pi} = 0^+$
and $2^+$, and to fifteen for $J^{\pi} = 1^+$.
The cross section for the DGDR excitation was obtained by summing over
the final magnetic substates of the square of the
occupation amplitudes and,
finally, by an integration over impact parameter.
We have chosen the minimum impact parameter, $b=15.54$~fm, corresponding
to the parameterization of Ref. \cite{BCV89}, appropriate for lead-lead
collisions.

The electromagnetic excitation cross sections for the reaction
$^{208}$Pb (640A MeV) $+^{208}$Pb with excitation of all our
basic 63 states is shown in Fig. \ref{vp-f6}.
The total cross sections for each multipolarity are presented
in Table \ref{vp-t4}, together with
the results of first-order (for one-phonon excitations)
and second-order (for two-phonon excitations) perturbation theory.
The coupled-channels calculation yields a non-zero cross section for the
$1^+$ DGDR state due to other possible routes (higher-order), not included
in second-order perturbation theory.
One observes a considerable reduction of the DGDR cross sections, as
compared to the predictions of the second-order perturbation theory.
The GDR cross sections are also reduced in magnitude.
However, the population of the $1^+$ DGDR states are not appreciable and
cannot be the source of the missing excitation
cross section needed to explain the experiments.
In general, the coupled-channels calculation practically does not change
the relative contribution
of different one-phonon $1^-_i$ and two-phonon states
$[1^-_i \otimes 1^-_{i'}]_{J^{\pi}}$
to the total cross section with given $J^{\pi} =$~$1^-$, $0^+$ and $2^+$.
But since the $1^+$ component of the DGDR, with its zero value of excitation
cross section in  second-order perturbation theory, has a special place
among the two other components, the main effect of coupled-channels
is to redistribute the total cross section between the $J^{\pi} =$~$0^+$,
$2^+$ and $J^{\pi} =$~$1^+$ components.

\begin{table}[tb]
\caption[ ]
{Cross section (in mb) for the excitation of the
GDR and the three components
with $J^{\pi} = 0^+, 2^+, 1^+$ of the DGDR in $^{208}$Pb (640A MeV)
$+^{208}$Pb collisions. Calculations are performed within
coupled-channels (CC) and within  first (PT-1) and second (PT-2)
order perturbation theory, respectively.
\label{vp-t4}}
\begin{center}
\begin{tabular}{ l c c c }
         \hline
                       &     CC    &     PT-1  &    PT-2\\
         \hline
          GDR  &            2830.  &    3275.  &    0.  \\
         \hline
         DGDR$_{0^+}$  &    33.0   &    0.     &    43.1 \\
         DGDR$_{2^+}$  &   163.0   &    0.11   &    201.8 \\
         DGDR$_{1^+}$  &     6.3   &    $<$0.01   &    0. \\
         \hline
          DGDR/GDR     &   0.071   &
\multicolumn{2}{c } {0.075 \hspace*{30mm}} \\
         \hline
\end{tabular}
\end{center}
\end{table}

The calculated cross section in coupled-channels for both GDR and DGDR
are somewhat smaller than reported in experimental findings
\cite{Bo96,Str96}.
This is not surprising since as mentioned above our chosen six $1^-$
states exhaust only 90.6\% of EWSR while the photo-neutron data \cite{Ve70}
indicate that this value equals to 122\%. Due to this underestimate
of exhaust of the EWSR the cross section for the DGDR excitation
reduces more strongly than the one for the single GDR. This is because
the GDR cross section is roughly proportional to the total $B(E1)$ value
while for the DGDR it is proportional to the square of it.
We will return back in more details to the problem of absolute cross
sections of the DGDR excitation in RHIC in the forthcoming subsection.

\subsection{Position, width and cross section of excitation in RHIC of
the DGDR in $^{136}$Xe and $^{208}$Pb}

To describe the width of two-phonon resonances it is necessary to take
into account a coupling of two-phonon configurations, which form these
resonances, with more complex ones. For that two types of calculations
have been performed.
In the first of them \cite{Pon94} the fine structure of the GDR
calculated with
the wave function which includes one- and two-phonon configurations and
presented in Fig.~\ref{vp-f2}b has been used.
The DGDR states have been constructed as a product of the GDR to
itself. In other words, following the Axel-Brink hypotheses on top of
each $1^-$ state in Fig.~\ref{vp-f2}b we have built the full set of the
same $1^-$ states.

The calculation has been performed for the nucleus $^{136}$Xe.
In the dipole case, $\lambda^{\pi} =1^-$, the one-phonon states
exhaust 107\% of the classical oscillator
strength and are displayed in the left part of Fig.~\ref{vp-f2}a.
Of these, 20 states have an oscillator strength which is at least 1\% of
the strongest
strength and together exhaust 104\% of the classical EWSR. We have used these
states in the coupling to two-phonon states.
We have included all the natural parity phonons
$\lambda^{\pi} = 1^- - 8^+$ with energy lower or equal
to 21 MeV, obtaining 2632 two--phonon configurations.
One obtains 1614 states described by the wave function which includes
one- and two-phonon configurations, in the energy
interval from 7 MeV to 19.5 MeV. Their photoexcitation cross
sections are shown in Fig.~\ref{vp-f2}b.
The $B(E1)$ value associated with each mixed state is
calculated through its admixture with one-phonon states,
as $|<\nu|| {\cal M}(E1)||0>|^2 =| \sum_{i} S^{\nu}_{i}(1^-)
<0||Q_{1^-i} {\cal M}(E1) || 0>|^2$.
Also shown by dashed curve in the left part of Fig.~\ref{vp-f2}b is the
result obtained adding an averaging parameter of 1.0~MeV.
This parameter represents in some average way
the coupling to increasingly more complicated states and eventually
to the compound
nuclear states. From the resulting smooth response it is easy to directly
extract the centroid and the full width at half maximum of the GDR.
The corresponding values are $E_{\mbox{\tiny GDR}}=$ 15.1 MeV and
$\Gamma_{\mbox{\tiny GDR}}=$ 4 MeV.
They can be compared with the values extracted from experiment,
$E_{\mbox{\tiny GDR}}$= 15.2 MeV and $\Gamma_{\mbox{\tiny GDR}}$ = 4.8
MeV.

The isoscalar and the isovector
giant quadrupole resonances (GQR) have also been calculated.
The centroid, width and percentage
of the EWSR associated with the isoscalar mode are 12.5 MeV, 3.2 MeV
and 75\% respectively. The corresponding quantities associated with the
isovector GQR are 23.1 MeV, 3.6 MeV and 80\%.

The differential Coulomb-excitation cross sections as a function of the
energy associated with the one-phonon GDR and GQR resonances and the
two-phonon DGDR in $^{136}$Xe (690A MeV) + $^{208}$Pb reaction
are displayed in Fig.~\ref{vp-f7}.
It is seen that the centroid of the two-phonon dipole excitation falls
at 30.6 MeV, about twice that of the one-phonon states, while the width is
$\Gamma \approx $ 6 MeV, the ratio to that of the one-phonon excitation
being 1.5.

The associated integrated values
are displayed in Table~\ref{vp-t5}, in comparison with the experimental
findings. The cross sections depend strongly on the choice of the value
of $b_{min} = r_o (A_p^{1/3} + A_t^{1/3})$. In keeping with the standard
``safe distance", that is,
the distance beyond which nuclear excitation can be safely neglected,
we have used $r_o $ = 1.5 fm. Because their values essentially do
not depend on the width of the GDR, we view the
calculated cross section of 1650 mb as a rather accurate value and if anything
an upper limit for the one-phonon Coulomb excitation cross section.
It is satisfactory that the measured cross section is rather
close to this value. Also shown in Table~\ref{vp-t5} are the predictions
associated with the sequential excitation of the DGDR.
This result is essentially not modified evaluating the direct Coulomb
excitation of the double GDR. In fact, the cross section associated with this
process is a factor $10^{-3}$ smaller than that associated with the two-step
process.
The calculated value of 50 mb is a factor of 0.25 smaller than experimentally
observed.

\begin{table}[tb]
\caption[ ]
{Calculated (for two values of $r_o$) and experimental cross section (in
mb) for the excitation of giant resonances in $^{136}$Xe in $^{136}$Xe
(690A MeV) + $^{208}$Pb reaction.
In the last row, the experimental cross sections for Coulomb excitation
of one- and two-phonon states from Ref.~\protect\cite{Sc93} are shown.
The value of the
integrated cross section reported in Ref.~\protect\cite{Sc93} is $1.85
\pm 0.1$ b.
The nuclear contribution has been estimated in Ref.~\protect\cite{Sc93}
to be about 100
mb, while about 3\%  (50 mb) of the cross section is found at higher energy.
Subtracting these two contributions and the two-phonon  cross section,
leads to the value 1485 $\pm 100$ mb shown in
the Table.
\label{vp-t5}}
\begin{center}
\begin{tabular}{ c c c c c c } \hline
& GDR & GQR$_{is}$  & GQR$_{iv}$ & GDR + GQR & DGDR \\
 \hline
$r_o = 1,2$ fm
& 2180          &  170 &   120 &  2470         & 130  \\
$r_o = 1,5$ fm
& 1480          &  110 &    60 &  1650         &  50  \\
Experiment
& 1024$\pm$100  &  $-$ &   $-$ & 1485$\pm$100  & 215 $\pm$ 50 \\
 \hline
\end{tabular}
\end{center}
\end{table}

Two other processes are possible within the sequential excitation of
the giant modes which can lead to an excitation energy similar to that
of the two--phonon GDR. They are the excitation of the isoscalar GQR mode
followed by a GDR mode and vice versa. The resulting cross section is
estimated to be an order of magnitude smaller, cfr. Table~\ref{vp-t5},
and does not change qualitatively this result.
In order to make clearer the seriousness of this discrepancy,
we have recalculated all the cross sections using $r_o = $ 1.2~fm,
namely with a much smaller radius
than that prescribed in order to respect the safe Coulomb excitation
distance of closest approach. The calculated value of 130 mb is still a
factor of 0.6 smaller than the reported experimental cross section.
At the same time the cross section of the one-phonon states has become a
factor
1.7 larger than the empirical value. This factor becomes 1.5 when
the coupling to higher multiphonon states is included according to the
standard Poisson distribution for the excitation probabilities
\cite{AW75}.

The main shortcoming of the above discussed theoretical scheme to
treat the DGDR, when the DGDR states are obtained by folding of the fine
structure of two GDR's, is the fact that the DGDR states obtained this
way are not eigenstates of the used microscopic Hamiltonian.
To overcome this shortcoming another calculations have been performed
in which two-phonon $[1^- \otimes 1^-]$ DGDR states are coupled directly
to more complex ones \cite{Pon96a,Pon96b,Pon97}. From rather general
arguments \cite{Ber83}, the most important
couplings leading to real transitions of the double giant resonances and
thus to a damping width of these modes are to
configurations built out by promoting three nucleon across the Fermi
surface.
That is, configurations containing three holes in the Fermi sea and
three particles above the Fermi surface ($3p3h$ configurations).
We use the wave function (\ref{wf1}) to describe the DGDR states and
their coupling to $1p1h$ and to $3p3h$ doorway configurations.

The spectrum of excited states which form the DGDR is obtained by
solving the secular equation (\ref{det}) and the wave function
coefficient S, D and T are calculated from Eq.~(\ref{st}).
Pauli principle corrections,
the coefficients $\tilde{K}_J(\beta_2 \alpha_2 | \alpha_2' \beta_2')$
and anharmonicity shifts $\Delta \omega_{\alpha_2 \beta_2}^J$,
were omitted in calculations presented in Ref.~\cite{Pon96a} and
accounted for in Ref.~\cite{Pon96b}.
While they are small, they produce shifts
in the energy centroid of the double giant resonance.
Similar coefficients appear also in connection with the term arising from
the ``doorway states" containing three phonons in Eq.~(\ref{wf1}).
We have neglected them because they again are small and furthermore act
only in higher order as compared to the previous term,
in defining the properties of the double giant dipole resonance.
Finally, the corresponding $\tilde{K}$-coefficient
associated with the first term in Eq.~(\ref{wf1}) is proportional to
the number of quasiparticles present in the ground state of the system,
a quantity which is assumed to be zero within linear response theory.

In keeping with the fact that the Q-value dependence of the Coulomb excitation
amplitude is rather weak at relativistic energies \cite{Bee93},
the cross section associated with the two-step excitation of the double giant
dipole resonance is proportional to
\begin{eqnarray}
&&[B(E1) \times B(E1)]=| \sum_{\nu_1}
 \langle \Psi^{\nu}_{0^+(2^+)}
| {\cal M}(E1) | \Psi^{\nu_1}_{1^-} \rangle  \cdot
 \langle \Psi^{\nu_1}_{1^-} | {\cal M}(E1) | \Psi_{g.s.} \rangle |^2
\label{trans} \\
&=&  \left|~2 \cdot \sum_{\alpha_2 \beta_2}
D^{\nu}_{\alpha_2 \beta_2} (J) \cdot \left[
\frac{M_{\alpha_2} M_{\beta_2}}{\sqrt{1+\delta_{\alpha_2, \beta_2}}}
+ \sum_{\alpha_2' \beta_2'}
 M_{\alpha_2'} M_{\beta_2'} \sqrt{1+\delta_{\alpha_2', \beta_2'}}
\tilde{K}_J(\beta_2 \alpha_2 | \alpha_2' \beta_2') \right] \right|^2~,
\nonumber
\end{eqnarray}

\vspace*{5mm}
\noindent
where $M_{\alpha} = \langle
Q_{\alpha} ||{\cal M}(E1) || 0^+_{g.s.}\rangle$ is
the reduced matrix element of the $E1$-operator which acting on the
ground state $\mid \rangle_{ph}$ excites the one-phonon state with
quantum numbers $\alpha = (1^-, i)$.

Making use of the elements discussed above
we calculated the distribution
of the quantity Eq.~(\ref{trans}) over the states Eq.~(\ref{wf1}) in
$^{136}$Xe.
We considered only $J^{\pi}$~=$0^+$ and $2^+$ components of the two-phonon
giant dipole resonance.
As already discussed above its $J^{\pi}$~=$1^+$ component
cannot be excited in the second order perturbation theory and is
sufficiently quenched in coupled-channels calculation.
The fifteen configurations $\{ 1^- i, 1^- i'\}$ =
$\{ \alpha_2, \beta_2 \}$ displaying the largest $[B(E1) \times B(E1)]$
values were used in the calculation.
They are built up out of the five most collective RPA roots associated
with the one-phonon giant dipole resonance carrying the largest $B(E1)$
values and exhausting 77$\%$ of energy weighted sum rule (EWSR).
Two-phonon states of collective character and with quantum
numbers different from $1^-$ lie, as a rule, at energies few MeV away from
the double giant dipole states and were not included in the
calculations. The three-phonon states
$\{{\alpha_3 \beta_3 \gamma_3}\}$ were built out of phonons with angular
momentum and parity 1$^-$, 2$^+$, 3$^-$ and 4$^+$.
Only those configurations where either
$\alpha_3$, $\beta_3$ or $\gamma_3$ were equal to $\alpha_2$ or $\beta_2$
were chosen.
This is because other configurations lead to matrix elements
$U_{\alpha_3 \beta_3 \gamma_3}^{\alpha_2 \beta_2} (J)$
of the interaction,
which are orders of magnitude smaller than those associated with the above
mentioned three-phonon configurations, and which contain in the present
calculation 5742 states up to an excitation energy 38~MeV.
The single particle continuum has been approximated in the present
calculation by quasibound states.
This approximation
provides rather good description of the single GDR properties in $^{136}$Xe.
This means that our ($2p2h)_{[1^- \times 1^-]}$ spectrum is also
rather
complete for the description of the DGDR properties although it is
located at higher energies.

If one assumes a pure boson picture to describe the phonons, without
taking into account their fermion structure, the three-phonon configurations
omitted in the present calculation do not couple to two-phonon states
under consideration.
Furthermore, although the density of $3p3h$ configurations is quite
high
in the energy region corresponding to the DGDR, a selection of the important
doorway configurations in terms of the efficiency with which configurations
couple to the DGDR, can be done rather easily.
The above considerations testify to the advantage of employing a microscopic
phonon picture in describing the nuclear excitation spectrum, instead of
a particle-hole representation.
One can more readily identify the regularities typical of the collective
picture of the vibrational spectrum, and still deal with the fermion
structure of these excitations.
As far as the one-phonon term appearing in Eq.~(\ref{wf1}) is concerned,
essentially all phonons with angular momentum and parity $0^+$ and $2^+$
were taken into account within the energy interval 20--40~MeV.

A rather general feature displayed by the results of the present calculation
is that all two-phonon configurations of the type $\{ 1^- i, 1^- i'\}$
building the DGDR in the ``harmonic" picture are fragmented over a few
MeV due to the coupling to $3p3h$ ``doorway states".
Fragmentation of the most collective one is presented in the bottom part
of Fig.~\ref{fragm}. For a comparison the fragmentation of the most
collective one-phonon $1^-$ configuration due to the coupling to $2p2h$
``doorway states" is plotted in the top part of the same figure.
The results have been averaged with the aid of a Breit-Wigner
distribution of width 0.2~MeV.
The maximum amplitude with which each two-phonon configuration enters in
the wave function (\ref{wf1}) does not exceed a few percent.
Two-phonon configurations made out of two different $1^-$ phonons are
fragmented stronger then two-phonon configurations made out of two identical
$1^-$ phonons.
This in keeping with the fact that, as a rule, states of the type
$\{ 1^- i, 1^- i'\}$ with $i \ne i'$ are less harmonic than states with
$i=i'$ and consequently are coupled to a larger number of three-phonon
configurations.

In Figs.~\ref{vp-f8}b-c, the $[B(E1) \times B(E1)]$ quantity of Eq.
(\ref{trans}) associated with Coulomb excitation of the almost
degenerate $J^{\pi} = 0^+$ and $J^{\pi} = 2^+$
components of double giant dipole resonance are shown.
For comparison, the $B(E1)$ quantity
associated with the Coulomb
excitation of the one-phonon giant dipole resonance is also shown in
Fig.~\ref{vp-f8}a.
The reason why the two angular momentum components of the DGDR are
almost degenerate can be traced back to the fact that the density of
one-phonon configurations to which the DGDR couple and which are
different for $J^{\pi} = 0^+$ and $J^{\pi} = 2^+$ type states is much
lower than the
density of states associated with $3p3h$ ``doorway states", density of
states which is the same in the present calculation
for the two different angular momentum and parity.
Effects associated with the $J$-dependence of the $\tilde{K}_J$ and
$\Delta_J$ coefficients are not able to remove the mentioned degeneracy,
because of the small size of these coefficients.
These coefficients can also affect the excitation probability
with which the $J^{\pi} = 0^+$ and $J^{\pi} = 2^+$ states are excited
(cf. Eq.~(\ref{trans})).
The effect however is rather small, leading to a decrease of the order of
2-3\% in both cases.
The $J$-degeneracy would be probably somehow broken if one goes beyond
a one-boson exchange picture in the present approach of interaction
between different nuclear modes.
The next order term of interaction would couple the
DGDR states to many other $3p3h$ configurations,
not included in the present studies, some of these $3p3h$
configurations would be different for different $J^{\pi}$ values.
Unfortunately, such calculation is not possible the present moment.

\begin{table}[tb]
\caption[ ]
{Position, width
and the ratio values $R$, Eq.~(\protect\ref{R-val}), for
of $J = 0^+$ and $J = 2^+$ components of the DGDR
in respect to the ones of the single GDR in $^{136}$Xe.
The third row corresponds
to pure harmonic picture.
\label{vp-t6}}
\begin{center}
\begin{tabular}{cccc}
\hline
$J$ &\hspace*{3mm}$\langle E_{\mbox{\tiny DGDR}} \rangle -
2 \cdot \langle E_{\mbox{\tiny GDR}} \rangle$, keV\hspace*{3mm} &
$\Gamma_{\mbox{\tiny DGDR}} / \Gamma_{\mbox{\tiny GDR}}$ & $R$\\
\hline
$0^+$ &$-120$&1.44&1.94\\
$2^+$ & $-90$&1.45&1.96\\
  &  0&$\sqrt{2}$&2 \\
\hline
\end{tabular}
\end{center}
\end{table}

The calculated excitation functions displayed in Figs.~\ref{vp-f8}b-c
yield
the following values for the centroid and width of the DGDR in $^{136}$Xe:
$<E_{0^+}>$~=~30.68~MeV and $\Gamma_{0^+}$~=~6.82~MeV for the $0^+$
component of the DGDR and
$<E_{2^+}>$~=~30.71~MeV and $\Gamma_{2^+}$~=~6.84~MeV for the $2^+$
component.
These values have to be compared to
$<E_{1^-}>$~=~15.40~MeV and $\Gamma_{1^-}$~=~4.72~MeV for the single
GDR in this nucleus from our calculation.
The correspondence between these values is presented in
Table~\ref{vp-t6}
in comparison with the prediction of the harmonic model.
Also shown is the ratio
\begin{equation}
R = \frac{  \sum_{\nu}
\sum_{\nu_1}
 \langle \Psi^{\nu}_{0^+(2^+)}
|{\cal M}(E1) | \Psi^{\nu_1}_{1^-} \rangle  \cdot
 \langle \Psi^{\nu_1}_{1^-} | {\cal M}(E1) | \Psi_{g.s.} \rangle |^2 }
{ | \sum_{\nu_1}
 \langle \Psi^{\nu_1}_{1^-} | {\cal M}(E1) | \Psi_{g.s.} \rangle |^4 }~,
\label{R-val}
\end{equation}
between the two-step excitation probability of the DGDR normalized
to the summed excitation probability of the one-phonon GDR.
The numerical results lie quite close to the predictions of the harmonical
model (see also a discussion of this problems in Ref~\cite{Yan85}).
While the on-the-energy-shell transitions are easier to identify and
calculate properly, off-the-energy shell corrections are considerably more
elusive.
In fact, it may be argued that the calculated shift of the energy centroid
of the DGDR with respect to that expected in the harmonic picture is somewhat
underestimated, because of the limitations used in selecting
two-phonon basis states used in the calculation.
Our calculated value $\Delta E = 2 \langle E_{\mbox{\tiny GDR}} \rangle-
\langle E_{\mbox{\tiny DGDR}} \rangle$ shown in Table~\ref{vp-t6} can be
compared to the ones in $^{40}$Ca \cite{Cat89} and $^{208}$Pb
\cite{Lan97} in calculations with Skyrme forces.
One of the purposes of the last calculations was to consider the
anharmonic properties of the DGDR with the wave function which includes
collective $1p1h$ and $2p2h$ states. Thus, an interaction not only
between
the two-phonon DGDR states, $[1^- \times 1^-]$ among themselves, but
with other two-phonon states made up of collective $2^+$ and $3^-$
phonons was taken into account. The reported value of $\Delta E$ in
these studies is of the order of $-200$~keV in consistency with our
results. It should be pointed out that the calculation with Skyrme
forces also yield somewhat larger anharmonicity shifts for low-lying
two-phonon states as compared to the QPM calculations \cite{Pon98s}.
The most complete basis of the $2p2h$ configurations has been used
in the second-RPA calculations of the DGDR properties in $^{40}$Ca
\cite{Nis95} and $^{208}$Pb \cite{Nis98} which includes not only
``collective phonons" but non-collective as well. The authors of
Refs.~\cite{Nis95,Nis98} obtained the values of $\Delta E$ equal to
$-670$ ($-40$) and $-960$ ($-470$)~keV for $0^+$ and $2^+$ components of
the DGDR, respectively, in $^{40}$Ca ($^{208}$Pb).
Recently, the problem of anharmonicity for the DGDR has been also
studied within macroscopic approaches in Refs.~\cite{BF97,Den98}.
In Ref.~\cite{BF97} it has been concluded that the $A$-dependence of it
should be as $A^{-4/3}$ while in Ref.~\cite{Den98} it is $A^{-2/3}$ in
consistency with Ref.~\cite{BM75}.

The fragmentation of the DGDR due to the coupling to three-phonon
configurations has been also calculated in $^{208}$Pb \cite{Pon97}.
The fine structure of the GDR as a result of interaction with two-phonon
$1^-$ configurations in this nucleus is presented in Fig.~\ref{vp-f9}a.
The $[B(E1) \times B(E1)]$ values for the DGDR states described by the
wave
function which includes two- and three-phonon configurations are plotted
in Fig.~\ref{vp-f9}b.
In this calculation we have used the same basis of six the most
collective one-phonon $1^-$ states for the GDR and twenty-one the most
collective two-phonon $[1^- \times 1^-]$ states for $0^+$ and $2^+$
components of the DGDR as in the coupled-channels calculation in the
previous subsection (see, Fig.~\ref{vp-f6}).
For a description of the GDR width a coupling to 1161 two-phonon $1^-$
configurations was taken into account.
In calculation of the DGDR strength distribution we have neglected the
interaction with one-phonon configurations and Pauli principle
corrections since these effects are weaker in double-magic nucleus
$^{208}$Pb as compared to the ones in $^{136}$Xe. As a results the
$0^+$ and $2^+$ components of the DGDR are completely degenerated in
this calculation. The DGDR width is determined by the coupling of the
selected 21 two-phonon configurations with 6972 three-phonon ones and is
very close to $\sqrt{2}$ times the width of the GDR.
This is a natural result for a folding of two independent phonons in
microscopic treatment of the problem.
As already discussed in Sect.. 3.3, when damping width of giant resonances is
described phenomenologically by Breit-Wigner strength
distribution one obtains the value 2 for the quantity 
$r = \Gamma_{\mbox{\tiny DGDR}} /  \Gamma_{\mbox{\tiny GDR}}$.
On the other hand, when the Gaussian strength distribution is used, it
yields the value $r = \sqrt{2}$. 
This is due to the different behavior of the wings of the 
abovementioned strength functions at the infinity.
In a microscopic picture, collective resonance state(s) couples to some
finite number of doorway configurations and the strength distribution, as a
result of this coupling, is always concentrated in a definite energy region. 
It results in  $r = \sqrt{2}$.

The square of the amplitude $a^{(1)}_{J_f,M_f;J_i,M_i} (E)$ of the
Coulomb excitation of one-phonon resonances in RHIC in the first order
perturbation theory has a smooth exponential energy dependence.
This rather simplifies a calculation of the excitation cross section
in RHIC of the states of Eq.~(\ref{wf1}) which form single giant
resonances. Although a large number of the states of Eq.~(\ref{wf1}), 
the giant resonance excitation cross section in this reaction can be easily
calculated as a product of the $B(E1)$ values of each state, presented
in Figs.~\ref{vp-f8}a and \ref{vp-f9}a, and an interpolated value of the
tabulated function $|a^{(1)}_{J_f,M_f;J_i,M_i} (E)|^2$ at
$E=E_{\nu}^{J_f}$.
The cross sections of the GDR and GQR excitation in $^{136}$Xe (see,
Fig.~\ref{vp-f7}) have been calculated this way.
A similar procedure may be applied for calculation of
the cross section of the DGDR$(\nu_{0^+,2^+})$ states excitation via the
GDR$(\nu_{1^-})$ states. In the second order perturbation theory it
equals to
\begin{eqnarray}
\sigma_{\nu_{0^+,2^+}}
&=&    \big|      \sum_{\nu_{1^-}}
A(E_{\nu_{1^-}},E_{\nu_{0^+,2^+}})
<1^-(\nu_{1^-})||E1||0^+_{g.s.}>\nonumber \\
&\times&
<[1^- \otimes 1^-](\nu_{0^+,2^+})||E1||1^-(\nu_{1^-})>
\big|^2
\end{eqnarray}
where $A(E_1,E_2)$ is the reaction amplitude which has a very smooth
dependence on both arguments.
This function was tabulated and used in the final calculation of
the DGDR Coulomb excitation cross section in relativistic heavy ion
collisions.

Let us consider the excitation of the DGDR in the projectile for a
$^{208}$Pb (640A~MeV) $+^{208}$Pb collision, according to the
experiment in Ref.~\cite{Bo96}, and use the minimum value of the
impact parameter, b= 15.54~fm, corresponding to the parameterization of
Ref.~\cite{BCV89}.
The cross section for Coulomb excitation of the DGDR is presented in
Fig.~\ref{vp-f10} by the short-dashed curve as a strength function
calculated with an averaging parameter equal to 1~MeV.
The contribution of the background of the
two-phonon $1^-$ states to the total cross section is shown by a long-dashed
curve in the same figure. It was calculated in first order perturbation
theory. The role of the background in this reaction is much less important
than in photoexcitation studies. First, it is because in heavy ion
collisions we have a special mechanism to excite selected two-phonon states
in the two-step process. Second, the Coulomb excitation amplitude is
exponentially decreasing with the excitation energy, while the
$E1$-photoexcitation amplitude is linearly increasing. Nonetheless,
Fig.~\ref{vp-f10} shows that the direct excitation of two-phonon $1^-$
states cannot
be completely excluded from consideration of this reaction. Integrated over
the energy interval from 20 to 35~MeV these states give a cross section of
50.3~mb which should be compared to the experimental cross section in the
DGDR region for the $^{208}$Pb (640A~MeV) $+^{208}$Pb reaction which
is equal to 380~mb \cite{Bo96}.

The solid line in Fig.~\ref{vp-f10} is the sum of DGDR and two-phonon
background excitations in relativistic heavy ion collisions.
The first and second moments of excitation functions, displayed
by the short-dashed and solid curves in Fig.~\ref{vp-f10}, indicate that the
centroid of the total strength is 200~keV lower and the width is 16\% larger
than the same quantities for the pure DGDR.
We point out that this 200~keV shift is even somewhat larger than the
one due to the anharmonicities studied in $^{208}$Pb \cite{Lan97}.

Direct excitation of the two-phonon $1^-$ states in
$^{208}$Pb (640A~MeV) $+^{208}$Pb reaction was also investigated
in Ref.~\cite{Lan97} in calculation with Skyrme forces.
The reported effect (a difference between 5.07 and 3.55 mb for
22 $<$ E$_x$ $<$ 28 MeV) is much weaker than in our calculation
because of a rather limited two-phonon space.
Another source of the DGDR enhancement in \cite{Lan97} is due to
anharmonicity effects.
We also checked the last by coupling one-phonon GDR states to
(the most important) 1200 two-phonon $1^-$ states in the DGDR region.
Due to the constructive interference between one- and two-phonon states at
DGDR energies we got an additional enhancement of
24 mb, which is again larger than the difference between 6.42 and 3.55 mb
obtained in Ref.~\cite{Lan97} for the same reason.

The absolute value of the total cross section of the DGDR excitation in
RHIC in $^{208}$Pb is somewhat small in our calculation (cf.
Table~\ref{vp-t4}) as compared to experimental findings. For example,
the experimental value of the total DGDR excitation in the reaction
$^{208}$Pb (640A~MeV) $+^{208}$Pb, for which the calculations have been
performed, equals to 0,38(4)~b.
As mentioned above our chosen six $1^-$
states exhaust only 90.6\% of EWSR while the photo-neutron data \cite{Ve70}
indicate that this value equals to 122\%. Due to this underestimate
of exhaust of the EWSR the cross section for the DGDR excitation
reduces more strongly than the one for the single GDR. This is because
the GDR cross section is roughly proportional to the total $B(E1)$ value
while for the DGDR it is proportional to the square of it.

If we apply a primitive scaling to obtain
the experimental value 122\% of EWSR
the ratio $R = \sigma_{\mbox{\tiny (DGDR)}}/\sigma_{\mbox{\tiny (GDR)}}$,
the last line of Table~\ref{vp-t4},
changes into 0.096 and 0.101 for the coupled-channels calculation and
for the perturbation theory, respectively. The experimental findings
\cite{Bo96} yield the value $R_{exp}$~=~0.116$\pm$0.014.
The reported \cite{Bo96} disagreement
$R_{exp}/R_{calc}$~=1.33$\pm$0.16 is the result of
a comparison with $R_{calc}$ obtained within a folding model, assuming
122\% of the EWSR.
We get a somewhat larger value of $R_{calc}$ (taking into account
our scaling procedure) because the $B(E1)$ strength distribution
over our six $1^-$ states is not symmetrical
with respect to the centroid energy, $E_{\mbox{\tiny GDR}}$: the lower
part is enhanced.
A weak energy dependence in the excitation amplitude, which is also
squared for  the DGDR, enhances the DGDR cross section for a
non-symmetrical distribution with respect to the symmetrical one, or
when the GDR is treated as a single state.
The effect of the energy dependence is demonstrated for a single GDR
in the top part of Fig.\ref{vp-f6} where the excitation cross
sections are compared to the $B(E1)$ strength distribution.
It produces a shift to lower energies of the centroid of the
GDR and the DGDR
cross sections with respect to the centroid of the $B(E1)$ and
the $[B(E1) \times B(E1)]$ strength distribution, respectively.
In our calculation this shift equals to 0.26~MeV for the GDR
and to 0.33, 0.28~MeV for the DGDR within coupled-channels and
perturbation theory, respectively.

Of course, this scaling procedure has no deep physical meaning but we
have included this discussion to indicate that the disagreement between
experiment and theory for the DGDR excitation cross sections in
$^{208}$Pb reached the stage when theoretical calculations have to
provide a very precise description of both the GDR and the DGDR to draw
up final conclusions.

The situation with the absolute values of cross sections of the DGDR
excitation in $^{136}$Xe in RHIC is much less clear than in $^{208}$Pb.
The experimental value for the reaction $^{136}$Xe (700A~MeV) $+^{208}$Pb
is reported to be equal to 215$\pm$50 mb \cite{Sc93}. This value is
sufficiently larger as compared to any theoretical predictions available
(cf. Table~\ref{vp-t5}).
But it should be pointed out that a comparison between of experimental
data for xenon \cite{Sc93} and lead \cite{Bo96} reveal some essential
contradiction.
While for $^{208}$Pb the discussed above quantity of the ratio between
the total cross sections of the DGDR and GDR excitation
$R_{exp}(^{208}\mbox{Pb})$~=~0.116$\pm$0.014, its value for
$^{136}$Xe: $R_{exp}(^{136}\mbox{Xe})$~=~0.21$\pm$0.05 \cite{Sc93}.
Taking into account that experiments for both nuclei have been performed
at close projectile energies (per nucleon) and the cross section of the
GDR excitation in $^{136}$Xe is about three times less as compared
to the one in $^{208}$Pb, the ratio $R_{exp}(^{136}\mbox{Xe})$ should be
sufficiently smaller than $R_{exp}(^{208}\mbox{Pb})$ and not vice
versa. Probably, the problem with the absolute value of the DGDR
excitation in $^{136}$Xe is related to uncertainties in separating
of the contribution of single resonances, the characteristics of which
are unknown experimentally for this nucleus and the results of
interpolation have been used in evaluating of the experimental data.
Recently, the experiment for $^{136}$Xe has been repeated by LAND
collaboration \cite{Eml95}. The analysis of the new data in the
nearest future should clear up the situation.

\subsection{The role of transitions between complex configurations of
the GDR and the DGDR}

In the previous subsection considering the excitation properties of the
DGDR, $[B(E1) \times B(E1)]$ values or excitation cross sections in RHIC
in  second order perturbation theory, we have taken into account only
the transition matrix elements between simple one-phonon $1^-$ GDR and
two-phonon $0^+$ or $2^+$ DGDR configurations for the second step of the
excitation process $g.s.\rightarrow\mbox{GDR}\rightarrow\mbox{DGDR}$.
In fact, as already discussed above these configurations couple to more
complex ones to produce the widths of single and double resonances
and in principle, additional transitions between complex configurations
of the GDR and the DGDR, together with interference effects, may alter
the predicted values of excitation probabilities. This problem will be
considered in the present subsection (see, also Ref.~\cite{Pon98a}).
It will be concluded that their role is
marginal in the process under consideration although a huge amount of the
$E1$-strength is hidden in the GDR$\rightarrow $DGDR transition. This
negative result ensures that calculations, in which only transitions
between simple components of the GDR and DGDR are taken into account and
which are much easier to carry out, require no further corrections.

In microscopic approaches the strength of the GDR is split among several
one-phonon $1_\alpha ^{-}$ states (due to the Landau damping). The wave
function $|1_\alpha ^{-}>$ couples to complex configurations $|1_\beta ^{-}>$
yielding the GDR width. We use the index $\alpha $ for simple configurations
and the index $\beta $ for complex ones, respectively. Thus, the wave
function of the $i^{th}$ $1^{-}$ state in the GDR energy region can
be schematically written as:
\begin{equation}
|1_i^{-}>=\sum_\alpha S_i^{\mbox{\tiny GDR}}(\alpha )|1_\alpha
^{-}>+\sum_\beta C_i^{\mbox{\tiny GDR}}(\beta )|1_\beta ^{-}>  \label{gdr}
\end{equation}
where coefficients $S_i^{\mbox{\tiny GDR}}(\alpha )$ and
$C_i^{\mbox{\tiny GDR}}(\beta)$ may be obtained by diagonalizing the
nuclear model Hamiltonian on the set of wave functions (\ref{gdr}).

The total $E1$-strength of the GDR excitation from the ground state,
$$B_{\mbox{\tiny GDR}}(E1)=\sum_i|<1_i^{-}||E1||0_{g.s.}^{+}>|^2,$$ remains
practically the same as in the one-phonon RPA calculation because the direct
excitation of complex configurations from the ground state is a few order of
magnitude weaker as compared to excitation of one-phonon states. However
these complex configurations play a fundamental role for the width of the
GDR.

The wave function of the $2^{+}$ component of the DGDR states can be written
in the similar fashion:
\begin{eqnarray}
|2_f^{+}>= &&\sum_{\tilde{\alpha}=\{\alpha _1\times \alpha _2\}}
S_f^{\mbox{\tiny DGDR}}(\tilde{\alpha})|[1_{\alpha _1}^{-}\times
1_{\alpha _2}^{-}]_{2^{+}}> \nonumber \\
+ &&\sum_{\alpha ^{\prime \prime }}{\tilde{S}}_f^{\mbox{\tiny DGDR}}(\alpha
^{\prime \prime })|2_{\alpha ^{\prime \prime }}^{+}>+\sum_{\beta ^{\prime
}}C_f^{\mbox{\tiny DGDR}}(\beta ^{\prime })|2_{\beta ^{\prime }}^{+}>~.
\label{dgdr}
\end{eqnarray}
In this equation we separated in the first term the $[1^{-}\times
1^{-}]$ DGDR configurations from other two-phonon configurations (second
term) and complex configurations (the last term). The same equation as
(\ref{dgdr}) is valid for the $0^{+}$ DGDR states.

The total $E1$-transition strength between the GDR and DGDR,
$$\sum_f\sum_i|<2^{+}(0^{+})_f||E1||1_i^{-}>|^2,$$ is much larger as
compared
to that for the GDR excitation, $\sum_i|<1_i^{-}||E1||0_{g.s.}^{+}>|^2,$
 from the ground state. This is because the former includes transitions
not only between simple GDR and DGDR states but also between complex
configurations as well. The enhancement factor should be the ratio between
the density of simple and complex configuration in the GDR energy
region.
But in the two-step excitation process the sum over intermediate GDR states
in Eq.~(\ref{cs}) reduces the total transition strength for $g.s.\rightarrow
\mbox{GDR}\rightarrow \mbox{DGDR}$ to $\sim 2\cdot |B_{\mbox{\tiny
GDR}}(E1)|^2$ (the
factor 2 appears due to the bosonic character of the two phonons which also
holds if Landau damping is taken into account).
To prove this let us consider
the excitation probability of the DGDR
\begin{eqnarray}
P_{\mbox{\tiny DGDR}}(E_f,\;b) =\frac 14\sum_{M_f}\Bigg|
\sum_{i,M_i}a_{0(0)\rightarrow
1_i^{-}(M_i)}^{E1(\mu )}(E_i,\;b) \times
a_{1_i^{-}(M_i)\rightarrow [1^{-}\times 1^{-}]_f(M_f)}^{E1(\mu
^{\prime })}(E_f-E_i,\;b)\Bigg| ^2\nonumber \\
 \label{cs}
\end{eqnarray}
where the index $i$ labels intermediate states belonging to the GDR, and
$a_{J_1(M_1)\rightarrow J_2(M_2)}^{E1(\mu )}$ is the first-order $E1$
excitation amplitude for the transition $J_1(M_1)\longrightarrow J_2(M_2)$
in a collision with impact parameter $b$. For each state, $J$ and $M$ denote
the total angular momentum and the magnetic projection, respectively.

The amplitude $a_{J_1(M_1)\rightarrow J_2(M_2)}^{E1(\mu )}$ is given by
\begin{equation}
a_{J_1(M_1)\rightarrow J_2(M_2)}^{E1(\mu )}(E,\;b) =(J_1M_11\mu |J_2M_2)
\times  <J_2||E1||J_1>f_{E1(\mu )}(E,\;b)\;.
\end{equation}
It is a product of the reduced matrix element $<J_2||E1||J_1>$ for the
$E1$-transition between the states $J_1(M_1)$ and $J_2(M_2)$ which
carries nuclear structure information and the reaction function
$f_{E1(\mu )}(E,\;b)$.
The latter depends on the excitation energy, charge of the target, beam
energy, and is calculated according to Ref. \cite{WA79}. Except for the
dependence on the excitation energy, it does not carry any nuclear structure
information. The cross section for the DGDR is obtained from Eq. (\ref{cs})
by integration over impact parameters, starting from a minimal value $b_{min}
$ to infinity. This minimal value is chosen according to Ref. \cite{BCH96}.

Now we substitute the wave functions of the GDR and DGDR states given by
Eqs. (\ref{gdr},\ref {dgdr}) in expression (\ref{cs}).
We obtain two terms. The first one
corresponds to transitions between simple GDR and DGDR states (after the
GDR is excited from the ground state through its simple component):
\begin{eqnarray}
A_{\mu \mu ^{\prime }} &=&\sum_i\sum_{\alpha \alpha ^{\prime
}\tilde{\alpha} }S_i^{\mbox{\tiny
GDR}}(\alpha )f_{E1(\mu )}(E_i,\;b)<1_\alpha ^{-}||E1||0_{g.s.}^{+}>
\label{sim} \\
&\times  &S_i^{\mbox{\tiny GDR}}(\alpha ^{\prime })S_f^{\mbox{\tiny
DGDR}}({
\tilde{\alpha}})f_{E1(\mu ^{\prime })}(E_f-E_i,\;b)
  <[1_{\alpha _1}^{-}\times 1_{\alpha _2}^{-}]_f||E1||1_{\alpha
^{\prime }}^{-}>\delta _{\alpha _2,\alpha ^{\prime }}
\nonumber
\end{eqnarray}
and the second one accounts transitions between complex configurations in
the wave functions of Eqs.~(\ref{gdr},\ref{dgdr}):
\begin{eqnarray}
B_{\mu \mu ^{\prime }} &=&\sum_i\sum_{\alpha \alpha ^{\prime }\beta \beta
^{\prime }}S_i^{\mbox{\tiny GDR}}(\alpha )f_{E1(\mu )}(E_i,\;b)<1_\alpha
^{-}||E1||0_{g.s.}^{+}>  \label{com}   \\
&\times& C_i^{\mbox{\tiny GDR}}(\beta )C_f^{\mbox{\tiny DGDR}}(\beta
^{\prime })f_{E1(\mu ^{\prime })}(E_f-E_i,\;b)
<[1_{\alpha ^{\prime }}^{-}\times 1_\beta ^{-}]_f||E1||1_\beta
^{-}>\delta _{\beta ^{\prime },[\alpha ^{\prime }\times \beta ]}~.
\nonumber
\end{eqnarray}

The second reduced matrix element in the above equations is proportional to
the reduced matrix element between the ground state and the simple
one-phonon configuration (see, Eq.~(\ref{eq:v3})).

For a given impact parameter $b$, the function $f_{E1(\mu )}(E,\;b)$ can be
approximated by a constant value $f_{E1(\mu )}^0$ \cite{BB88} for the
relevant values of the excitation energies. Then the energy dependence can
be taken out of summations and orthogonality relations between different
components of the GDR wave functions can be applied \cite{Pon96a}. The
orthogonality relations between the wave functions imply that
\begin{equation}
\sum_i S_i^{\mbox{\tiny GDR}}(\alpha )C_i^{\mbox{\tiny
GDR}}(\beta )\equiv 0~.
\end{equation}
This means that the term $B_{\mu \mu ^{\prime }}$
vanishes. The term $A_{\mu \mu ^{\prime }}$ summed over projections and all
final states yields a transition probability to the DGDR,
$P_{\mbox{\tiny DGDR}}(E_f,\;b),$ which is proportional to $2\cdot
|B_{\mbox{\tiny GDR}}(E1)|^2$ in second
order perturbation theory. This argument was the reason for neglecting the
term $B_{\mu \mu ^{\prime }}$ in previous calculations of DGDR
excitation
where the coupling of simple GDR and
DGDR states to complex configurations was taken into account.

In Fig.~\ref{vp-f11} we plot the value of $\chi _{E1}(E)=2\pi \int
db\;b\sum_\mu |f_{E1(\mu )}(E,\;b)|^2$ as a function of energy calculated
for the $^{208}$Pb (640A~MeV)~+~$^{208}$Pb reaction. This value
corresponds to $\sigma _{\mbox{\tiny GDR}}$ if $B_{\mbox{\tiny GDR}}(E1)=1$.
The square in this figure indicates the location of the GDR in $^{208}$Pb.
This figure demonstrates that the function $\chi _{E1}(E)$ changes by 60\%
in the GDR energy region. The role of this energy dependence for other
effects has been considered in Refs.~\cite{BCH96,Lan97}. Taking into
account that one-phonon $1_\alpha ^{-}$ configurations are fragmented over a
few MeV \cite{Pon96b}, when a sufficiently large two-phonon basis is
included in the wave function given by Eq.~(\ref{gdr}), the role of the
$B_{\mu \mu ^{\prime }}$ term in the excitation of the DGDR should be studied
in more detail.

To accomplish this task we have performed firstly a simplified calculation
in which we used the {\it boson type }Hamiltonian:
\begin{equation}
H =\sum_\alpha \omega _\alpha Q_\alpha ^{\dagger }Q_\alpha +\sum_\beta
\widetilde{\omega }_\beta \widetilde{Q}_\beta ^{\dagger }\widetilde{Q}_\beta
+\sum_{\alpha ,\beta }U_\beta ^\alpha (Q_\alpha ^{\dagger }
\widetilde{Q}_\beta +h.c.)
\label{hb}
\end{equation}
where $Q_\alpha ^{\dagger }$ is the phonon creation operator and $\omega
_\alpha $ is the energy of this one-phonon configuration;
$\widetilde{Q}_\beta ^{\dagger }$ is the operator for creation of a
complex configuration
with energy $\widetilde{\omega }_\beta $ and $U_\beta ^\alpha $ is the
matrix element for the interaction between these configurations. We have
assumed that the energy difference between two neighboring one-phonon
configurations is constant and equals to $\Delta \omega $. An equidistant
spacing with the energy $\Delta \widetilde{\omega }$ was assumed for the
complex configurations. We also have used a constant value $U$ for the
matrix elements of the interaction. The $B_{\mbox{\tiny GDR}}(E1)$ value was
distributed symmetrically over one-phonon configurations. Thus, the
free parameters of this model are: $\Delta \omega $,
$\Delta \widetilde{\omega }$, $U$, the number of one-phonon and complex
configurations, and the
distribution of the $B_{\mbox{\tiny GDR}}(E1)$ value over the simple
configurations. The only condition we want to be satisfied is that the
energy spectrum for the GDR photoexcitation is the same as the one known
 from  experiment.

After all parameters are fixed we diagonalize the model Hamiltonian of
Eq.~(\ref{hb}) on the set of wave functions of Eq. (\ref{gdr}) for the
GDR and on
the set of Eq. (\ref{dgdr}) for the DGDR. The diagonalization procedure
yields the information on eigen energies of the $1_i^{-}$ GDR states and on
the coefficients $S_i^{\mbox{\tiny GDR}}(\alpha )$ and
$C_i^{\mbox{\tiny GDR} }(\beta ),$ respectively. One also obtains
information on eigen energies of
the $2_f^{+}$ or $0_f^{+}$ DGDR states and the coefficients
$S_f^{ \mbox{\tiny DGDR}}(\tilde{\alpha})$ and $C_f^{\mbox{\tiny
DGDR}}(\beta ^{\prime }),$
respectively. With this information we are able to study the role of the
$B_{\mu \mu ^{\prime }}$ term in the excitation of the DGDR in RHIC.

The big number of free parameters allows an infinite number of suitable
choices. In fact, not all of the parameters are really independent. For
example, the increase in the number of simple or complex configurations goes
together with the decreasing of the value of $U.$ This is necessary for a
correct description of the GDR photoabsorption cross section. This makes it
possible to investigate the role of the $B_{\mu \mu ^{\prime }}$ term in
different conditions of weak and strong Landau damping and for different
density of complex configurations. In our calculations we vary the number of
collective simple states from one to seven and the number of complex
configurations from 50 to 500. The value of $U$ then changes from about 100
to 500~keV. The results of one of these calculations for the excitation of
the $2^{+}$ component of the DGDR in $^{208}$Pb
(640A~MeV)~+~$^{208}$Pb collisions are presented in
Fig.~\ref{vp-f12}. For a better visual
appearance the results are averaged with a smearing parameter equal to
1~MeV. The dashed curve shows the results of a calculation in which $\sigma
_{\mbox{\tiny DGDR}}^A(E)\equiv \sigma _{\mbox{\tiny DGDR}}(E)\sim \int
db\;b|A_{\mu \mu ^{\prime }}|^2$ and the results of another one in which
$\sigma _{\mbox{\tiny DGDR}}^{A+B}(E)\equiv \sigma _{\mbox{\tiny
DGDR}}(E)\sim \int db\;b\left| A_{\mu \mu ^{\prime }}+B_{\mu \mu ^{\prime
}}\right| ^2$ are represented by a solid curve.

Our calculation within this simple model indicates that the role of the
$B_{\mu \mu ^{\prime }}$ term in  second order perturbation theory is
negligibly small, although the total $B(E1)$ strength for transitions
between complex GDR and DGDR configurations, considered separately, is more
than two orders of magnitude larger than the ones between simple GDR and
DGDR configurations. The value $\Delta \sigma =(\sigma _{\mbox{\tiny
DGDR} }^{A+B}-\sigma _{\mbox{\tiny DGDR}}^A)/\sigma _{\mbox{\tiny
DGDR}}^A$, where
$\sigma _{\mbox{\tiny DGDR}}^{A(A+B)}=\int \sigma _{\mbox{\tiny
DGDR}}^{A(A+B)}(E)dE$, changes in these calculations from 1\% to 2.5\%. The
results practically do not depend on the number of complex configurations
accounted for. The maximum value of $\Delta \sigma $ is achieved in a
calculation with a single one-phonon GDR state (no Landau damping). This
is
because the value of $U$ is the larger in this case and the fragmentation of
the one-phonon state is stronger. Thus, in such a situation, the energy
dependence of the reaction amplitude modifies appreciably the orthogonality
relations. But in general the effect is marginal.

We also performed a calculation with more realistic wave functions for the
GDR and DGDR states taken from our studies presented in the previous
subsection.
These wave functions include 6 and 21 simple states for the GDR and
DGDR, respectively. The complex configurations are two-phonon states for
the GDR and three-phonon states for the DGDR.
The value $\Delta
\sigma $ equals in this realistic calculation to 0.5\%. This result is not
surprising because realistic calculations with only two-phonon complex
configurations, and a limited number of them, somewhat underestimate the GDR
width which is crucial for the modification of the orthogonality relations.

We have proved that the transitions between complex GDR and DGDR
configurations within second-order perturbation theory for the DGDR
excitation in RHI collisions play a marginal role in the process under
consideration and it is sufficient to take into account only transitions
between the ground state and one-phonon GDR and two-phonon DGDR
configurations.

\subsection{The DGDR in deformed nuclei}

The possibility to observe two-phonon giant resonances in deformed nuclei
with the present state of art experimental techniques is still questionable.
This is mainly due to the fact that one has to expect a larger width of
these resonances as compared to spherical nuclei. Also, the situation with
the low-lying two-phonon states in deformed nuclei is much less clear than
in spherical ones.

The first experiment with the aim to observe the double giant dipole
resonance (DGDR) in $^{238}$U in relativistic heavy ion collisions (RHIC)
was performed recently at the GSI/SIS facility by the LAND collaboration
\cite{Eml95}. It will take some time to analyze the experimental data
and to present the first experimental evidence of the DGDR in deformed
nuclei, if any. Thus, we present here the first theoretical
predictions of the properties of the DGDR in deformed
nuclei based on microscopic study \cite{Pon98b}. The main attention will
be paid to the width of the DGDR and its shape.

In a phenomenological approach the GDR is considered as a collective
vibration of protons against neutrons. In spherical nuclei this state is
degenerate in energy for different values of the spin $J=1^{-}$ projection
$M=0,\pm 1$. The same is true for the $2^{+}$ component of the DGDR with
projection $M=0,\pm 1,\pm 2$. In deformed nuclei with an axial symmetry like
$^{238}$U, the GDR is spit into two components $I^\pi (K)=1^{-}(0)$ and
$I^\pi (K)=1^{-}(\pm 1)$ corresponding to vibrations against two different
axes. In this approach one expects a three-bump structure for the DGDR with
the value $K=0$, $K=\pm 1$ and $K=0,\pm 2$, respectively (see,
Fig.~\ref{vp-f13}). Actually, the GDR
possesses a width and the main mechanism responsible for it in deformed
nuclei is the Landau damping. Thus, the conclusion on how three bumps
overlap and what is the real shape of the DGDR in these nuclei, i.e., either
a three-bump or a flat broad structure, can be drawn out only from some
consistent microscopic studies.

We use in our calculations for $^{238}$U the parameters of
Woods-Saxon potential for the average field and monopole pairing from
Ref.~\cite{Iva76}. They were adjusted to reproduce the
properties of the ground state and low-lying excited states. The average
field has a static deformation with the deformation parameters
$\beta_2=0.22$ and $\beta_4=0.08$. To construct the phonon basis for
the $K=0$ and
$K=\pm 1$ components of the GDR we use the dipole-dipole residual
interaction (for more details of the QPM application to deformed nuclei,
see e.g. Ref.~\cite{Sol92}). The strength parameters of this interaction
are taken from Ref.~\cite{Aku77} where they have been fitted
to obtain the centroid of the $B(E1, 0_{g.s.}^{+}\rightarrow 1^{-}(K=0,\pm
1))$ strength distribution at the value known from experiment \cite{Gur76}
and to exclude the center of mass motion. In this approach, the information
on the phonon basis (i.e. the excitation energies of phonons and their
internal fermion structure) is obtained by solving the RPA equations. For
electromagnetic $E1$-transitions we use the values of the effective
charges, $e_{eff}^{Z(N)}=eN(-Z)/A$ to separate the center of mass motion.

The results of our calculation of 
$B(E1)$ strength distribution over
$\left| 1_{K=0}^{-}(i)\right\rangle $
and $\left| 1_{K=\pm 1}^{-}(i^{\prime})\right\rangle$ GDR states are
presented in Fig.~\ref{vp-f14}, together with
experimental data. The index $i$ in the wave function stands for the
different RPA states. All one-phonon states with the energy lower than
20~MeV and with the $B(E1)$ value larger than $10^{-4}\ e^2$ fm$^2$ are
accounted
for. Their total number equals to 447 and 835 for the $K=0$ and $K=\pm 1$
components, respectively. Only the strongest of them with $B(E1)\ge 0.2\
e^2$ fm$^2$ are shown in the figure by vertical lines. Our phonon basis
exhausts 32.6\% and 76.3\% of the energy weighted sum rules, $14.8\cdot NZ/A$
$e^2$ fm$^2$ MeV, by the $K=0$ and $K=\pm 1$ components, respectively. For a
better visual appearance we also present in the same figure the strength
functions averaged with a smearing parameter, which we take as 1~MeV. The
short (long) dashed-curve represent the $K=0$ ($K=\pm 1)$ components of the
GDR. The solid curve is their sum. The calculation reproduces well the
two-bump structure of the GDR and the larger width of its $K=\pm 1$
component. The last is consistent with the experiment \cite{Gur76} which is
best fitted by two Lorentzians with widths equal to $\Gamma _1=2.99$~MeV and
$\Gamma _2=5.10$~MeV, respectively. The amplitudes of both maxima in the
calculation are somewhat overestimated as compared to the experimental data.
This happens because the coupling of one-phonon states to complex
configurations is not taken into account which can be more relevant for the
$K=\pm 1$ peak at higher energies. But in general the coupling matrix
elements are much weaker in deformed nuclei as compared to spherical ones
and the Landau damping describes the GDR width on a reasonable level.

The wave function of the $0^{+}$ and $2^{+}$ states belonging to the DGDR
are constructed by the folding of two $1^{-}$ phonons from the previous
calculation. When a two-phonon state is constructed as the product of two
identical phonons its wave function gets an additional factor $1/\sqrt{2}$.
The $1^{+}$ component of the DGDR is not considered here for the same
reasons as in spherical nuclei.
The anharmonicity effects which arise from interactions
between different two-phonon states are also not included in the present
study.

The folding procedure yields three groups of the DGDR states: 
\vspace*{-01mm}
\begin{eqnarray}
&&a)~~| [1_{K=0}^{-}(i_1)\otimes
1_{K=0}^{-}(i_2)]_{0_{K=0}^{+},2_{K=0}^{+}}\rangle ,  \nonumber \\
&&b)~~| [1_{K=0}^{-}(i)\otimes 1_{K=\pm 1}^{-}(i^{\prime })]_{2_{K=\pm
1}^{+}} 
\rangle ~~\mbox{and}  \nonumber \\
&&c)~~| [1_{K=\pm 1}^{-}(i_1^{\prime })\otimes 1_{K=\pm
1}^{-}(i_2^{\prime })]_{0_{K=0}^{+},2_{K=0,\pm 2}^{+}}\rangle .
\label{wf}
\end{eqnarray}
The total number of non-degenerate two-phonon states equals to about
$1.5\cdot 10^6$. The energy centroid of the first group is the lowest
and of
the last group is the highest among them. So, we also obtain the three-bump
structure of the DGDR. But the total strength of each bump is fragmented
over a wide energy region and they strongly overlap.

Making use of the nuclear structure elements discussed above, we have
calculated the excitation of the DGDR in $^{238}$U projectiles
(0.5~GeV$\cdot $A) incident on $^{120}$Sn and $^{208}$Pb targets,
following the
conditions of the experiment in Ref.~\cite{Eml95}. These calculations
have
been performed in  second order perturbation theory \cite{BB88}, in
which the DGDR states of Eq.~(\ref{wf}) are excited within a two-step
process: g.s.$\rightarrow $GDR$\rightarrow $DGDR. As intermediate states,
the full set of one-phonon $\left| 1_{K=0}^{-}(i)\right\rangle $ and $\left|
1_{K=\pm 1}^{-}(i^{\prime })\right\rangle $ states was used. We have also
calculated the GDR excitation to first order for the same systems. The
minimal value of the impact parameter, which is very essential for the
absolute values of excitation cross section has been taken according to
$b_{min}=1.28\cdot (A_t^{1/3}+A_p^{1/3})$.

The results of our calculations are summarized in Fig.~\ref{vp-f15} and
Table~\ref{vp-t7}. In Fig.~\ref{vp-f15} we present the cross sections of
the GDR
(part a) and the DGDR (part b) excitation in the $^{238}$U (0.5~GeV$\cdot
$A) + $^{208}$Pb reaction. We plot only the smeared strength functions
of the
energy distributions because the number of two-phonon states involved is
numerous. The results for $^{238}$U (0.5~GeV$\cdot $A) + $^{120}$Sn reaction
look very similar and differ only by the absolute value of cross sections.
In Table~\ref{vp-t7} the properties of the GDR and the DGDR, and their
different $K$ components are given. The energy centroid $E_c$ and the second
moment, $m_2=\sqrt{\sum_k\sigma _k\cdot (E_k-E_c)^2/\sum_k\sigma _k},$ of
the distributions are averaged values for the two reactions under
consideration.

\begin{table}[tb]
\caption[ ]
{The properties of the different components of the GDR and the DGDR
in $^{238}$U. The energy centroid $E_c$, the second moment of the strength
distribution $m_2$ in RHIC, and the cross sections $\sigma $ for the
excitation of the projectile are presented for:
a) $^{238}$U (0.5A~GeV) + $^{120}$Sn, and b) $^{238}$U
(0.5A~GeV) + $^{208}$Pb.
\label{vp-t7}}
\begin{center}
\begin{tabular}{lccrr}
\hline
& $E_{c}$ & $m_2$ & \multicolumn{2}{c}{$\sigma$ [mb]} \\
& [MeV] & [MeV] & a)\hspace*{2mm} & b)\hspace*{2mm} \\ \hline
GDR($K=0$) & 11.0 & 2.1 & 431.2 & 1035.4 \\
GDR($K=\pm1$) & 12.3 & 2.6 & 1560.2 & 3579.1 \\
GDR(total) & 12.0 & 2.6 & 1991.4 & 4614.5 \\ \hline
DGDR$_{0^+}$($K=0$) & 25.0 & 3.4 & 18.3 & 88.9 \\
DGDR$_{2^+}$($K=0$) & 24.4 & 3.5 & 11.8 & 58.7 \\
DGDR$_{2^+}$($K=\pm1$) & 23.9 & 3.2 & 22.7 & 115.4 \\
DGDR$_{2^+}$($K=\pm2$) & 25.3 & 3.4 & 49.7 & 238.3 \\
DGDR(total) & 24.8 & 3.4 & 102.5 & 501.3
\\ \hline
\end{tabular}
\end{center}
\end{table}

The two-bump structure can still be seen in the curve representing the cross
section of the GDR excitation in $^{238}$U in RHIC as a function of the
excitation energy. But its shape differs appreciably from the $B(E1)$
strength distribution (see Fig.~\ref{vp-f15}a in comparison with
Fig.~\ref{vp-f14}). The
reason for that is the role of the virtual photon spectra. First, for the
given value of the excitation energy and impact parameter it is larger for
the $K=\pm 1$ component than that for the $K=0$ one (see also the first two
lines in Table~\ref{vp-t7}). Second, for both components it has a
decreasing
tendency with an increase of the excitation energy \cite{BB88}. As a
result, the energy centroid of the GDR excitation in RHIC shifts by the
value 0.7~MeV to lower energies as compared to the same value for the
$B(E1)$
strength distribution. The second moment $m_2$ increases by 0.2~MeV.

The curves representing the cross sections of the excitation of the $K=\pm 1$
and $K=\pm 2$ components of the DGDR in $^{238}$U in RHIC have typically a
one-bump structure (see the curves with squares and triangles in
Fig.~\ref{vp-f15}b, respectively). It is because they are made of
two-phonon $2^{+}$
states of one type: the states of Eq.~(\ref{wf}b) and Eq.~(\ref{wf}c),
respectively. Their centroids should be separated by an energy approximately
equal to the difference between the energy centroids of the $K=0$ and $K=\pm
1$ components of the GDR. They correspond to the second and the third bumps
in a phenomenological treatment of the DGDR. The $K=0$ components of the
DGDR include two group of states: the states represented by Eq.~(\ref{wf}a)
and those of Eq.~(\ref{wf}c). Its strength distribution has two-bumps (see
the curve with circles for the $2^{+}(K=0)$ and the dashed curve for the
$0^{+}(K=0)$ components of the DGDR, respectively). The excitation of the
states given by Eq.~(\ref{wf}a) in RHIC is enhanced due to their lower
energies, while the enhancement of the excitation of the states given by
Eq.~(\ref{wf}c) is related to the strongest response of the $K=\pm 1$
components to the external $E1$ Coulomb field in both stages of the
two-step process.

Summing together all components of the DGDR yields a broad one-bump
distribution for the cross section for the excitation of the DGDR in
$^{238}$U, as a function of excitation energy. It is presented by the
solid curve in
Fig.~\ref{vp-f15}b. Another interesting result of our calculations is
related
to the position of the DGDR energy centroid and to the second moment of the
DGDR cross section. The centroid of the DGDR in RHIC is shifted to the
higher energies by about 0.8~MeV from the expected value of two times the
energy of the GDR centroid. The origin for this shift is in the energy
dependence of the virtual photon spectra and it has nothing to do with
anharmonicities of the two-phonon DGDR states. In fact, the energy centroid
of the $B(E1,~g.s.\rightarrow 1_i^{-})\times  B(E1,~1_i^{-}\rightarrow
$DGDR$_f)$ strength function appears exactly at twice the energy of the
centroid of the $B(E1,~g.s.\rightarrow $GDR) strength distribution
because
the coupling between different two-phonon DGDR states are not accounted for
in the present calculation. The same shift of the DGDR from twice the energy
position of the GDR in RHIC also takes place in spherical nuclei. But the
value of the shift is smaller there because in spherical nuclei the GDR and
the DGDR strength is less fragmented over their simple configurations
due to the Landau damping. But the
larger value of the shift under consideration in deformed nuclei should
somehow simplify the separation of the DGDR from the total cross section in
RHIC.

Another effect which also works in favor of the extraction of the DGDR from
RHIC excitation studies with deformed nuclei is its smaller width than
$\sqrt{2}$ times the width of the GDR, as observed with spherical
nuclei. Our
calculation yields the value 1.33 for the ratio $\Gamma {\mbox {\tiny
DGDR}}/\Gamma {\mbox {\tiny GDR}}$ in this reaction. The origin for this
effect is
in the different contributions of the GDR $K=0$ and $K=\pm 1$ components to
the total cross section, due to the reaction mechanism. It should be
remembered that only the Landau damping is accounted for the width of both
the GDR and the DGDR.
But since the effect of narrowing of the DGDR width
is due to the selectivity of the reaction mechanism
it will still hold if the coupling to complex configurations is included in
the calculation.

It may be argued that the procedure of independent excitations of two
RPA phonons applied here is not sufficient for a consistent
description of the properties of the two-phonon giant resonances.
This is true for the case of spherical nuclei where only the coupling
of two GDR phonons to more complex, $3p3h$, configurations allows one
to describe the DGDR width as discussed above.
But the typical matrix element of
this coupling in deformed nuclei does not exceed the value of 200~keV
\cite{Sol94} while in spherical nuclei it is an order of magnitude
larger.
It means that due to the coupling, the strength of each GDR RPA-phonon
will fragment within the energy interval of 100-200~keV in deformed
nuclei. The last value should be compared to the second moment, $m_2$,
presented in Table~\ref{vp-t7} which is the result of the Landau damping
accounted for in our calculation. Taking into account that the
reaction amplitude has very weak energy dependence and that mixing of
different RPA phonons in the GDR wave function does not change the
total strength \cite{Pon98a}, the total cross sections of the GDR and
DGDR excitation in RHIC will be also conserved.

\bigskip\bigskip

{\bf Acknowledgments}

We thank our colleagues G. Baur, P.F. Bortignon, R.A. Broglia,
L. F. Canto,
M. Hussein, A.F.N. de Toledo Piza,
A.V. Sushkov, V.V. Voronov for fruitful
collaboration. We also thank H. Emling for many fruitful and
stimulative discussions.
This work was supported in part by
the Brazilian agencies CNPq, FINEP, FAPERJ and FUJB,  the Russian Fund 
for Basic Research (grant No. 96-15-96729) and the Research Council of the 
University of Gent.

\newpage

\vspace*{20mm}

\centerline{\Large \bf Figure captions}

\begin{figure}[ht]
\caption[ ]
{Experimental cross sections in arbitrary units for the excitation
of $^{208}$Pb targets by $^{17}$O (22A MeV and 84A MeV) and by $^{36}$Ar
(95A MeV), as a function of the excitation energy.
\label{a-f1}}
\caption[ ]
{A nuclear target is Coulomb excited by a fast moving
projectile.  The coordinates used in text
are shown.\label{a-f2}}
\caption[ ]
{Electric dipole number of equivalent photons per unit
area $d^2b \equiv 2 \pi
b \, db$, with energy of 10~MeV, incident on $^{208}$Pb in a
collision
with $^{16}$O at impact parameter $b= 15$~fm, and as a function
of the bombarding energy in MeV per nucleon. The dotted line and the
dashed line correspond to calculations performed with the non-relativistic
and with the relativistic approaches, respectively. The solid line represents
a more correct calculation, as described in the text.
\label{a-f3}}
\caption[ ]
{Same as Fig.~\protect\ref{a-f3}, but for the
$E2$ multipolarity. 
\label{a-f4}}
\caption[ ]
{Same as Fig.~\protect\ref{a-f3}, but for the $M1$ mul\-tipolarity.
\label{a-f5}}
\caption[ ]
{Equivalent photon numbers per unit area incident on $^{208}$Pb,
in a collision with $^{16}$O at $100$A~MeV and with impact
parameter $b=15$~fm, as a function of the photon energy $\hbar
\omega$.
The curves for the $E1$, $E2$ and $M1$ multipolarities are shown.
\label{a-f6}}
\caption[ ]
{Total cross sections for the excitation of giant electric
dipole ($E1$) and quad\-rupole ($E2$) resonances in $^{208}$Pb
by means of the Coulomb interaction with
$^{16}$O, as a function of the laboratory energy.
\label{a-f7}}
\caption[ ]
{Ratio to the Rutherford cross section of the
    elastic cross section for the $^{17}$O + $^{208}$Pb reaction at
    84A MeV, as a function of the center-of-mass scattering
    angle.  Data are from Ref.~\protect\cite{Bar88}.
\label{a-f8}}
\caption[ ]
{Differential cross section for the excitation of the isovector
    giant dipole resonance in $^{208}$Pb by means of $^{17}$O projectiles
    at 84A MeV, as a function of the center-of-mass scattering
    angle. Data are from Ref.~\protect\cite{Bar88}.
\label{a-f9}}
\caption[ ]
{Virtual photon numbers for the electric dipole multipolarity
    generated by 84A MeV $^{17}$O projectiles incident
    on $^{208}$Pb, as a function of the center-of-mass scattering
    angle. The solid curve is a semiclassical calculation. The
    dashed and dotted curves are eikonal calculations
    with and without relativistic corrections, respectively.
\label{a-f10}}   
\caption[ ]
{    Differential cross sections for the excitation of the giant dipole
    resonance (GDR), the isoscalar giant quadrupole reso\-nance
(GQR$_{is}$), and
    the isovector giant quadrupole reso\-nance (GQR$_{iv}$), in Pb for
    the collision $^{208}$Pb+$^{208}$Pb at 640A MeV.
    The solid (dotted) [dashed-dotted] line is the differential
    cross section for
    the excitation of the GDR (GQR$_{is}$) [GQR$_{iv}$]. The dashed line
    is the result of a semiclassical calculation.
\label{a-f11}}
\end{figure}
\begin{figure}[th]
\caption[ ]
{    Calculated cross section for the excitation followed by $\gamma$-decay
of
    $^{208}$Pb induced by $^{17}$O projectiles at 84A MeV.
    The photoabsorption cross section was parameterized by a simple
Lorentzian
    representing the GDR,
    and the statistical component of the photon decay was neglected.
    The solid curve uses the formalism described in the text
(Eq.~\protect\ref{dsig3})
    while the dashed curve uses a constant branching ratio for photon decay
    (Eq.~\protect\ref{dsig2}).
\label{a-f12}}
\caption[ ]
{     Cross section for the excitation of
the GDR without the detection of the decay photon. Data are from
Ref.~\protect\cite{Bee90}.
\label{a-f13}}
\caption[ ]
{Cross section for excitation followed by $\gamma$-decay of
    $^{208}$Pb
    by $^{17}$O projectiles at 84A MeV.
    The solid (dashed) line includes (excludes) the Thomsom scattering
    amplitude. Data are from Ref.~\protect\cite{Bee90}.
\label{a-f14}}   
\caption[ ]
{The GDR Coulomb excitation probabilities as functions of the
               impact parameter, for sharp and smooth absorptions.
               The system is $^{208}$Pb (640A MeV) + $^{208}$Pb.
\label{a-f15}}
\caption[ ]
{Nuclear excitation probabilities, as  functions of the
               impact parameter, of the isoscalar giant monopole resonance
(GMR$_{is}$), the GDR$_{iv}$, and the GQR$_{is}$,
in $^{208}$Pb for the collision $^{208}$Pb+$^{208}$Pb at 640A MeV.
\label{a-f16}}
\caption[ ]
{Experimental 1n- and 3n- removal cross sections
for $^{197}$Au bombarded with relativistic projectiles (from
Ref.~\protect\cite{abs}) in comparison with theoretical calculations
from this work (solid curve: "soft- spheres" model; dotted curve:
"sharp-cutoff" model with $b^{BCV}_{min}$ from
Eq. (\protect\ref{b_bcv})).
For completeness, we also show a sharp-cutoff calculation
with $b^{Kox}_{min}$ (Eq. (\protect\ref{b_kox})) used in
Ref. \protect\cite{Au93}.
\label{a-f17}}
\end{figure}
\begin{figure}[h]
\caption[ ]
{A nuclear target is Coulomb excited by a fast moving
deformed projectile.  Besides the
angle $\theta$,  the orientation of the projectile also
includes an azimuthal angle $\phi$ which can rotate
its symmetry axis
out the scattering plane. For simplicity, this is not shown.
$\chi$ is the angular position of the c.m. of the projectile with
respect to the target.
\label{a-f18}}
\caption[ ]
{Percent  increase of the Coulomb excitation cross section
of dipole states in $^{208}$Pb due to the  dependence of
the minimum impact parameter on the deformation. The effect is shown for
$^{238}$U projectiles at
100A MeV, 1A GeV and 10A GeV, respectively, and
as a function of the deformation parameter $\beta$. The solid (dashed)
[dotted] line corresponds to  an excitation energy of 1 (10) [25] MeV.
For the actual deformation of $^{208}$U, $\beta \simeq 0.3$, the
effect is small.
\label{a-f19}}
\caption[ ]
{Coulomb excitation cross section
of a  giant dipole resonance  in $^{208}$Pb due to the  quadrupole-dipole
interaction with 100A MeV uranium projectiles, as a
function of the deformation parameter $\beta$.
These cross section are averaged over
all possible orientations of the projectile.
\label{a-f20}}
\caption[ ]
{Experimental results for $^{136}$Xe projectile excitation (at 690A~MeV)
on a Pb target (squares) and a C target (circles). The spectrum for
the C target is multiplied by a factor 2 for better presentation.
The resonance energies for one- and two-phonon giant resonances are
indicated. The dashed curve reflects the results of a first-order
calculation for the Pb target. Fig. is taken from
Ref.~\protect\cite{Sc93}.
\label{a-f21}}
\end{figure}
\begin{figure}[th]
\caption[ ]
{Compilation of experimental findings with heavy ion (full symbols)
and pion induced (open symbols) reactions for the energy, width, and cross
sections of the double giant resonance. The data are compared to the
energies and widths of the giant dipole resonance, respectively, and to the
theoretical values of excitation cross sections.
\label{a-f22}}
\caption[ ]
{Schematic representation of the excitation of
                 Giant Resonances, populated in heavy ion collisions.
\label{a-f23}}
\caption[ ]
{Time-dependence of the occupation probabilities
               $|a_0|^2$ and $|a_1|^2$, in a collision with impact
               parameter $b=15$~fm. The time is measured in terms of
               the dimensionless variable $\tau=(v\gamma/b)\ t$.
               The system is $^{208}$Pb (640A MeV) + $^{208}$Pb.
\label{a-f24}}
\caption[ ]
{Excitation energy spectra  of the main Giant Resonances
               for both Breit-Wigner and Lorentzian line shapes.
               The system is $^{208}$Pb (640A MeV) + $^{208}$Pb.
\label{a-f25}}
\caption[ ]
{Ratio between the DGDR and the GDR cross sections in
               $^{208}$Pb + $^{208}$Pb collisions, as a function of the
               bombarding energy.
\label{a-f26}}
\caption[ ]
{Dependence of $\sigma_{\mbox{\tiny GDR}}$ and $\sigma_{\mbox{\tiny
DGDR}}$ on the
               GDR width, treated as a free parameter. For details
               see  the text.
               The system is $^{208}$Pb (640A MeV) + $^{208}$Pb.
\label{a-f27}}
\caption[ ]
{Cross sections of the direct photoexcitation of two-phonon
configurations
$[1^{-} \otimes 1^{-}]_{2^{+}}$,
$[1^{-} \otimes 2^{+}]_{1^{-}}$ and
$[2^{+} \otimes 3^{-}]_{1^{-}}$ from the ground state in $^{136}$Xe and
$^{208}$Pb.
\label{vp-f1}}
\caption[ ]
{Photoexcitation cross section of the GDR in $^{136}$Xe and $^{208}$Pb.
Calculations are performed: a) within one-phonon approximation and
b) with taking into account of the coupling between one- and two-phonon
configurations.
Continues curves in the bottom part are the strength functions
calculated with a smearing parameter $\Delta = 1$~MeV; dashed
curve corresponds to electromagnetic transitions to one-phonon   
$1^-$ states, solid curve -- to one- and two-phonon $1^-$ states.
\label{vp-f2}}
\caption[ ]
{Photoneutron cross sections in $^{208}$Pb. Solid
curve is the result of calculation with the wave function
including one- and two-phonon terms presented with a smearing parameter
$\Delta = 1$~MeV; vertical lines (in arbitrary units) -- within
one-phonon approximation. Experimental data are plotted by experimental
error bars.
\label{vp-f3}}
\caption[ ]
{Photo-neutron cross section for $^{208}$Pb. Experimental data (dots
with experimental errors) are from Ref.~\protect\cite{Bel95}. The
long-dashed curve is the high energy tail of the GDR, the short-dashed curve
is the GQR$_{iv}$ and the curve with squares is their sum. The contribution
of two-phonon states is plotted by a curve with triangles. The solid curve
is the total calculated cross section.
\label{vp-f4}}
\caption[ ]
{The possible paths to the excitation of
a given magnetic substate of the $1^+$ component of
the DGDR are displayed.
The  transitions caused by the different projections of the operator
$V_{E1}$ are shown by: (a) dashed lines for $m=0$, (b) dashed-dotted
lines for $m=-1$, and (c) solid-lines are $m=+1$.
\label{vp-f5}}
\end{figure}
\begin{figure}[th]
\caption[ ]
{The electromagne\-tic excitation cross sections for the reaction
$^{208}$Pb (640A MeV) $+^{208}$Pb calculated in
coupled-channels.
It is shown the excitation of the GDR (top) and the three components
$J^{\pi}$~=~$0^+$,
$2^+$ and $1^+$ of the DGDR. The $B(E1)$ strength distribution (in
arbitrary
units) over $1^-$ states is shown by dashed lines. For a visuality
it is shif\-ted up by 100~keV.
\label{vp-f6}}
\caption[ ]
{The cross section for Coulomb excitation of the one-phonon GDR
(continuous curve), of the isoscalar GQR (dash-dotted), of the
isovector GQR (long dashed)
as well as for the double-phonon GDR (short dashed) are shown.   
They have been calculated at $E_{lab} =681$A~MeV, taking into
account the energy reduction of the beam in the target
\protect\cite{Sc93}.
The one-phonon GDR cross section  has been reduced in the
figure by a factor 10.
\label{vp-f7}}
\caption[ ]
{Fragmentation of the most collective a) one-phonon
1$^-$ and b) two-phonon [$1^- \otimes 1^-$] configurations in $^{136}$Xe
due to the coupling to more complex configurations.
The result are presented with a smearing parameter $\Delta = 0.2$~MeV.
\label{fragm}}
\caption[ ]
{a) $B(E1)$ values for the GDR and b,c) $[B(E1) \times B(E1)]$ values
Eq.~(\protect\ref{trans})
for the DGDR associated
with Coulomb excitation in $^{136}$Xe
in relativistic heavy ion collision.
b) and c) correspond to $J = 0^+$ and $J = 2^+$ components of the DGDR,
respectively.
A smooth curve is a result of averaging over all states
with a smearing parameter $\Delta$~=~0.5~MeV.
See text for details.
\label{vp-f8}}
\caption[ ]
{a) $B(E1)$ values for the GDR and b) $[B(E1) \times B(E1)]$ values  
Eq.~(\protect\ref{trans})
for the DGDR ($J = 0^+ + 2^+$) associated
with Coulomb excitation in $^{208}$Pb
in relativistic heavy ion collision.
A smooth curve is a result of averaging over all states
with a smearing parameter $\Delta$~=~0.5~MeV.
\label{vp-f9}}
\caption[ ]
{The contribution for the excitation of two-phonon $1^-$ states
(long-dashed curve) in first order perturbation theory, and for
two-phonon $0^+$ and $2^+$ DGDR states in second order (short-dashed
curve). The total cross section (for $^{208}$Pb (640A~MeV)
$+^{208}$Pb) is shown by the solid curve.
\label{vp-f10}}
\caption[ ]
{The energy dependence of the $^{208}$Pb
(640A~MeV)~+~$^{208}$Pb reaction function calculated within
first order perturbation theory. The square indicates the location of
the GDR in $^{208}$Pb.
\label{vp-f11}}
\caption[ ]
{The cross section for the excitation of the $2^{+}$ component of
the DGDR in the reaction $^{208}$Pb (640A MeV) + $^{208}$Pb,
calculated within second order perturbation theory. The dashed curve shows
the contribution of the $E1$-transition between simple GDR and DGDR
configurations only. The solid curve is a sum of the above result and the
contribution of the $E1$-transitions between complex GDR and DGDR
configurations. See text for details.
\label{vp-f12}}
\caption[ ]
{The possible paths to the excitation of
a given magnetic substate of the $0^+$ and $2^+$ components of
the DGDR in spherical and deformed nuclei.
The notations are the same as in
Fig.~\protect\ref{vp-f5}.
\label{vp-f13}}
\end{figure}
\begin{figure}[th]
\caption[ ]
{The $B(E1)$ strength distribution over $K=0$ (short-dashed curve) and
$K=\pm 1$ (long-dashed curve) $1^{-}$ states in $^{238}$U. The solid curve
is their sum. The strongest one-phonon $1^{-}$ states are shown by vertical
lines, the ones with $K=0$ are marked by a triangle on top. Experimental
data are from Ref.~\protect\cite{Gur76}.
\label{vp-f14}}
\caption[ ]
{The strength functions for the excitation: a) of the GDR, and b) of
the DGDR in $^{238}$U in the $^{238}$U (0.5A GeV) + $^{208}$Pb
reaction. In a), the short-dashed curve corresponds to the GDR ($K=0$) and
the long-dashed curve to the GDR ($K=\pm 1$). In b) the dashed curve
corresponds to the DGDR$_{0^{+}}$ ($K=0$), the curve with circles to the
DGDR$_{2^{+}}$($K=0$), the curve with squares to the DGDR$_{2^{+}}$
($K=\pm 1$),
and the curve with triangles to the DGDR$_{2^{+}}$ ($K=\pm 2$) . The solid
curve is the sum of all components. The strength functions are calculated
with the smearing parameter equal to 1 MeV.
\label{vp-f15}}
\end{figure}

\end{document}